\documentclass{jfm}
\usepackage{graphicx}
\usepackage{epstopdf, epsfig}
\usepackage{float}
\usepackage{amsmath}
\usepackage{MnSymbol}
\usepackage[percent]{overpic}
\usepackage{xcolor}
\usepackage{lineno}
\usepackage{ulem}
\DeclareMathOperator{\sech}{sech}

\newcommand\rev[1]{{#1}}
\newcommand\rgm[1]{{#1}}
\newcommand\finalrev[1]{{#1}}
\newcommand\rms{\mathrm{rms}}
\newcommand{\mylab}[3]{\raisebox{#2}[0mm][0mm]{\makebox[0mm][l]{\hspace*{#1}#3}}}

\shorttitle{Turbulence over porous and rough substrates}
\shortauthor{Z. Hao and R. Garc{\'i}a-Mayoral}

\title{Turbulent flows over porous and rough substrates}

\author{Zengrong Hao\aff{1}
\and Ricardo Garc{\'i}a-Mayoral \aff{1} \corresp{\email{r.gmayoral@eng.cam.ac.uk}}}

\affiliation{\aff{1}Department of Engineering, University of Cambridge, Trumpington St., Cambridge CB2 1PZ, UK}

\begin{document}

\maketitle

\begin{abstract}
Turbulent flows over porous substrates are studied via a systematic exploration of the dependence of the flow properties on the substrate parameters, including permeability $K$, grain \rgm{pitch} $L$, and depth $h$. The study uses direct numerical simulations mainly for staggered-cube substrates with 
$L^+\approx10$ - $50$, $\sqrt{K}/L\approx0.01$ - $0.25$, and depths from $h=O(L)$ to $h\gg L$, ranging from typical impermeable rough surfaces to deep porous substrates. The results indicate that the permeability has significantly greater relevance than the grain size \rev{and microscale  topology} for the properties of the overlying flow, including the mean-flow slip and the shear across the interface, the drag increase relative to smooth-wall flow, and the statistics and spectra of the overlying turbulence, whereas the direct effect of grain size is only noticeable near the interface as grain-coherent flow fluctuations.
The substrate depth also has a significant effect, with shallower substrates suppressing the effective transpiration at the interface.
\rgm{Based on the direct-simulation results,} we propose an empirical `equivalent permeability' $K_{eq}^t$, that incorporates this effect and scales well the overlying turbulence for substrates with different depths, \rgm{permeabilities, etc. This result suggests that wall normal transpiration driven by pressure fluctuations is the leading contributor to the changes in the drag and the overlying turbulence}. Based on this, we propose a conceptual $h^+$-$\sqrt{K^+}$ regime diagram where\rgm{, for any given substrate topology,} turbulence transitions smoothly from that over impermeable rough surfaces with ${h=O(L)}$ to that over deep porous substrates with $h^+\gtrsim50$, with the latter limit determined by the typical lengthscale of the overlying pressure fluctuations.
\end{abstract}


\section{Introduction}
\label{Sec:Intro}
Turbulent flows over porous substrates are prevalent in both nature and engineering. They play a central role in diverse problems in environmental science (e.g. forest winds, soil evaporation, sediment transport in water, and riverbed/seabed erosion), aerospace engineering (e.g. surface treatment for drag and noise reduction and boundary layer control), chemical engineering (e.g. heat/mass transfer enhancement in catalyst layers, fluidized beds, and nuclear reactors), metallurgical engineering (e.g. industrial painting and metal foam processing), and light industry (e.g. food dehydration). This subject is characterised by a coupling of two systems originally with distinctively different nature -- the turbulent boundary layer, which features vigorous fluctuations, inertia-dominated inter-scale energy transfer, and self-organised flow structures, and the porous medium, in which flow is relatively creeping and viscosity-dominated, and large flow structures spanning many pores are strongly impeded. Such contrast indicates an acute transition of flow behaviour across the interface between the overlying and the subsurface flow, which involves multiple mechanisms with a broad range of characteristic scales. In recent decades, this subject has been increasingly attracting experts from various traditional communities of fluid mechanics, including wall turbulence, flow instability, free shear turbulence, low-Reynolds-number flows, transport phenomena, and chaotic systems. Some typical effects of porous substrates, such as drag increase and mixing enhancement, have been extensively studied in diverse scenarios. However, due to the complicated interplay between different mechanisms, our understanding of the general dependence of the flow behaviour on different characteristics of a porous substrate is still vague.

The most distinctive characteristic of a porous substrate is permeability, i.e.~ the ability for the overlying flow to penetrate into the substrate. \citet{Jimenez01} represented permeability using a boundary condition where the transpiration, i.e.~ the wall-normal velocity at a notional interface plane, was proportional to the local instantaneous overlying pressure, with a `porosity coefficient' of proportionality. Their direct numerical simulations (DNSs) showed that such boundary conditions cause the onset of large spanwise rollers associated to a
Kelvin-Helmholtz-like (K-H) instability, resulting in an increase in mixing and drag. This boundary condition is a reasonable characterisation of substrates where
the flow can travel freely through a plenum below the substrate \citep{Kawano2021}.

A practical parameter to characterise substrate permeability is the bulk permeability $K$ of the porous medium, which is defined as $K\equiv-\nu U/\partial_xP$, where $U$ is the mean velocity induced by a uniform mean pressure gradient $\partial_xP$ and $\nu$ is the kinematic viscosity \citep{Darcy56}.
\rgm{Note that here, and elsewhere throughout the paper, any pressure we refer to is the kinematic pressure, that is, the actual, static pressure divided by the density. The above expression for $K$} can be obtained by volume-averaging the corresponding pore-resolved solution assuming Stokes flow, and $\sqrt{K}$ is a characteristic permeability lengthscale.
\citet{Breugem06} systematically investigated the influence of $K$ on the overlying turbulence. They conducted DNSs for overlying flows while using a volume-averaged-Navier-Stokes (VANS) approach to model subsurface flows. Their results showed that turbulence differs little from smooth-wall flows for $\sqrt{K^+}\lesssim0.3$, where the superscript~`$+$' denotes wall-unit scaling. As $\sqrt{K^+}$ increases up to $\sqrt{K^+}\approx9$, typical near-wall structures like low-speed streaks and quasi-streamwise vortices, are gradually destroyed. The significance of $\sqrt{K^+}$ to near-wall structures was confirmed by the experiments of \citet{Suga10,Suga11} for $\sqrt{K^+}\approx1$ - $11$. They found that the increase of $\sqrt{K^+}$ tends to intensify sweep events and weaken ejection ones near the interface, and presented a conceptual model for the associated destruction of structures. A broader range $\sqrt{K^+}\approx2$ - $17$ was experimentally investigated by \citet{Manes11}. Based on the evolution of near-wall vortical structures with $\sqrt{K^+}$, they proposed a theory to describe the competition between two types of eddies: the typical smooth-wall-like eddies, and the spanwise-elongated eddies induced by the K-H instability. The onset and development of this instability was theoretically modeled by \citet{Abderrahaman17}, \citet{Sharma17}, and \citet{GmezdeSegura2018} in a more general situation with anisotropic permeability, establishing a criterion for its onset of $\sqrt{K^+_y}\gtrsim0.3$ - $0.4$, where $K_y$ is wall-normal permeability. 
\rgm{\citet{Khorasani2024} have recently corroborated this value in texture-resolving DNSs of mesh-like anisotropic permeable substrates.}
Focusing on low permeability in a range $\sqrt{K^+}\approx0.05$ - $0.7$, \citet{Rosti15} conducted VANS-based simulations and suggested a similar critical value, $\sqrt{K^+}\approx0.2$, below which almost all the flow statistics are indistinguishable from those of a smooth wall. \citet{Voermans17} measured more flow details across the substrate interface for $\sqrt{K^+}\approx0.3$ - $6$.
\rgm{They could obtain detailed measurements of the flow near the interface and in the subsurface region, not usually 
accessible in experiments, by matching the refraction index between the flow and the solid inclusions of the substrate
\citep{Rousseau2020}.}
The results confirmed the strong dependence of various interfacial flow properties and penetration depths on $\sqrt{K^+}$. More recently, \citet{Wang21} investigated the transfer of information across the interface. They found strong asymmetry between top-down and bottom-up transfer in terms of both scale and strength for $\sqrt{K^+}\sim O(1)$, which provided a novel perspective on understanding the turbulence-subsurface flow interaction.

Since real-world porous media are composed of grains or inclusions of finite size 
\rgm{and pitch} $L$, the free-flow/substrate interface has an irregular topography, and thus a porous substrate also exhibits some features of surface roughness. Some research has aimed to understand the role of surface roughness in the problem of turbulence over porous substrates. A natural strategy to approach this issue is to compare between porous substrates and impermeable rough surfaces with analogous interfacial topography. Such discussions can be at least traced back to \citet{Zagni76}, \citet{Kong82}, and \citet{Zippe83}. In their experiments, the corresponding rough surface was obtained by placing a flat and smooth plate, which we term `floor' in this paper, just below the first layer of grains that constitute a porous substrate, so the former has a depth $h\sim L$ in contrast to the latter where $h\gg L$. They all observed that a porous substrate induces higher drag than its rough counterpart. 

The role of surface roughness in turbulence over porous substrates has drawn more attention in recent decades. \citet{Manes09} explored this issue using particle image velocimetry (PIV). They considered porous substrates and rough surfaces consisting of regularly packed spheres with $\sqrt{K^+}\approx31$ - $45$, $L^+\approx260$ - $370$, and $h/L\approx5$ and $1$. Their measurements confirmed that porous substrates had higher drag coefficients, and they proposed that the intense downward transport of turbulent kinetic energy by pressure fluctuations is an important feature distinguishing porous substrates from impermeable rough surfaces. A similar argument has been proposed by \citet{Karra2023} using fully-resolved DNSs for substrates constituted by randomly packed spheres with  $\sqrt{K^+}\approx3$-$9$, $L^+\approx80$-$300$, and $h/L\approx1$-$4$. Also using fully-resolved DNSs, \citet{Kuwata16a,Kuwata16b} compared the flow structures over porous and rough surfaces consisting of staggered cubes with $\sqrt{K^+}\approx3$, $L^+\approx50$, and $h/L\approx5$ and $1$. They argued that the porous surface has stronger spanwise eddies originating from the K-H instability but weaker near-wall streaks. This is reminiscent of the `competing mechanism' in \citet{Manes11}, where they discussed the influence of $\sqrt{K^+}$. 
\rgm{\citet{Cooper2017} conducted experiments of turbulence over permeable substrates and impermeable surfaces with replicated interfacial topography, suggesting that momentum transfer was more efficient in the permeable case, and Reynolds stresses higher.}
\citet{Fang18} considered substrates composed of very large grains with $\sqrt{K^+}\approx1$ - $100$, $L^+\approx250$ - $3000$, and $h/L\approx3$ and $0.5$ using large eddy simulations (LES). Their results suggested that the flow behaviour near the interface depends more on $\sqrt{K^+}$ than $L^+$ even for such large grains. To separate the permeability and roughness effects, \citet{Esteban22} provided more experimental data in the ranges $\sqrt{K^+}\approx1$ - $60$, $L^+\approx10$ - $500$, and $h/L\approx O(10)$, $3$, and $1$. A generic formulation to predict drag increase was proposed based on an analogy between the roles of $\sqrt{K^+}$ and $L^+$ in the problem. This generic formulation was partially verified more recently by \citet{Wangsawijaya23}. They overlaid external roughness over the surface of a permeable substrate. The substrate had large permeability $\sqrt{K^+}\approx10$ - $30$ but small grains, while the mesh-like roughness had very large mesh \rgm{pitch} $L^+\gtrsim 5000$. Their results suggested that the drag increase for the composite permeable-and-rough surface could be characterised by the scale $\sqrt{K^+}L^+$.

To understand the role of surface roughness in turbulence over porous substrates, \citet{Kim20} instead polished the substrate interface, in essence eliminating its rough character. For $\sqrt{K^+}\approx50$ and $L^+\approx1000$, the comparison between the original, unpolished substrate and the polished one showed that the latter induces higher drag than the former. They attributed the difference primarily to the roughness-coherent flow present in the former. 
A similar strategy was used in \citet{Shen20}. For $\sqrt{K^+}\approx3$ and $L^+\approx80$. They compared two substrates with a regularly packed and a randomly packed surface layer, respectively, suggesting that the interfacial topographic details affect the flow dynamics.

Overall, the works cited above suggest that turbulence over porous substrates is affected by three characteristics of the substrate. The first is the permeability of the porous medium, characterised by the bulk parameter $K$. This is a macroscale property, in the sense
that it emerges from a volume-average over scales larger than the grain \rgm{pitch}, and therefore does not reflect the details of
the geometry at the grain size, or microscale.
 The permeability controls the general degree of penetration of the overlying turbulent fluctuations into the substrate. The second characteristic is the granularity of the porous medium, which refers to those microscale features directly associated with the geometrical detail of individual grains. The granularity induces grain-coherent fluctuations in the flow, especially near the interface, with characteristic length scale $L$. The third characteristic is the substrate depth, $h$. A finite depth tends to suppress the penetration of the overlying flow, thus counteracting the effect of the bulk permeability. For the three characteristics, the literature generally suggests that the permeability and the granularity of a porous medium have some similar phenomenological effects on its overlying flow, such as intensifying the near-wall turbulence and increasing the drag, while these effects are attenuated if the substrate is not sufficiently deep.
 
 In the present paper we aim to characterise the above general trends quantitatively.
One difficulty lies in separating the effects of permeability and granularity. In previous studies, the grain topology,
which determines the ratio $\sqrt{K}/L$, has typically little variation. $K$ and $L$ are then varied in synchrony,
making it difficult to separate their effects.
In addition, transpiration is known to be important not only for porous substrates, but also for roughness \rgm{\citep{Orlandi06,Orlandi06JoT,Orlandi08}}. Although strictly speaking rough surfaces are impermeable, in the sense that they do not allow flow through, they allow transpiration in the sense of non-zero wall-normal velocity at a notional interface plane at top of the roughness crests. This transpiration does not only occur at the microscale $L$, but also at the macroscale, for the typical sizes of the overlying turbulent eddies. The latter would be more intense for porous substrates,
but the question arises of whether the nature of the transpiration effect is different for porous and rough walls, or whether the difference is only in intensity, and a smooth transition can be observed between a porous substrate with $h\gg L$ and a corresponding rough surface, with $h\sim L$. 

To address these questions, we systematically explore the parameter space of $K$, $L$, and $h$ using DNS, aiming to understand the effects of permeability, granularity, and  substrate depth as independent parameters on the overlying turbulence. We limit the scope of this study to substrates composed of relatively small grains, $10\lesssim L^+\lesssim50$, i.e.~essentially in the transitionally rough regime. In this range, the overlying turbulence deviates from smooth-wall-like behaviour but the near-wall turbulent structures are not fully disrupted by the granularity of the substrate \citep{Abderrahaman19}. At $L^+\approx$ \rev{$20$-}$50$, the length scales of the overlying turbulence and the grain-coherent flow become comparable, and microscale and macroscale cannot be clearly separated
\finalrev{\citep{Fairhall2019,Xie2024}}. 

The paper is organised as follows. \S \ref{Sec:Methods} describes the  substrate configurations, numerical methods and simulation setup, and techniques for post-processing. \S \ref{Sec:Results} reports the general dependence of flow properties on the geometrical parameters of substrates. \S \ref{Sec:FurtherDiscussion} discusses the scaling of turbulence with substrate parameters. \S \ref{Sec:unification} investigates the transition from porous substrates to typical rough surfaces. \S \ref{Sec:Conclusions} concludes this paper.

\section{Methods}
\label{Sec:Methods}

\subsection{Configurations of substrates}
\label{SubSec:Configuration}

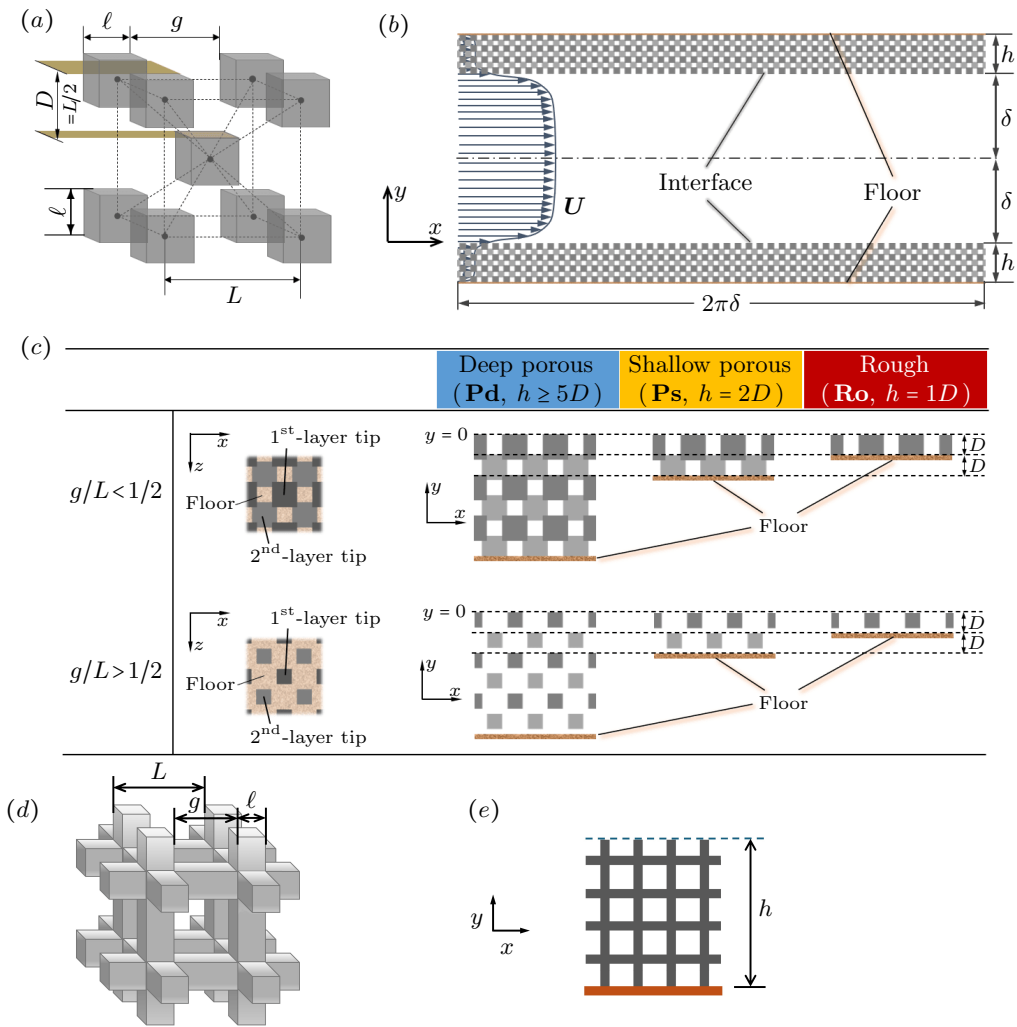
\begin{figure}
\centering
\vspace{6pt}
\begin{overpic}[width=0.96\textwidth]{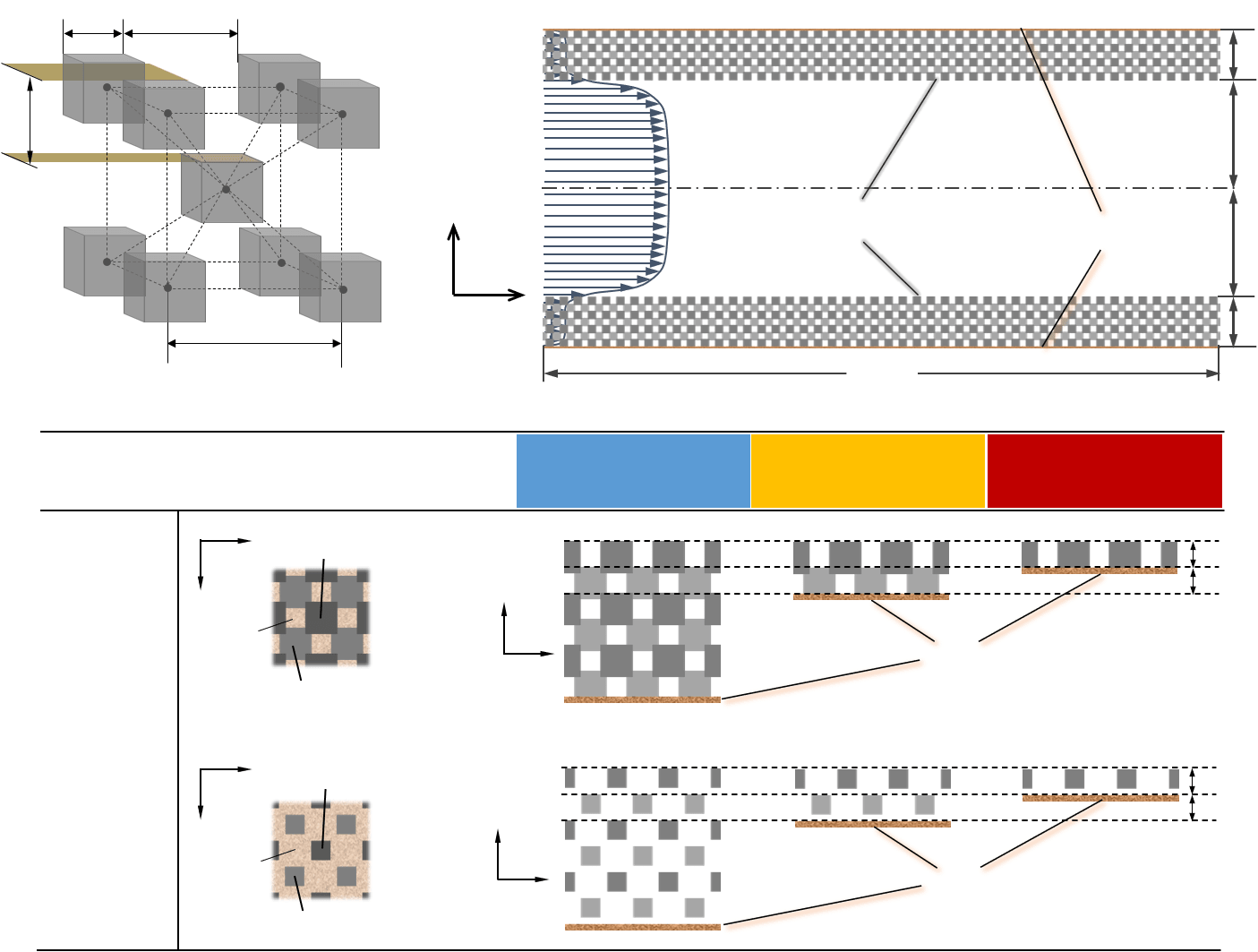}

 \put (-1.5, 74.0) {\footnotesize$(a)$}
 \put (13.9, 74.0) {\footnotesize$g$}
 \put (6.9, 73.8) {\footnotesize$\ell$}
 \put ( 0.2, 65.5) {\footnotesize\rotatebox{90}{$D$}}
 \put ( 2.8, 64.1) {\scriptsize\rotatebox{90}{$=\!\!L\!/2$}}
 \put (19.5, 46.0) {\footnotesize$L$}
 \put(1.0, 57.6) {\line(1,0){5}}
 \put(1.0, 52.9) {\line(1,0){4.1}}
 \put(3.7,52.9){\vector(0,1){4.7}}
 \put(3.7,57.6){\vector(0,-1){4.7}}
 \put (1.5, 54.7) {\footnotesize\rotatebox{90}{$\ell$}}

 \put (35.0, 73.5) {\footnotesize$(b)$}
 \put (36.7, 56.8) {\footnotesize$y$}
 \put (40.2, 53.0) {\footnotesize$x$}
 \put (54.0, 55.0) {\footnotesize$\boldsymbol{U}$}
 \put (98.6, 49.3) {\footnotesize$h$}
 \put (68.1, 45.2) {\footnotesize$2\pi\delta$}
 \put (98.6, 70.5) {\footnotesize$h$}
 \put (98.6, 64.0) {\footnotesize$\delta$}
 \put (98.6, 55.5) {\footnotesize$\delta$}
 \put (63.7, 57.5) {\footnotesize Interface}
 \put (84.6, 56.5) {\footnotesize Floor}
 
 \put (-1.5, 40.8) {\footnotesize$(c)$}
 \put ( 3.6, 26.2) {\footnotesize$g/L\!<\!1/2$}
 \put ( 3.6,  7.9) {\footnotesize$g/L\!>\!1/2$}
 \put (18.6, 31.4) {\scriptsize$x$}
 \put (16.2, 28.9) {\scriptsize$z$}
 \put (18.6, 13.2) {\scriptsize$x$}
 \put (16.2, 10.7) {\scriptsize$z$}
 \put (42.8, 24.1) {\scriptsize$x$}
 \put (40.4, 26.9) {\scriptsize$y$}
 \put (42.3,  6.1) {\scriptsize$x$}
 \put (39.9,  8.9) {\scriptsize$y$}
 \put (39.8, 32.1) {\scriptsize$y=0$}
 \put (39.6, 14.1) {\scriptsize$y=0$}
 \put (95.3, 30.9) {\scriptsize$D$}
 \put (95.3, 28.8) {\scriptsize$D$}
 \put (95.2, 12.9) {\scriptsize$D$}
 \put (95.2, 10.8) {\scriptsize$D$}
 
 \put (24.0, 31.8) {\scriptsize $1^\mathrm{st}$-layer tip}
 \put (22.0, 19.7) {\scriptsize $2^\mathrm{nd}$-layer tip}
 \put (15.4, 25.1) {\scriptsize Floor}
 \put (24.0, 13.4) {\scriptsize $1^\mathrm{st}$-layer tip}
 \put (22.0,  1.3) {\scriptsize $2^\mathrm{nd}$-layer tip}
 \put (15.4,  6.7) {\scriptsize Floor}
 
 \put (73.9, 22.7) {\scriptsize Floor}
 \put (73.9,  4.7) {\scriptsize Floor}

 \put (43.3, 39.0) {\footnotesize Deep porous}
 \put (42.6, 36.0) {\footnotesize ($\,\textbf{Pd},\;h\geq5D\,$)}

 \put (60.6, 39.0) {\footnotesize Shallow porous}
 \put (61.6, 36.0) {\footnotesize ($\,\textbf{Ps},\;h=2D\,$)}

  \put (84.2, 39.0) {\footnotesize \textcolor{white}{Rough}}
 \put (80.2, 36.0) {\footnotesize \textcolor{white}{($\,\textbf{Ro},\;h=1D\,$)}}
\end{overpic}

\includegraphics[width=0.75\linewidth]{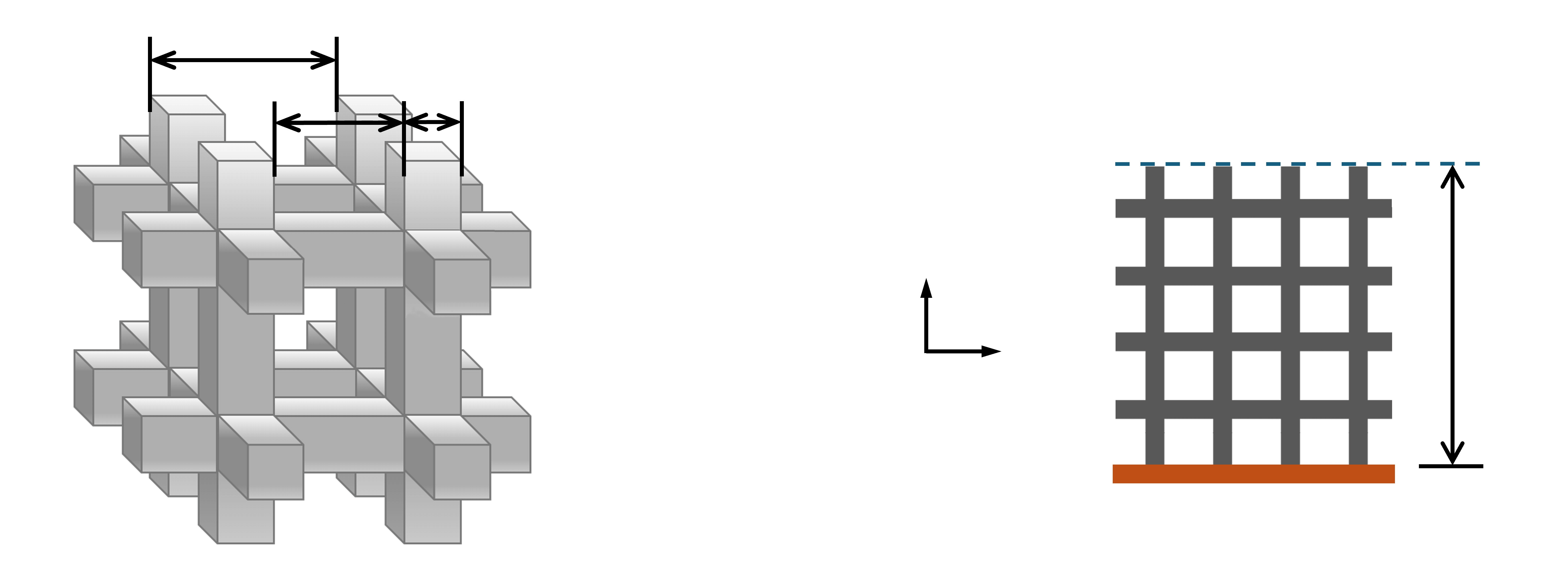}\hspace*{18mm}
\mylab{-125mm}{29mm}{$(d)$}
\mylab{-65mm}{29mm}{$(e)$}
\mylab{-108mm}{34mm}{$L$}
\mylab{-104.1mm}{30.2mm}{$g$}
\mylab{-97.5mm}{30.2mm}{$\ell$}
\mylab{-65mm}{11mm}{$x$}
\mylab{-70mm}{15mm}{$y$}
\mylab{-33mm}{16mm}{$h$}

\caption{Sketches of porous substrates and computational domain used in the DNSs. $L$ 
and $g$ are the grain \rgm{pitch} and the gap size, and $h$ is the substrate
depth. The wall-normal coordinate is set to $y=0$ at the interface of the bottom substrate with the
free flow, the plane of the tips of the top layer of cubes.
\rgm{$(a)$ Staggered-cube topology, where $D\equiv L/2$ is the thickness of one cube layer. $(b)$ Computational domain, with  dimensions $2\pi\delta$, $2(\delta+h)$, and $\pi\delta$ in $x$, $y$, and $z$, respectively.
$(c)$ Plan and side views of staggered-cube substrates of different depths.
$(d)$ Mesh topology.
$(e)$ Side view of a mesh substrate.
}
}
\label{fig:Configuration}
\end{figure}

The substrate configurations considered in this study are arrays of staggered solid cubes with grain \rgm{pitch $L$, inclusion width $\ell$, and gap size $g=L-\ell$}, as shown in figure \ref{fig:Configuration}, similar to those of \citet{Kuwata16a}.
Compared with collocated arrays, such as those in \citet{Breugem05}, the staggered arrays are more representative of randomly packed grains that are prevalent in realistic scenarios, where large gaps between grains would be occupied and blocked by other grains.
The gap-to-pitch ratio $g/L$ controls the connectivity of pores, which can be regarded as partially connected for $g/L<1/2$ and fully connected for $g/L>1/2$.
The porosity is \rgm{$\varepsilon=1-2(\ell/L)^3+\max[(2\ell/L-1)^3,0]=1-2(1-g/L)^3+\max[(1-2g/L)^3,0]$. 
Regular topologies as these have been used often in the literature to study porous substrates}
\rev{\citep{Breugem05,Breugem06,Manes09,Chandesris2009,Zhang2009,Liu2011,Jin2015,Kuwata16a,Kuwata16b,Fang18,Kim20,Wang21,Rao2022,Khorasani2024}.}
\rgm{They present the 
advantage that in macroscopically homogeneous flow only one pore unit of size $L$ is needed to obtain volume averaged quantities \citep{Breugem05}.}
\rgm{Concerns were raised by the reviewers of the paper about substrates with $g/L>1/2$ being made up of `frozen suspensions' of inclusions,
and thus not being realisable experimentally. This is indeed the case, but such substrates have nevertheless been used widely in the literature as idealised high-permeability topologies 
\citep{Prinos2003,Breugem05,Chandesris2009,Zhang2009,Liu2011,Jin2015,Lacis20,Naqvi21,Sudhakar2021,Wang21,Rao2022,AghaeiJouybari2024}.
In appendix~\ref{Sec:MicroFlow}, we show that there is no fundamental difference between the flow in these and in substrates made up
of interconnected inclusions, and that there is a continuum in the flow properties as $g/L$ increases above $1/2$ and the
inclusions cease to be interconnected. We have nevertheless included some additional simulations with mesh-like substrates,
similar to those of \citet{Khorasani2024}, which could also yield the desired high permeabilities, albeit at an increased
computational cost. These substrates were designed so that they exhibited a rough interface with protrusions, like our cube topologies, as shown in figure \ref{fig:Configuration}$(d)$.}

\begin{figure}
  \centerline{\includegraphics[width=0.5\linewidth]{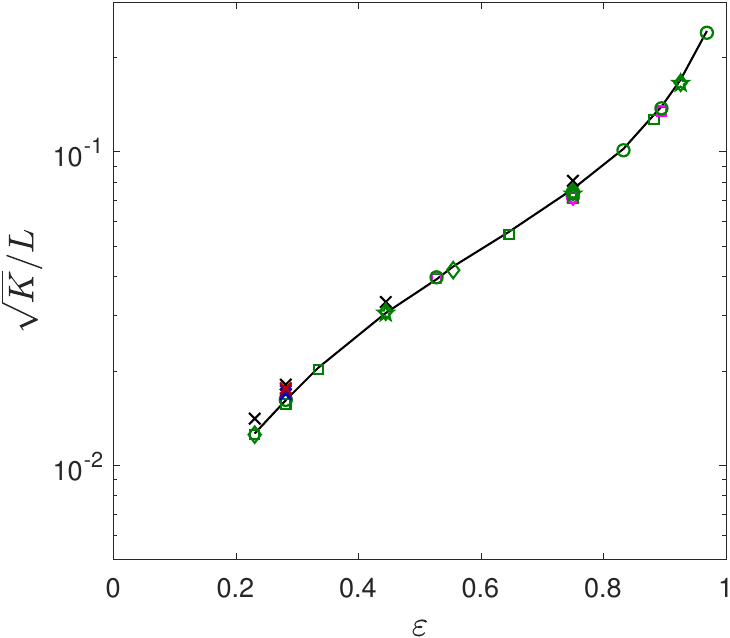}}
  \caption{Porosity-permeability relationship for the staggered-cube configuration considered in this study. The \rev{green} markers represent the \textit{a posteriori} values resulting from the DNSs for all deep porous (Pd) substrates, \rev{with those at $Re_\tau\approx360$ and $Re_\tau\approx550$ in magenta and purple.}
  The black line represents the \textit{a priori} values obtained from Stokes-flow simulations \rev{at the same resolution}.
  \rev{The blue to red crosses are \textit{a priori} results for increasing resolution and the black ones values kindly provided by one of the manuscript reviewers, as detailed in Appendix \ref{Sec:GridDependence}.}
  Symbols: $\medstar$ for $L^+\approx12$; $\medcircle$ for $L^+\approx24$; $\meddiamond$ for $L^+\approx36$; $\medsquare$ for $L^+\approx48$.}
\label{fig:sqrtK_epsilon}
\end{figure}

\rgm{For each topology and} value of $g/L$ or $\varepsilon$, the bulk permeability $K\equiv-\nu U/\partial_xP$ is obtained via a simulation of Stokes flow driven by a uniform pressure gradient $\partial_xP$ and yielding a mean velocity $U$, \rev{as in \citet{Sharma20dense}. We refer to this as the \textit{a priori} permeability. Alternatively, the value of $K$ can also be estimated from the Darcy region in a DNS. We refer to this as the \textit{a posteriori} permeability.
In the literature, the former is often denominated the intrinsic permeability, and the latter the effective permeability. For consistency, the Stokes computations were conducted
with the same resolution per pitch as the DNSs -- see Appendix \ref{Sec:GridDependence} for a discussion on spatial resolution. Both Stokes-flow and DNS values of $K$}
are displayed \rgm{for our staggered-cube topologies} against the porosity $\varepsilon$ in figure \ref{fig:sqrtK_epsilon}, showing no significant discrepancies between the two. The figure also shows that a change of $g/L$ in the range $2/9$ - $3/4$ or $\varepsilon$ in $0.23$ - $0.97$ changes $\sqrt{K}/L$ in $0.013$ - $0.243$, more than one order of magnitude.

The interface of a staggered-cube configuration can be defined as the plane through the tips of the first layer of cubes and set to have $y=0$. The thickness $D$ of a grain layer, which is defined as the distance between the tips of two adjacent layers, is $D=L/2$. The bottom of the substrate is a smooth wall, termed `floor',  at $y=-h$, where $h$ is the substrate depth. In this paper, we primarily consider three categories of substrates based on the ratio of the depth $h$ to the grain-layer thickness $D$: deep porous substrates (Pd), with $h/D\geq5$, shallow porous substrates (Ps), with $h/D=2$, and rough surfaces (Ro), with $h/D=1$. The floor of our rough surfaces is thus at $y=-D=-L/2$, exactly through the tips of the second grain layer for the corresponding deep and shallow porous layouts, where the flow first perceives blockage after penetrating the interface from above.  Note that this corresponding rough surface is slightly different from the one
in \rgm{\citet{Breugem05} or} \citet{Kuwata16a}, who set the floor at the bottom of the first layer of cubes.
\rev{The \textit{a posteriori} permeability was obtained from our deep-substrate DNSs, Pd, from the mean pressure gradient and the bulk velocity in the third layer of grains from the floor.}
\rgm{Our aim is to investigate if and how the character of a given substrate topology changes gradually from being simply 
rough to being fully permeable, as its depth progressively increases. Although our interest is mainly academic, there are 
practical cases where substrates would exhibit an intermediate character, being permeable yet thin, e.g. for thin sediment beds
or in porous surface coatings or paints.}
\rev{We note that, as detailed in \S \ref{Sec:Intro}, porous substrates and rough walls with equivalent grain topology or superficial texture have been 
previously compared extensively, likewise for regular 
\citep{Manes09,Kuwata16a,Kuwata16b,Fang18} and irregular topologies \citep{Zagni76,Zippe83,Cooper2017,Karra2023}, and both experimentally and numerically.
One of the paper's reviewers raised concerns, especially for our rough and shallow substrates, about regular topologies having a uniform
superficial texture, while irregular ones would have unique, heterogeneous distributions that would make generalisations
questionable. It is, however, well established for rough walls, which is what our substrates with $h/D=1$-$2$ essentially are, that
heterogeneities at the grain level do not result in significant differences in the flow over different samples of the same surface
\citep{Jimenez04,Chung21}. Surface heterogeneity only becomes relevant when it occurs over significantly larger lengthscales, 
typically comparable to the boundary layer thickness, as studied recently in \citet{Li2021}, \citet{Wangsawijaya2020} and
\citet{Stroh2020} and reviewed in \citet{BouZeid2020} and \citet{GarciaMayoral2024}.
}

\subsection{Numerical methods and simulation setup}
\label{SubSec:Numerical}

The direct numerical simulations in this paper are conducted in channels bounded by a pair of parallel substrates with identical configurations. The two substrates are symmetric about the channel central plane and the distance between their interfaces is $2\delta$, as portrayed in figure \ref{fig:Configuration}$(b)$. The size of the channel is $2\pi\delta$ in the streamwise ($x$) direction and $\pi\delta$ in the spanwise ($z$) direction. Periodic boundary conditions are applied in $x$ and $z$. A constant mean pressure gradient $\partial_xP<0$ is imposed in $x$ in the entire domain to drive the channel flow.  The substrate parameters in wall units are based on the kinematic viscosity $\nu$ and the friction velocity $u_\tau$ measured at the interface $y=0$, i.e.~$u_\tau = \sqrt{-\partial_xP\,\delta}$. We note that this value of $u_\tau$ is not strictly the one that is expected to scale the flow, which would
be set at the virtual origin at $y=-\Delta y_d$ perceived by the overlying turbulence, i.e.~$u_\tau = \sqrt{-\partial_xP(\delta+\Delta y_d)}$ \citep{Luchini96,Ibrahim21}.
Nevertheless, as we will see in \S\ref{SubSec:ResultProcessing}, the difference between the two values of $u_\tau$ is never larger than $1\%$.

We use the computational code of \citet{Sharma20sparse,Sharma20dense}, which is briefly summarised here. The code solves the incompressible Navier-Stokes equations using a pseudo-spectral discretisation in the $x$ and $z$ directions and a finite difference discretisation in the $y$ direction. A three-step Runge-Kutta scheme with a fractional-step method is used for the time discretisation. No-slip conditions are imposed on all solid surfaces of substrates through an immersed boundary method. The code features a multi-block structure \citep{Garcia11} that allows local refinement of near-wall grids in $x$ and $z$ to resolve small texture, while retaining a coarser resolution away from substrates, sufficient to fully resolve the turbulence. Full details of the numerical method and its validation can be found in \citet{Sharma20thesis}.

In the substrate blocks, \rgm{for the staggered cubes} the resolution is set to $24$-$32$ grid points in $x$ and $z$ per grain period $L$.  The number of grid points covering one gap size $g$ is \rev{13} or greater, except for cases Pd-12-33, Ro/Pd-24-25, Pd-36-22, Ro/Pd-48-22,  Pd-48-25, and Ro/Pd-48-28 (nomenclatures to be introduced below), for which it is \rev{9-11}. For these cases, both the porosity and the permeability are low, and thus the interfacial turbulent fluctuations are weak. In this situation, the grid-dependence investigation in \citet{Sharma20dense} showed that further refinement had no significant influence.
\rgm{For the mesh substrates, the resolution is set to 36 grid points per grain period $L$, with 8-16 points resolving the mesh ligaments of width $\ell$, depending on the ratio $g/L$.}
The fine-resolution blocks reach up to $y\approx2L$ away from the substrate interface into the channel core, sufficient for the fine scales induced by the grain geometry to have vanished already. In the channel-core block, the resolution is set lower to $\Delta x^+\approx6$ and $\Delta z^+\approx3$. In the $y$ direction, the grid is finest with $\Delta y^+\approx0.35$ near the interface $y=0$, where the shear is higher, and is gradually stretched away from this plane, as in \citet{Chen23}. For $y>0$, the grid spacing is stretched to $\Delta y^+\approx3.5$ near the channel centre; for $y<0$, the grid spacing is stretched to $\Delta y\approx\Delta x=\Delta z$ at $y=-2D=-L$ and becomes uniform for $y<-2D$.

\begin{figure}
\vspace*{-1mm}
  \centerline{\includegraphics[width=0.95\linewidth]{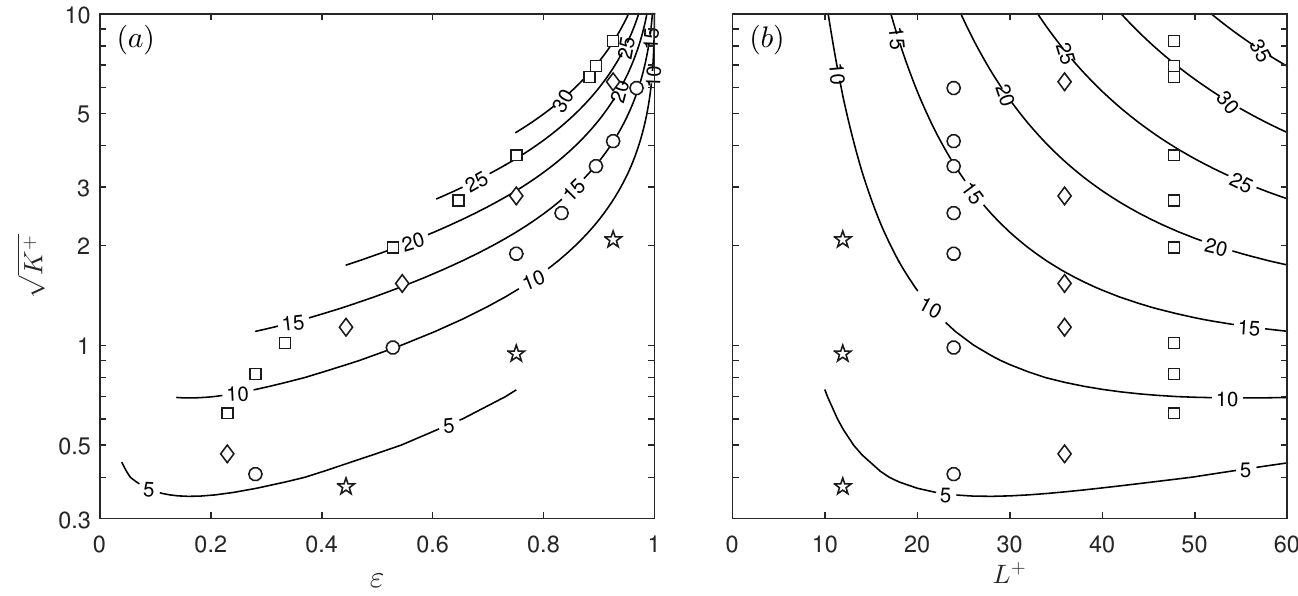}}
  \vspace{-1.5mm}
  \caption{DNS cases in this study represented in ($a$) $\sqrt{K^+}$-$\varepsilon$ and ($b$) $\sqrt{K^+}$-$L^+$  parameter spaces. The isolines with embedded numbers represent constant values of the gap size $g^+$. Symbols: $\medstar$ for $L^+\approx12$; $\medcircle$ for $L^+\approx24$; $\meddiamond$ for $L^+\approx36$; $\medsquare$ for $L^+\approx48$.}
\label{fig:TheCases}
\vspace*{-2.5mm}
\end{figure}

Table \ref{tab:cases1} lists the basic parameters of all 58 DNSs presented in this paper, among which 52 cases have friction Reynolds number $Re_\tau\equiv u_\tau\delta/\nu\approx180$, 4 cases with suffix `HR' have $Re_\tau\approx360$, and 2 cases with suffix `HHR' have $Re_\tau\approx550$.
\rgm{The cases at higher Reynolds numbers were run to verify that the effect of the texture is Reynolds-number independent in viscous scaling, with results compared in Appendix \ref{Sec:ResHR}.} 
The two numbers in the label of a case indicate its pitch $L^+$ and gap-to-pitch ratio $g/L$, respectively, e.g.~ `Pd-24-56' has $L^+\!\approx24$ and $g/L\approx0.56$. The prefix `Ro/Ps/Pd', as introduced in \S \ref{SubSec:Configuration} and figure \ref{fig:Configuration}$(c)$, indicates the depth $h=1D$, $2D$, or $\geq5D$; there is also one very deep case with suffix `VD', with $h=9D$. \rgm{A letter `M' precedes the name for cases with interconnected-mesh substrates.}
The parameter space considered is portrayed in figure  \ref{fig:TheCases}. The porosity $\varepsilon$, pitch $L^+$, and permeability $K^+$ are in the ranges $\varepsilon\approx0.23$ - $0.97$, $L^+\approx12$ - $48$, and $\sqrt{K^+}\approx0.4$ - $8.1$, respectively.

\begin{table}
 \vspace*{-2mm}
 \begin{center}
\def~{\hphantom{0}}
\fontsize{7.8}{8.7}\selectfont
\renewcommand{\arraystretch}{0.94}
\setlength{\tabcolsep}{5.3pt}
\vspace*{-1mm}
  \begin{tabular}{lccccclrrrr}
        $\quad$Case & $\ Re_\tau\ \ $ & $L^+$ & $g/L$ &$h/\!D$ & $\quad\varepsilon\quad$ & $\quad\sqrt{K^+}$ & $N_x\ $ & $N_z\ $ & $N_{yc}$ & $N_{ys}$  \\[2pt]
     Pd-12-33    & 182.7 & 12.0 & 1/3 & 5 & 0.44 & 0.37(0.37) & 2304 & 1152 & 176 &  64 \\
     Ro-12-50    & 182.7 & 12.0 & 1/2 & 1 & 0.75 & 0.91       & 2304 & 1152 & 176 &  16 \\
     Ps-12-50    & 182.7 & 12.0 & 1/2 & 2 & 0.75 & 0.91       & 2304 & 1152 & 176 &  28 \\
     Pd-12-50    & 182.7 & 12.0 & 1/2 & 5 & 0.75 & 0.91(0.88) & 2304 & 1152 & 176 &  64 \\
     Pd-12-67    & 182.7 & 12.0 & 2/3 & 7 & 0.93 & 2.03(1.97) & 2304 & 1152 & 176 &  88 \\
     Ro-24-25    & 182.7 & 23.9 & 1/4 & 1 & 0.28 & 0.39       & 1536 &  768 & 176 &  28 \\
     Pd-24-25    & 182.7 & 23.9 & 1/4 & 7 & 0.28 & 0.39(0.39) & 1536 &  768 & 176 & 126 \\
     Ro-24-38    & 182.7 & 23.9 & 3/8 & 1 & 0.53 & 0.94       & 1536 &  768 & 176 &  28 \\
     Ps-24-38    & 182.7 & 23.9 & 3/8 & 2 & 0.53 & 0.94       & 1536 &  768 & 176 &  46 \\
     Pd-24-38    & 182.7 & 23.9 & 3/8 & 7 & 0.53 & 0.94(0.95) & 1536 &  768 & 176 & 126 \\
     Ro-24-50    & 182.7 & 23.9 & 1/2 & 1 & 0.75 & 1.82       & 1536 &  768 & 176 &  28 \\
     Ps-24-50    & 182.7 & 23.9 & 1/2 & 2 & 0.75 & 1.82       & 1536 &  768 & 176 &  46 \\
     Pd-24-50    & 182.7 & 23.9 & 1/2 & 5 & 0.75 & 1.82(1.76) & 1536 &  768 & 176 &  94 \\
     Pd-24-50-HR & 360.0 & 23.6 & 1/2 & 5 & 0.75 & 1.79(1.69) & 2304 & 1152 & 352 &  74 \\
     Pd-24-50-VD & 182.7 & 23.9 & 1/2 & 9 & 0.75 & 1.82(1.78) & 1536 &  768 & 176 & 158 \\
     Pd-24-56    & 182.7 & 23.9 & 9/16& 7 & 0.83 & 2.44(2.42) & 1536 &  768 & 176 & 126 \\
     Ro-24-62    & 182.7 & 23.9 & 5/8 & 1 & 0.89 & 3.33       & 1536 &  768 & 176 &  28 \\
     Ps-24-62    & 182.7 & 23.9 & 5/8 & 2 & 0.89 & 3.33       & 1536 &  768 & 176 &  46 \\
     Pd-24-62    & 182.7 & 23.9 & 5/8 & 7 & 0.89 & 3.33(3.29) & 1536 &  768 & 176 & 126 \\
     Ro-24-67    & 182.7 & 23.9 & 2/3 & 1 & 0.93 & 4.06       & 1728 &  864 & 176 &  29 \\
     Ro-24-75    & 182.7 & 23.9 & 3/4 & 1 & 0.97 & 5.80       & 1536 &  768 & 176 &  28 \\
     Ps-24-75    & 182.7 & 23.9 & 3/4 & 2 & 0.97 & 5.80       & 1536 &  768 & 176 &  46 \\
     Pd-24-75    & 182.7 & 23.9 & 3/4 & 7 & 0.97 & 5.80(5.73) & 1536 &  768 & 176 & 126 \\
     Pd-36-22    & 182.7 & 35.9 & 2/9 & 5 & 0.23 & 0.45(0.45) & 1152 &  576 & 176 & 112 \\
     Ps-36-33    & 182.7 & 35.9 & 1/3 & 2 & 0.44 & 1.10       & 1152 &  576 & 176 &  51 \\
     Pd-36-33    & 182.7 & 35.9 & 1/3 & 5 & 0.44 & 1.10(1.11) & 1152 &  576 & 176 &  96 \\
     Pd-36-39    & 182.7 & 35.9 & 7/18& 7 & 0.55 & 1.55(1.50) & 1152 &  576 & 176 & 133 \\
     Ro-36-50    & 182.7 & 35.9 & 1/2 & 1 & 0.75 & 2.73       & 1152 &  576 & 176 &  33 \\
     Ps-36-50    & 182.7 & 35.9 & 1/2 & 2 & 0.75 & 2.73       & 1152 &  576 & 176 &  49 \\
     Pd-36-50    & 182.7 & 35.9 & 1/2 & 6 & 0.75 & 2.73(2.67) & 1152 &  576 & 176 & 105 \\
     Pd-36-50-HR & 360.0 & 35.3 & 1/2 & 6 & 0.75 & 2.69(2.54) & 1536 &  768 & 352 &  91 \\
     Ro-36-67    & 182.7 & 35.9 & 2/3 & 1 & 0.93 & 6.10       & 1152 &  576 & 176 &  34 \\
     Ps-36-67    & 182.7 & 35.9 & 2/3 & 2 & 0.93 & 6.10       & 1152 &  576 & 176 &  51 \\
     Pd-36-67    & 182.7 & 35.9 & 2/3 & 6 & 0.93 & 6.10(5.95) & 1152 &  576 & 176 & 111 \\
     Ro-48-22    & 182.7 & 47.8 & 2/9 & 1 & 0.23 & 0.60       &  864 &  432 & 176 &  44 \\
     Pd-48-22    & 182.7 & 47.8 & 2/9 & 5 & 0.23 & 0.60(0.60) &  864 &  432 & 176 & 117 \\
     Pd-48-25    & 182.7 & 47.8 & 1/4 & 5 & 0.28 & 0.77(0.75) &  768 &  384 & 176 & 106 \\
     Ro-48-28    & 182.7 & 47.8 & 5/18& 1 & 0.33 & 0.98       &  864 &  432 & 176 &  44 \\
     Pd-48-28    & 182.7 & 47.8 & 5/18& 5 & 0.33 & 0.98(0.97) &  864 &  432 & 176 & 117 \\
     Ro-48-38    & 182.7 & 47.8 & 3/8 & 1 & 0.53 & 1.88       &  768 &  384 & 176 &  40 \\
     Ps-48-38    & 182.7 & 47.8 & 3/8 & 2 & 0.53 & 1.88       &  768 &  384 & 176 &  58 \\
     Pd-48-38    & 182.7 & 47.8 & 3/8 & 7 & 0.53 & 1.88(1.89) &  768 &  384 & 176 & 138 \\
     Pd-48-38-HR & 360.0 & 47.1 & 3/8 & 7 & 0.53 & 1.85(1.87) & 1536 &  768 & 352 & 138 \\
     Ro-48-44    & 182.7 & 47.8 & 7/16& 1 & 0.65 & 2.66       &  768 &  384 & 176 &  40 \\
     Pd-48-44    & 182.7 & 47.8 & 7/16& 6 & 0.65 & 2.66(2.60) &  768 &  384 & 176 & 122 \\
     Ro-48-50    & 182.7 & 47.8 & 1/2 & 1 & 0.75 & 3.64       &  768 &  384 & 176 &  40 \\
     Ps-48-50    & 182.7 & 47.8 & 1/2 & 2 & 0.75 & 3.64       &  768 &  384 & 176 &  58 \\
     Pd-48-50    & 182.7 & 47.8 & 1/2 & 6 & 0.75 & 3.64(3.50) &  768 &  384 & 176 & 122 \\
     Pd-48-50-HHR& 550.4 & 48.0 & 1/2 & 6 & 0.75 & 3.66(3.42) & 1728 &  864 & 484 & 107 \\
     Ro-48-61    & 182.7 & 47.8 &11/18& 1 & 0.88 & 6.28       &  864 &  432 & 176 &  44 \\
     Ps-48-61    & 182.7 & 47.8 &11/18& 2 & 0.88 & 6.28       &  864 &  432 & 176 &  63 \\
     Pd-48-61    & 182.7 & 47.8 &11/18& 6 & 0.88 & 6.28(6.05) &  864 &  432 & 176 & 135 \\
     Ro-48-62    & 182.7 & 47.8 & 5/8 & 1 & 0.89 & 6.66       &  768 &  384 & 176 &  40 \\
     Ps-48-62    & 182.7 & 47.8 & 5/8 & 2 & 0.89 & 6.66       &  768 &  384 & 176 &  58 \\
     Pd-48-62    & 182.7 & 47.8 & 5/8 & 7 & 0.89 & 6.66(6.42) &  768 &  384 & 176 & 138 \\
     Pd-48-62-HR & 360.0 & 47.1 & 5/8 & 7 & 0.89 & 6.56(6.35) & 1536 &  768 & 352 & 138 \\
     Pd-48-62-HHR& 550.4 & 48.0 & 5/8 & 7 & 0.89 & 6.69(6.47) & 2304 & 1152 & 484 & 137 \\
     Ro-48-67    & 182.7 & 47.8 & 2/3 & 1 & 0.93 & 8.13       &  864 &  432 & 176 &  44 \\
     \rgm{MPd-36-78}   & \rgm{182.7} & \rgm{35.9} & \rgm{7/9} &\rgm{10} & \rgm{0.87} & \rgm{5.34(5.16)}       &    \rgm{1152} &    \rgm{576} &   \rgm{176} &   \rgm{202}     \\
     \rgm{MRo-48-56}   & \rgm{182.7} & \rgm{47.8} & \rgm{5/9} & \rgm{1} & \rgm{0.58} & \rgm{3.30}       &    \rgm{864} &    \rgm{432} &   \rgm{176} &   \rgm{44}     \\
     \rgm{MPs-48-56}   & \rgm{182.7} & \rgm{47.8} & \rgm{5/9} & \rgm{2} & \rgm{0.58} & \rgm{3.30}       &    \rgm{864} &    \rgm{432} &   \rgm{176} &   \rgm{63}     \\
     \rgm{MPd-48-56}   & \rgm{182.7} & \rgm{47.8} & \rgm{5/9} &\rgm{10} & \rgm{0.58} & \rgm{3.30(3.18)}       &    \rgm{864} &    \rgm{432} &   \rgm{176} &   \rgm{207}     \\
     \rgm{MPd-48-72}   & \rgm{182.7} & \rgm{47.8} &\rgm{13/18}&\rgm{10} & \rgm{0.81} & \rgm{5.94(5.68)}       &    \rgm{864} &    \rgm{432} &   \rgm{176} &   \rgm{207}     \\
     \rgm{MPd-48-78}   & \rgm{182.7} & \rgm{47.8} &\rgm{7/9}&\rgm{10} & \rgm{0.87} & \rgm{7.12(6.71)}       &    \rgm{864} &    \rgm{432} &   \rgm{176} &   \rgm{207}     \\
  \end{tabular}
\vspace*{-1mm}
  \caption{Simulation parameters. $L$ is the grain pitch, $g$ the gap size, $h$ the substrate depth, $D=L/2$ the thickness of one grain layer, $\varepsilon$ the porosity, and $K$ the permeability, with \textit{a posteriori} values in parenthesis. The number of grid points is $N_x$ in $x$, $N_z$ in $z$, and $N_{yc}$ and $N_{ys}$ in $y$ in the free-flow region and for the substrates, respectively.}
  \label{tab:cases1}
  \end{center}
\vspace*{-1.5mm}
\end{table}

\subsection{Techniques for post-processing}
\label{SubSec:ResultProcessing}

\begin{table}
 \vspace*{-2mm}
 \begin{center}
\def~{\hphantom{0}}
\fontsize{7.8}{8.7}\selectfont
\renewcommand{\arraystretch}{0.94}
\vspace*{-1mm}
  \begin{tabular}{lccccccccrr}
        $\quad$Case & $\ \ U_s^+\ \ $ & $\ \ \ell_U^+\ \ $ & $\ \ r_\textit{sh}\ \ $ & $\ \ r_\nu\ \ $ & $\ U_{\mathrm{Da}}^+\ $ & $\sqrt{K^+}$ & $\sqrt{K_{eq}^{s+}}$ & $\sqrt{K_{eq}^{t+}}$ & $\Delta U_\delta^+$ & $\ \Delta U^+$ \\[3pt]
     Pd-12-33    & 0.43 & 0.42 & 0.24 & 0.12 & 0.00 & 0.37 & 0.37 & 0.31 & 0.25 & 0.25 \\
     Ro-12-50    & 0.81 & 0.80 & 0.40 & 0.35 & 0.00 & 0.91 & 0.91 & 0.34 & 0.50 & 0.50 \\
     Ps-12-50    & 0.81 & 0.80 & 0.40 & 0.34 & 0.00 & 0.91 & 0.91 & 0.51 & 0.80 & 0.79 \\
     Pd-12-50    & 0.80 & 0.80 & 0.40 & 0.34 & 0.00 & 0.91 & 0.91 & 0.77 & 0.98 & 0.97 \\
     Pd-12-67    & 1.23 & 1.58 & 0.55 & 0.51 & 0.02 & 2.03 & 2.02 & 1.86 & 5.26 & 5.25 \\
     Ro-24-25    & 0.43 & 0.43 & 0.20 & 0.06 & 0.00 & 0.39 & 0.39 & 0.22 & 0.40 & 0.39 \\
     Pd-24-25    & 0.42 & 0.43 & 0.20 & 0.07 & 0.00 & 0.39 & 0.39 & 0.38 & 0.43 & 0.41 \\
     Ro-24-38    & 0.80 & 0.80 & 0.32 & 0.17 & 0.00 & 0.94 & 0.94 & 0.51 & 0.99 & 1.04 \\
     Ps-24-38    & 0.80 & 0.81 & 0.32 & 0.16 & 0.00 & 0.94 & 0.94 & 0.72 & 1.29 & 1.31 \\
     Pd-24-38    & 0.80 & 0.81 & 0.32 & 0.16 & 0.00 & 0.94 & 0.94 & 0.93 & 1.28 & 1.27 \\
     Ro-24-50    & 1.35 & 1.41 & 0.44 & 0.34 & 0.02 & 1.82 & 1.81 & 0.94 & 1.70 & 1.71 \\
     Ps-24-50    & 1.28 & 1.49 & 0.43 & 0.24 & 0.02 & 1.82 & 1.81 & 1.35 & 3.64 & 3.71 \\
     Pd-24-50    & 1.24 & 1.52 & 0.43 & 0.20 & 0.02 & 1.82 & 1.81 & 1.75 & 4.32 & 4.28 \\
     Pd-24-50-HR & 1.20 & 1.45 & 0.42 & 0.21 & 0.01 & 1.79 & 1.78 & 1.72 & 4.05 & 4.05 \\
     Pd-24-50-VD & 1.23 & 1.53 & 0.43 & 0.20 & 0.02 & 1.82 & 1.81 & 1.80 & 4.55 & 4.54 \\
     Pd-24-56    & 1.35 & 2.03 & 0.46 & 0.24 & 0.03 & 2.44 & 2.41 & 2.40 & 6.37 & 6.36 \\
     Ro-24-62    & 1.96 & 2.33 & 0.55 & 0.69 & 0.05 & 3.33 & 3.29 & 1.64 & 3.38 & 3.44 \\
     Ps-24-62    & 1.58 & 2.56 & 0.53 & 0.29 & 0.06 & 3.33 & 3.27 & 2.33 & 6.45 & 6.44 \\
     Pd-24-62    & 1.42 & 2.66 & 0.51 & 0.24 & 0.06 & 3.33 & 3.25 & 3.23 & 7.66 & 7.65 \\
     Ro-24-67    & 2.16 & 2.77 & 0.59 & 0.84 & 0.08 & 4.06 & 3.99 & 1.95 & 4.01 & 4.01 \\
     Ro-24-75    & 2.46 & 3.69 & 0.65 & 0.98 & 0.14 & 5.80 & 5.53 & 2.54 & 5.12 & 5.13 \\
     Ps-24-75    & 2.01 & 4.31 & 0.61 & 0.38 & 0.16 & 5.80 & 5.51 & 3.68 & 7.84 & 7.85 \\
     Pd-24-75    & 1.69 & 4.53 & 0.59 & 0.19 & 0.18 & 5.80 & 5.36 & 5.30 & 9.67 & 9.66 \\
     Pd-36-22    & 0.46 & 0.46 & 0.19 & 0.05 & 0.00 & 0.45 & 0.45 & 0.45 & 0.51 & 0.53 \\
     Ps-36-33    & 0.89 & 0.92 & 0.29 & 0.11 & 0.01 & 1.10 & 1.09 & 0.96 & 1.68 & 1.69 \\
     Pd-36-33    & 0.89 & 0.92 & 0.29 & 0.10 & 0.01 & 1.10 & 1.09 & 1.09 & 1.79 & 1.79 \\
     Pd-36-39    & 1.13 & 1.30 & 0.33 & 0.09 & 0.01 & 1.55 & 1.53 & 1.52 & 3.27 & 3.23 \\
     Ro-36-50    & 1.70 & 2.17 & 0.43 & 0.20 & 0.04 & 2.73 & 2.67 & 1.60 & 3.80 & 3.81 \\
     Ps-36-50    & 1.46 & 2.25 & 0.42 & 0.13 & 0.04 & 2.73 & 2.66 & 2.25 & 5.81 & 5.80 \\
     Pd-36-50    & 1.40 & 2.27 & 0.41 & 0.11 & 0.04 & 2.73 & 2.65 & 2.64 & 6.48 & 6.46 \\
     Pd-36-50-HR & 1.32 & 2.11 & 0.41 & 0.12 & 0.02 & 2.69 & 2.61 & 2.61 & 5.85 & 5.79 \\
     Ro-36-67    & 2.30 & 4.13 & 0.57 & 0.50 & 0.15 & 6.10 & 5.73 & 3.21 & 6.06 & 6.02 \\
     Ps-36-67    & 1.81 & 4.39 & 0.55 & 0.20 & 0.18 & 6.10 & 5.58 & 4.44 & 8.44 & 8.45 \\
     Pd-36-67    & 1.61 & 4.44 & 0.54 & 0.15 & 0.20 & 6.10 & 5.48 & 5.46 & 9.71 & 9.73 \\
     Ro-48-22    & 0.54 & 0.54 & 0.19 & 0.05 & 0.00 & 0.60 & 0.60 & 0.46 & 0.80 & 0.85 \\
     Pd-48-22    & 0.55 & 0.55 & 0.19 & 0.05 & 0.00 & 0.60 & 0.60 & 0.60 & 0.87 & 0.84 \\
     Pd-48-25    & 0.62 & 0.63 & 0.22 & 0.07 & 0.00 & 0.77 & 0.77 & 0.77 & 0.90 & 0.91 \\
     Ro-48-28    & 0.78 & 0.79 & 0.25 & 0.07 & 0.01 & 0.98 & 0.97 & 0.73 & 1.11 & 1.14 \\
     Pd-48-28    & 0.78 & 0.80 & 0.25 & 0.07 & 0.01 & 0.98 & 0.97 & 0.97 & 1.48 & 1.50 \\
     Ro-48-38    & 1.28 & 1.49 & 0.32 & 0.07 & 0.02 & 1.88 & 1.82 & 1.27 & 2.98 & 3.00 \\
     Ps-48-38    & 1.22 & 1.54 & 0.31 & 0.05 & 0.02 & 1.88 & 1.81 & 1.66 & 4.02 & 4.01 \\
     Pd-48-38    & 1.20 & 1.55 & 0.31 & 0.05 & 0.02 & 1.88 & 1.81 & 1.81 & 4.35 & 4.36 \\
     Pd-48-38-HR & 1.14 & 1.54 & 0.31 & 0.05 & 0.01 & 1.85 & 1.78 & 1.78 & 4.43 & 4.44 \\
     Ro-48-44    & 1.57 & 2.10 & 0.37 & 0.08 & 0.03 & 2.66 & 2.53 & 1.70 & 4.24 & 4.26 \\
     Pd-48-44    & 1.37 & 2.15 & 0.36 & 0.06 & 0.04 & 2.66 & 2.53 & 2.52 & 5.82 & 5.81 \\
     Ro-48-50    & 1.82 & 2.78 & 0.42 & 0.12 & 0.06 & 3.64 & 3.41 & 2.23 & 4.96 & 4.97 \\
     Ps-48-50    & 1.52 & 2.80 & 0.41 & 0.10 & 0.07 & 3.64 & 3.42 & 3.09 & 6.99 & 7.01 \\
     Pd-48-50    & 1.46 & 2.80 & 0.40 & 0.11 & 0.07 & 3.64 & 3.44 & 3.44 & 7.34 & 7.32 \\
     Pd-48-50-HHR& 1.39 & 2.76 & 0.40 & 0.10 & 0.02 & 3.66 & 3.43 & 3.43 & 7.16 & 7.10 \\
     Ro-48-61    & 2.16 & 4.37 & 0.50 & 0.24 & 0.15 & 6.28 & 5.62 & 3.55 & 6.71 & 6.71 \\
     Ps-48-61    & 1.72 & 4.46 & 0.49 & 0.14 & 0.19 & 6.28 & 5.51 & 4.85 & 8.80 & 8.81 \\
     Pd-48-61    & 1.57 & 4.44 & 0.49 & 0.11 & 0.22 & 6.28 & 5.42 & 5.42 & 9.94 & 9.98 \\
     Ro-48-62    & 2.19 & 4.53 & 0.52 & 0.26 & 0.17 & 6.66 & 5.94 & 3.74 & 6.77 & 6.75 \\
     Ps-48-62    & 1.72 & 4.64 & 0.50 & 0.14 & 0.21 & 6.66 & 5.78 & 5.07 & 8.96 & 8.97 \\
     Pd-48-62    & 1.57 & 4.59 & 0.50 & 0.12 & 0.24 & 6.66 & 5.68 & 5.68 & 9.93 & 9.97 \\
     Pd-48-62-HR & 1.58 & 4.79 & 0.50 & 0.11 & 0.12 & 6.56 & 5.56 & 5.56 & 9.59 & 9.60 \\
     Pd-48-62-HHR& 1.58 & 5.00 & 0.50 & 0.10 & 0.08 & 6.69 & 5.58 & 5.58 & 9.72 & 9.75 \\
     Ro-48-67    & 2.35 & 5.40 & 0.55 & 0.27 & 0.21 & 8.13 & 6.81 & 4.15 & 7.31 & 7.33 \\
    \rgm{MPd-36-78}    & \rgm{2.55} & \rgm{7.78} & \rgm{0.68} & \rgm{0.30} & \rgm{0.16} & \rgm{5.34} & \rgm{5.11} & \rgm{5.11} & \rgm{8.74} & \rgm{8.73} \\
    \rgm{MRo-48-56}    & \rgm{2.18} & \rgm{4.11} & \rgm{0.48} & \rgm{0.09} & \rgm{0.05} & \rgm{3.30} & \rgm{3.08} & \rgm{1.97} & \rgm{6.05} & \rgm{6.03} \\
    \rgm{MPs-48-56}    & \rgm{2.16} & \rgm{4.11} & \rgm{0.48} & \rgm{0.10} & \rgm{0.06} & \rgm{3.30} & \rgm{3.13} & \rgm{2.84} & \rgm{6.37} & \rgm{6.37} \\
    \rgm{MPd-48-56}    & \rgm{2.02} & \rgm{4.10} & \rgm{0.48} & \rgm{0.09} & \rgm{0.06} & \rgm{3.30} & \rgm{3.12} & \rgm{3.12} & \rgm{7.30} & \rgm{7.30} \\
    \rgm{MPd-48-72}    & \rgm{2.55} & \rgm{7.97} & \rgm{0.62} & \rgm{0.23} & \rgm{0.19} & \rgm{5.94} & \rgm{5.54} & \rgm{5.54} & \rgm{8.74} & \rgm{8.76} \\
    \rgm{MPd-48-78}    & \rgm{2.62} & \rgm{9.91} & \rgm{0.67} & \rgm{0.27} & \rgm{0.28} & \rgm{7.12} & \rgm{6.53} & \rgm{6.53} & \rgm{9.52} & \rgm{9.55} \\
  \end{tabular}
\vspace*{-1mm}
  \caption{Substrate properties obtained from DNS. $U_s$ and $\ell_U$ are the mean slip velocity and slip length, respectively; $r_\textit{sh}$ is the inner/outer shear ratio across the interface and $r_\nu$ the effective viscosity ratio (see appendix \ref{Sec:Brinkman}); $K_{eq}^{s}$ and $K_{eq}^{t}$ are the shear- and transpiration-based equivalent permeabilities; $\Delta U_\delta^+$ is the velocity deficit at the channel centre, and $\Delta U^+$ the one obtained with a zero-plane displacement and optimal outer-layer matching.}
  \label{tab:cases2}
  \end{center}
\vspace*{-1.5mm}
\end{table}

Table \ref{tab:cases2} lists the main properties resulting from all the DNSs.  For each simulation, we obtain the flow statistics and spectra by averaging multiple instantaneous fields over a period of time at least $12\delta/u_\tau$ after turbulence reaches a statistically steady state. Unless otherwise stated, the statistics at a $y$ location below the interface are spatially averaged over the entire $x$-$z$ plane containing both fluid and solid areas, i.e.~the `superficial average'. 

We use the mean-velocity deficit in wall units, $\Delta U^+$, \rgm{commonly known as the roughness function,} to quantify how a substrate increases drag compared to a smooth wall.  Following classical turbulence theory, the mean velocity profile $U^+(y^+)$ in the logarithmic layer over a complex surface is shifted relative to a smooth-wall one, $U_\mathrm{Sm}^+(y^+)$,
\begin{equation}
\label{eq:DUplusDef}
    U^+(y^+) = \kappa^\text{-1}\ln{(y^++\Delta y_d^+)} + B - \Delta U^+,
\end{equation}
where $\kappa\approx0.4$ and $B\approx5$ are the constants characterising $U_\mathrm{Sm}^+(y^+)$ in the logarithmic layer and $\Delta y_d^+$ is the zero-plane displacement,
\rgm{the $y$-coordinate offset that results in a best collapse for the outer flow}.
We follow \citet{Chen23} and extend this relation above the logarithmic layer to include the wake region, for a robust estimation of $\Delta U^+$ by adjusting $\Delta y_d$ to maximise the region sufficiently above the surface where $U^+(y^+)$ and $U_\mathrm{Sm}^+(y^+\!+\!\Delta y_d^+)$ are parallel. Nevertheless, the resulting values of $\Delta U^+$ are almost identical to the velocity difference $\Delta U_\delta^+=U_\mathrm{Sm}^+(\delta^+)-U^+(\delta^+)$ measured at the channel centre $y=\delta$ without zero-plane offsetting, both listed in table \ref{tab:cases2}. For all the cases studied, the distance between the interface and the zero-plane-displacement height was under 3 wall units, and the resulting difference in $u_\tau$ under $1\%$, as mentioned above.

The mean pressure gradient $\partial_xP$ driving the flow induces a Darcy velocity $U_\mathrm{Da}$ within the substrate, which is not present in an external-flow application where $\partial_xP\simeq0$ \citep[see][]{Gomez19}. In an internal flow, $U_\mathrm{Da}^+$  scales with $Re_\tau^\text{-1}$, which is intrinsically different from other near-wall quantities that are essentially independent of $Re_\tau$ when normalised in wall units. To allow for direct application to external flows, we subtract $U_\mathrm{Da}^+$ from the mean velocity $U^+$ to evaluate the drag increase $\Delta U^+$, thus redefined by
\begin{equation}
\label{eq:DUplusDef2}
    U^+(y^+) = \kappa^\text{-1}\ln{(y^++\Delta y_d^+)} + B + U_\mathrm{Da}^+ - \Delta U^+.
\end{equation}
Similarly, we define the slip velocity $U_s^+$ as
\begin{equation}
\label{eq:DefUslip}
    U_s^+ = U^+\!|_0 - U_\mathrm{Da}^+,
\end{equation}
where $U^+\!|_0$ is the mean velocity at the interface. For a sufficiently deep substrate, the Darcy velocity is $U_\mathrm{Da}\approx-\nu^\text{-1}\!K\,\partial_xP$, while for a finite-depth substrate it can be approximated by equation (\ref{eq:BrkmMeanU}$a$) evaluated at $y=0$ -- note that both results converge for $h\gg\sqrt{K}$. In any event, table \ref{tab:cases2} shows that $U_{\text{Da}}^+$ is essentially negligible compared with $\Delta U^+$ or $U_s^+$.

\section{Effect of substrate parameters on turbulence}
\label{Sec:Results}

In this section, we report the dependence of flow properties on the substrate geometry.
\rgm{Research exploring the parameter space of the substrate properties, e.g. permeability, porosity, depth, grain size, etc, is typically limited
in which variations are possible. A sweep through Reynolds number for a fixed substrate, for instance, would vary $\sqrt{K^+}$, but in doing so would
also vary $L^+$ proportionately. The present set of simulations has been designed so that the effect of parameter pairs such as $\sqrt{K^+}$ and
$L^+$ could be analised separately. We therefore discuss the isolated effect of one parameter at a time} -- the grain spacing, or pitch, $L^+$, the porosity $\varepsilon$, or equivalently the gap-to-pitch ratio, $g/L$, and the relative depth, $h/D$.

\subsection{Effect of porosity under fixed grain spacing}
\label{subSec:statistics1}

\begin{figure}
\vspace*{-1mm}
  \centerline{\includegraphics[width=1.0\linewidth]{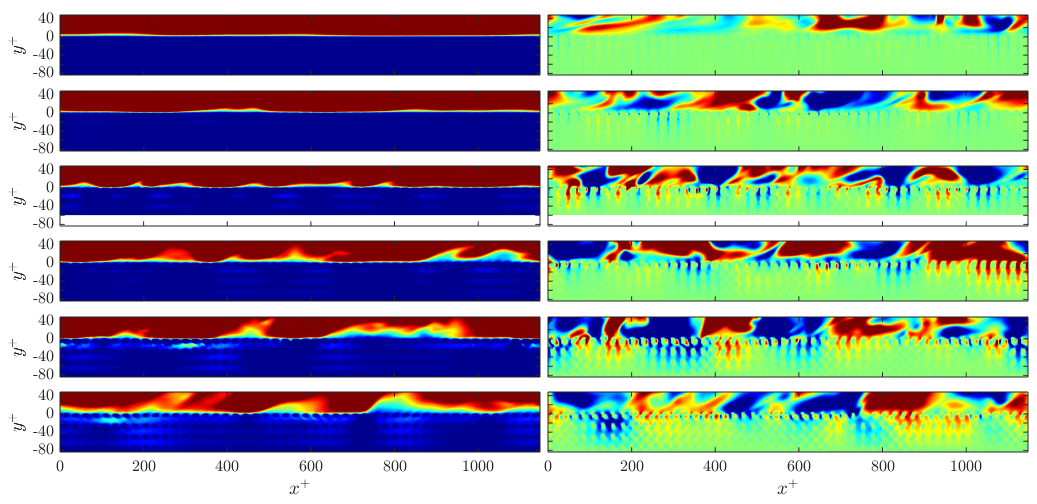}}
\vspace*{-1.5mm}
  \caption{Instantaneous fields of velocity components $u^+$ (left) and $v^+$ (right) on an $x$-$y$ plane for deep-porous substrates with identical pitch $L^+\approx24$ but different gap-to-pitch ratio $g/L=0.25\!$ - $\!0.75$, from top to bottom substrates Pd-24-25/38/50/56/62/75. Colours from blue to red correspond for $u^+$ to $[0:5]$ and for $v^+$ to $[-0.8:0.8]$.}
  \label{fig:Contours_L24_xy}
\vspace*{-1.5mm}
\end{figure}

First, let us consider deep porous (Pd) substrates with fixed $L^+$ but varying $\sqrt{K^+}$ by varying $g/L$. Figure \ref{fig:Contours_L24_xy} shows instantaneous velocity fields on the $x$-$y$ plane for six deep porous substrates with identical $L^+\approx24$ but varying $g/L$ from $1/4$ to $3/4$, corresponding to $\sqrt{K^+}$ from $0.4$ to $5.8$. For increasing $g/L$, the streamwise velocity $u^+$ of the subsurface flow generally increases, while regions of relatively low speed become more prevalent immediately above the interface. The changes in $u^+$ and $v^+$ suggest a gradually intensified penetration of the overlying flow into the substrate. There is strong impedance to the overlying turbulent eddies, which typically span multiple grains, penetrating
into the substrate, and their footprint below the interface is much attenuated, and dispersed lengthscale-wise, by the presence of the individual grains.

\begin{figure}
  \centerline{\includegraphics[width=1.0\linewidth]{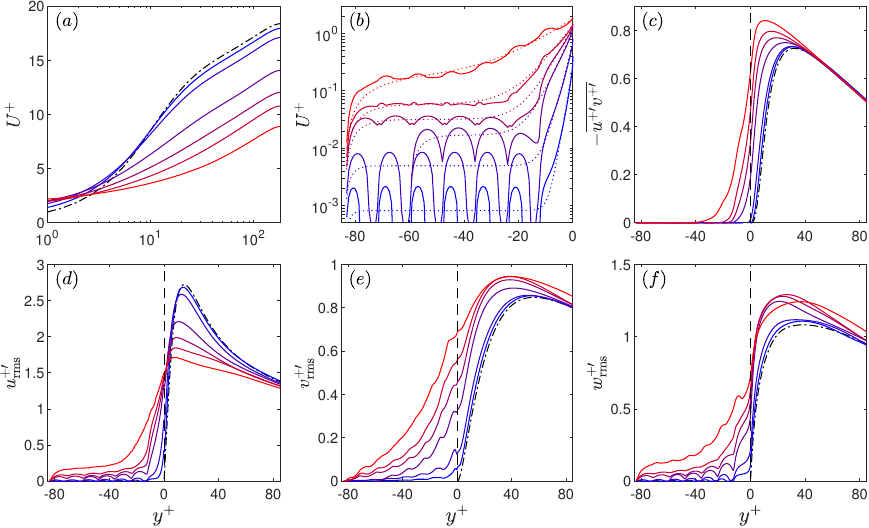}}
  \vspace*{-1mm}
  \caption{($a$-$b$) Mean velocity profile, ($c$) Reynolds shear stress, and ($d$-$f$) RMS velocity fluctuations for deep porous substrates with identical $L^+\!\approx\!24$ but different $g/L=0.25\!$ - $\!0.75$. Colours from blue to red are for cases Pd-24-25/38/50/56/62/75, and dash-dotted lines for smooth-wall data. The dotted lines in panel $(b)$ are Darcy-Brinkman analytical solutions, equation (\ref{eq:BrkmMeanU}), for the mean velocity within the substrate, and the dashed lines mark the location of the free-flow/substrate interface.
  }
\label{fig:prfls_L24}
\vspace*{-1mm}
\end{figure}

The profiles of mean velocity $U^+$, Reynolds shear stress $\overline{u^{+\prime}v^{+\prime}}$, and root-mean-square (RMS) velocity fluctuations $u^{+\prime}_{\rms}$, $v^{+\prime}_{\rms}$, and $w^{+\prime}_{\rms}$ for the above six substrates are portrayed in figure  \ref{fig:prfls_L24}.  For each case, the $U^+$ profile below the interface in figure \ref{fig:prfls_L24}$(b)$ shows a near-interface region with strong shear, i.e.~the Brinkman layer. Further below is a plateau where $U^+$ essentially results from the mean pressure gradient, i.e. the Darcy region. The Brinkman layer is thicker for larger $g/L$, indicating a deepened penetration of shear of the overlying flow.  This also results in a slight increase of the slip velocity, $U_s^+$, but generally decreases $U^+$ above the interface, as shown in figure \ref{fig:prfls_L24}$(a)$, which leads to an increase of the drag coefficient. Of the configurations discussed here, only those with smaller permeability, $\sqrt{K^+}\lesssim1$, exhibit a smooth-wall-like
character. Those with greater permeability experience significant departures from smooth-wall turbulence, exhibiting the usual decrease of $u'$ and increase in $v'$ and $w'$ near the wall, with all three 
converging towards similar peak values, together with an increase in near-wall Reynolds shear stress and the corresponding increase in $\Delta U^+$
and drag. Given that similar intense departures from the smooth-wall-like regime occur across the whole set of configurations 
studied, the virtual-origin framework proposed in \citet{Ibrahim21}
will not be used here, as it only applies to smooth-wall-like turbulence.

The drag increase with increasing $g/L$ is directly associated to the changes in the $\overline{u^{+\prime}v^{+\prime}}$ profiles 
\citep[see][\S 5.3]{Gomez19}, shown in figure \ref{fig:prfls_L24}$(c)$ for $y^+\approx0$ - $30$. Their magnitude increases significantly relative to smooth-wall values for  $g/L\gtrsim0.4$ or $\sqrt{K^+}\gtrsim1$. These profiles of $\overline{u^{+\prime}v^{+\prime}}$, together with $u^{+\prime}_{\rms}$, $v^{+\prime}_{\rms}$, and $w^{+\prime}_{\rms}$ in figure \ref{fig:prfls_L24}$(d,e,f)$, illustrate the gradually enhanced penetration of turbulence into the substrates as $g/L$ increases. The penetration of $u^{+\prime}_{\rms}$ is accompanied by a drop of its peak value above the interface, while such a drop is not observed for $v^{+\prime}_{\rms}$ or $w^{+\prime}_{\rms}$. Similar trends have been observed not only for porous substrates \citep{Breugem06} but also for rough surfaces \citep{Ligrani86,Abderrahaman19} and canopies \citep{Sharma20dense}. These trends have been interpreted by some authors \citep{Jimenez04,Flores06} as the flow losing some of the anisotropic characters of the near-wall cycle. It is also notable that for all cases studied, $v^{+\prime}_{\rms}$ decays with the depth into the substrate more slowly than $u^{+\prime}_{\rms}$ and $w^{+\prime}_{\rms}$, a feature consistent with the study for dense canopies by \citet{Sharma20dense}. This implies that, for a finite-depth porous substrate, the wall-normal velocity fluctuations in 
the subsurface flow are more likely to perceive the presence of the substrate floor than the tangential (i.e.~wall-parallel) fluctuations. This will be further discussed in \S\ref{Sec:unification}.

\subsection{Effect of grain spacing under fixed porosity}
\label{subSec:statistics2}

\begin{figure}
\vspace*{-.5mm}
  \centering
  \includegraphics[width=1.0\linewidth]{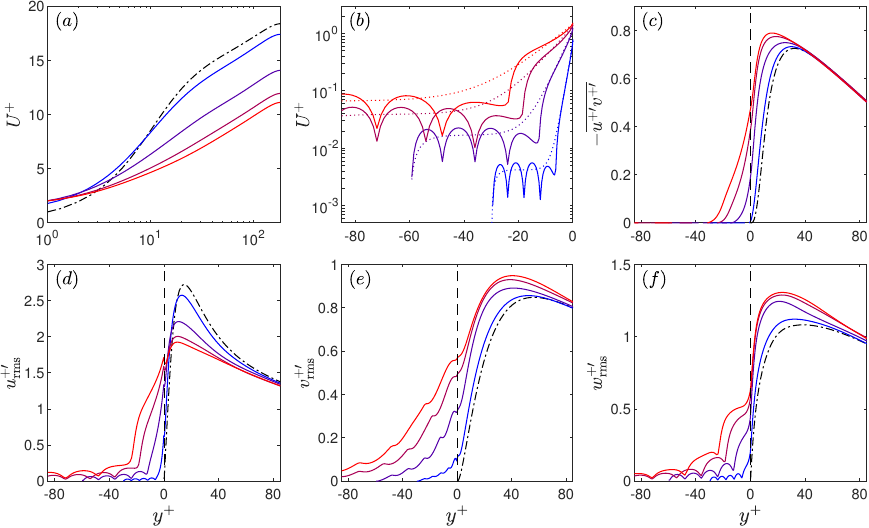}
\vspace*{-4mm}
  \caption{($a$-$b$) Mean velocity profile, ($c$) Reynolds shear stress, and ($d$-$f$) RMS velocity fluctuations for deep porous substrates with identical $g/L=0.50$ but different $L^+=12\!$ - $\!48$. Colours from blue to red are for cases Pd-12/24/36/48-50, and dash-dotted lines for smooth-wall data. The dashed lines mark the location of the free-flow/substrate interface.}
  \label{fig:prfls_Gr50}
\vspace*{-1mm}
\end{figure}

Next, we consider the deep porous (Pd) substrates with the same gap-to-pitch ratio $g/L=1/2$ but different pitch $L^+$. Figure \ref{fig:prfls_Gr50} displays the changes in flow statistics as the pitch $L^+$ increases from $12$ to $48$, which corresponds to $\sqrt{K^+}$ increasing from $0.9$ to $3.6$. The changes in terms of drag, subsurface mean flow and turbulence penetration are qualitatively similar to those changes with $g/L$ increasing illustrated in figure \ref{fig:prfls_L24}. 
The dominant factor that underlies both the $g/L$-induced changes and the $L^+$-induced changes will be discussed in \S\ref{subSec:deepPorous}. In both instances, we note that the differences in the value of the mean velocity at the interface is small, while the differences far above the substrate are significant. This can be traced to the increase in the shear Reynolds stress profile, as discussed above. The latter shows a good correlation with the fluctuating transpiration $v^{+\prime}_{\mathrm{\rms}}$ at the interface \citep{Abderrahaman19}, which in turn has been shown to correlate well with $\Delta U^+$ for rough surfaces
\rgm{\citep{Orlandi06,Orlandi06JoT,Orlandi08}}. This suggests that the increase in drag is more closely connected to the transpiration than to the tangential velocity at the interface.

\begin{figure}
  \vspace*{2mm}
  \includegraphics[width=1.0\linewidth]{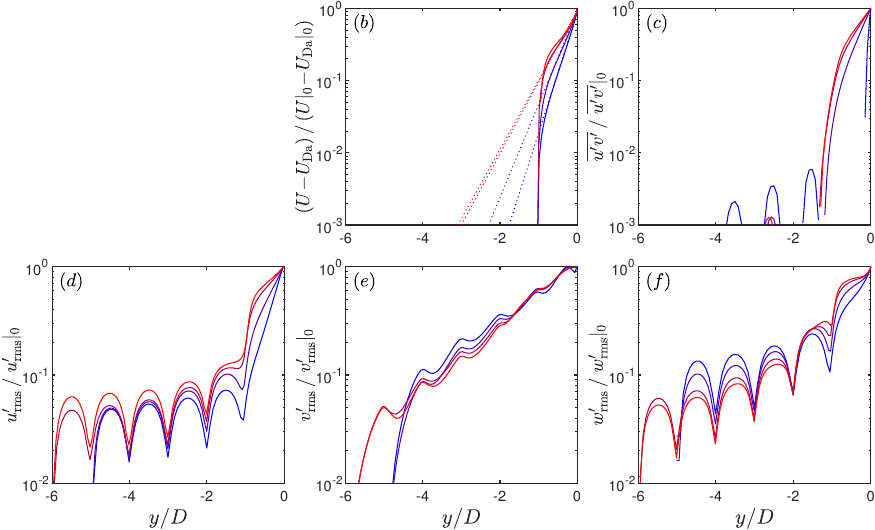}
  \caption{($a$) Shear-driven component of the mean velocity, ($b$) Reynolds shear stress, and ($c$-$e$) RMS velocity fluctuations for the flow within the substrate, normalised by the corresponding interfacial values and the thickness of one layer of cubes, $D$, for the same cases of figure \ref{fig:prfls_Gr50}. Colours are as in figure \ref{fig:prfls_Gr50}.}
  \label{fig:prfls_Gr50_Similar}
\end{figure}

In contrast to the cases with identical $L^+$ in figure \ref{fig:prfls_L24}, the four cases with identical $g/L=1/2$ in figure \ref{fig:prfls_Gr50} share a similarity in substrate geometry, and thus exhibit some degree of similarity in the decay of subsurface flow properties. In figure \ref{fig:prfls_Gr50_Similar}, where the flow statistics are normalised by the corresponding interfacial values and the wall-normal coordinate is normalised by $D$, the four cases have a similar decaying trend for $v^\prime_{\rms}$ in most of the subsurface region with $y<0$, while for $u^\prime_{\rms}$ and $w^\prime_{\rms}$ the similarity occurs only for $y\lesssim-1D$. This suggests that nonlinear inertial effects, which break the similarity of the subsurface flow, are largely limited to the near-interface region, and mainly influence tangential motions only. This would thus yield the difference in the decaying rate of near-interface $U$ among the four cases, shown in figure \ref{fig:prfls_Gr50_Similar}$(b)$.

\subsection{Effect of substrate depth}
\label{subSec:statistics3}

\begin{figure}
  \vspace*{2mm}
  \centerline{\includegraphics[width=1.0\linewidth]{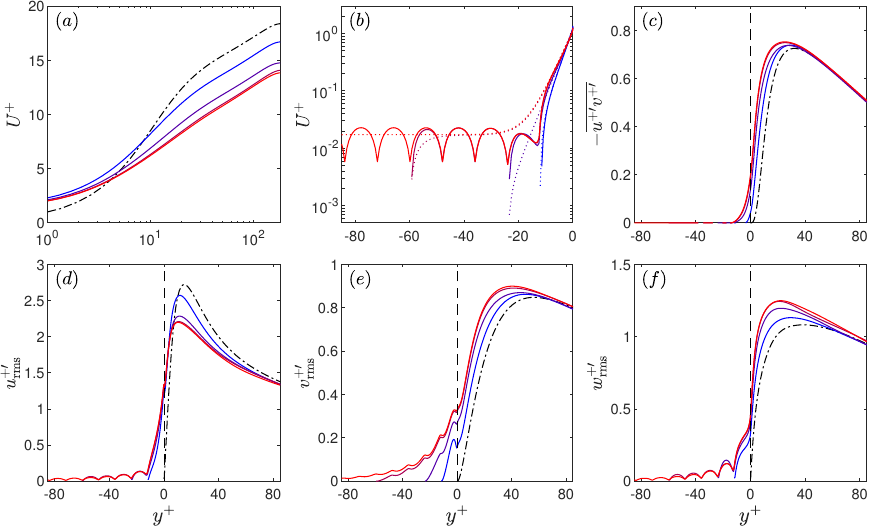}}
  \caption{($a$-$b$) Mean velocity profile, ($c$) Reynolds shear stress, and ($d$-$f$) RMS velocity fluctuations for substrates with identical $L^+\approx24$ and $g/L=0.50$ but different depth $h=1D$ - $9D$. Colours from blue to red are for cases Ro-24-50 ($h/D\!=\!1$), Ps-24-50 ($h/D\!=\!2$), Pd-24-50 ($h/D\!=\!5$), and Pd-24-50-VD ($h/D\!=\!9$), and dash-dotted lines for smooth-wall data. The dashed lines mark the location of the free-flow/substrate interface.}
\label{fig:prfls_PR2450}
\end{figure}

Lastly, for a fixed pitch $L^+\approx24$ and gap-to-pitch ratio $g/L=1/2$, we compare four cases with varying depths $h/D=1,2,5$, and $9$, which are respectively labeled as Ro-24-50, Ps-24-50, Pd-24-50, and Pd-24-50-VD. As shown in figure \ref{fig:prfls_PR2450}, as the depth $h$ increases from $1D$ to $5D$, we observe an increase in drag and a deeper penetration of turbulence, which are qualitatively similar to the changes with increasing $g/L$ in figure \ref{fig:prfls_L24} or with increasing $L^+$ in figure \ref{fig:prfls_Gr50}.  However, the present four cases demonstrate no substantial differences in $U^+$, $u_{\rms}^{+\prime}$, and $w_{\rms}^{+\prime}$ in the range $y\approx-1$ - $0D$, indicating that the change of $h$ has little influence on the penetration of tangential velocity components. Quantitative discussion on this phenomenon will be presented in \S\ref{subSec:SurfShear}.

The changes in flow statistics caused by the increase in $h$ diminish gradually. Eventually, for $h=5D$ and $9D$, all the statistics become essentially indistinguishable except for $v^{+\prime}_{\rms}$ deep inside the substrate, where the wall-normal fluctuations seem always able to penetrate to the floor, as discussed in \citet{Sharma20dense}. Nevertheless, $v_{\rms}^{+\prime}$  becomes ultimately negligible below $y\approx-4D$ even for the substrate with depth $h=9D$. Above $y\approx-4D$, an asymptotic state is already reached for $h\geq5D$. This suggests that $h=5D$ is a depth sufficient for the overlying turbulence to essentially no longer perceive the floor. This concept of `sufficient depth' will be further investigated in~\S \ref{Sec:unification}.

\section{Scaling of turbulence with substrate parameters}
\label{Sec:FurtherDiscussion}

This section discusses the scaling of the overlying flow properties with substrate parameters. 
\rgm{We compare the raw collapse of properties such as slip velocity, slip length
and, especially, roughness function, with different substrate parameters, without making any \textit{a priori} assumptions on which parameters will produce a better collapse.}
This is used to provide insight into the separate roles in the problem of the macroscale permeability, the microscale granularity, and the substrate depth.

\begin{figure}
  \centerline{\includegraphics[width=1.0\linewidth]{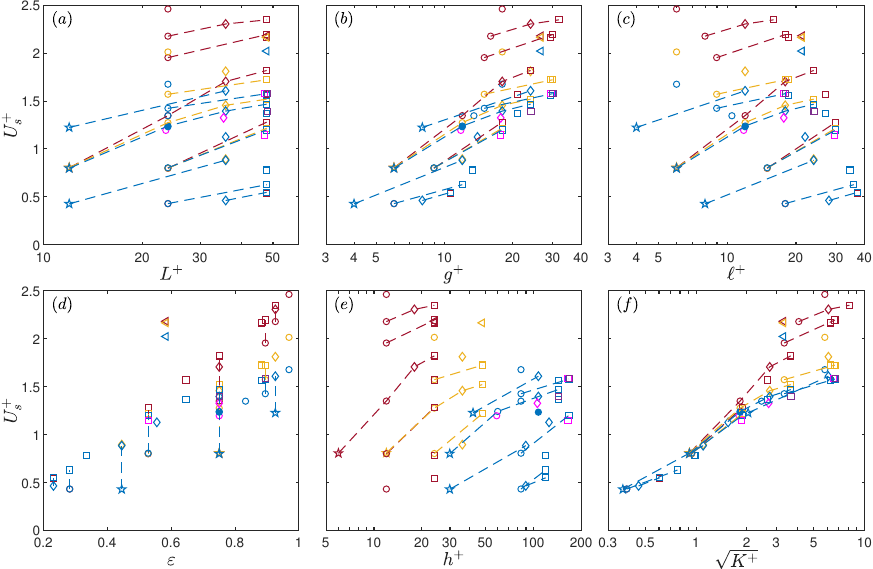}}
  \caption{Mean slip velocity, $U_s^+$, for all the substrates studied versus ($a$) pitch $L^+$, ($b$) gap size $g^+$, ($c$) inclusion size \rgm{$\ell^+$}, ($d$) porosity $\varepsilon$, ($e$) depth $h^+$, ($f$) permeability $\sqrt{K^+}$.
  $\medstar$, $L^+\approx12$; $\medcircle$,~$L^+\approx24$; $\meddiamond$, $L^+\approx36$; $\medsquare$, $L^+\approx48$; $\bullet$, very deep substrate Pd-24-50-VD; \rgm{$\medtriangleleft$, mesh substrates}. Blue, yellow and red colours are for deep porous (Pd), shallow porous (Ps), and rough (Ro) substrates at $Re_\tau\approx180$, respectively; magenta for $Re_\tau\approx360$; purple for $Re_\tau\approx550$. Symbols connected by dashed lines have the same gap-to-pitch ratio $g/L$ and porosity $\varepsilon$.}
  \label{fig:Us_vs_All6Len}
  \vspace*{-2mm}
\end{figure}

\subsection{Slip and shear across the interface}
\label{subSec:SurfShear}

The values of the mean slip velocity $U_s^+$ for all the cases in this study are portrayed versus different substrate parameters in figure \ref{fig:Us_vs_All6Len}. No apparent correlations are found between $U_s^+$ and any of the pitch $L^+$, the gap size $g^+$, the inclusion size \rgm{$\ell^+$}, the porosity $\varepsilon$, or the depth $h^+$. The values of $U_s^+$ for deep porous (Pd) substrates, however, tend to correlate well with $\sqrt{K^+}$, as shown in figure \ref{fig:Us_vs_All6Len}$(f)$. This suggests that the slip velocity for deep substrates is essentially determined by their macroscale permeability, and is not directly associated with the microscale details of the individual grains. Similar results were also observed by \citet{Efstathiou18}. 

\begin{figure} 
  \centerline{\includegraphics[width=1.0\linewidth]{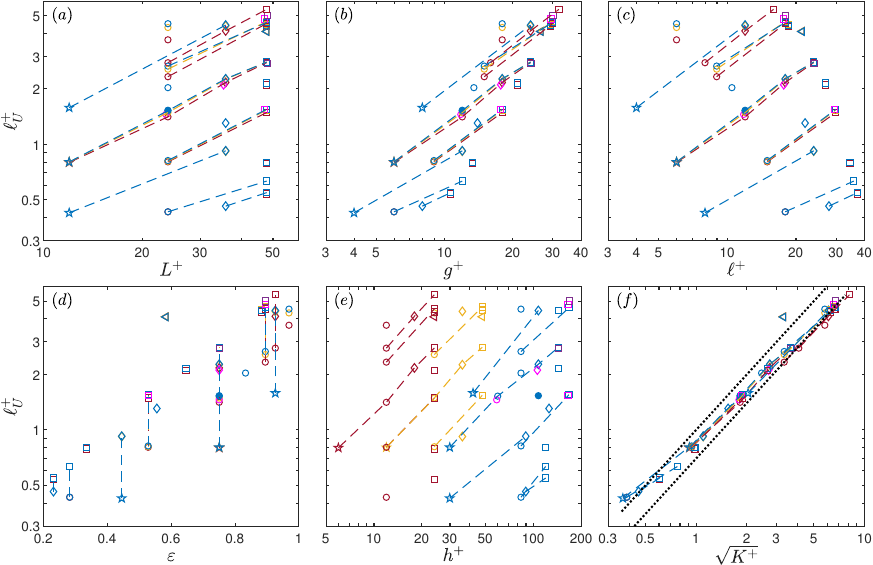}}
  \caption{Mean slip length, $\ell_U^+$, versus ($a$) pitch $L^+$, ($b$) gap size $g^+$, ($c$) inclusion size \rgm{$\ell^+$}, ($d$) porosity $\varepsilon$, ($e$) depth $h^+$, and ($f$) permeability $\sqrt{K^+}$. Symbols and colours are as in figure \ref{fig:Us_vs_All6Len}. The two dotted lines in $(f)$ are for $\ell_U^+=0.7\sqrt{K^+}$ and $\ell_U^+=1.0\sqrt{K^+}$.}
\label{fig:Ls_vs_All6Len}
\end{figure}

For $\sqrt{K^+}\lesssim2$, the values of  $U_s^+$ for rough surfaces in figure \ref{fig:Us_vs_All6Len}$(f)$ agree roughly with $U_s^+$ for their corresponding deep substrates. For $\sqrt{K^+}\gtrsim2$, however, the former are higher and more scattered than the latter.
This is in spite of the slip length having a strong correlation with the permeability,
as shown in figure \ref{fig:Ls_vs_All6Len}, with the slip length defined as
$\ell_U=U_s\,/\,\partial_{y}U|_0$, where $\partial_{y}U|_0$ is the mean shear in the free flow
at $y = 0$. Figure \ref{fig:Ls_vs_All6Len}$(f)$ shows that $\ell_U^+$ is roughly proportional to $\sqrt{K^+}$, with a constant of proportionality of order 0.7-1. This is consistent with the analysis \citep{Abderrahaman17,Gomez19} based on a homogenised model for the subsurface flow, which leads to $\ell_U^+\approx\sqrt{K^+}$. The results in \ref{fig:Ls_vs_All6Len}$(f)$ indicate that under a fixed value of shear at the interface, the slip length is essentially independent of the substrate depth, even for depths as shallow as $h=1D=L/2$.
The different behaviours of the slip length $U_s^+$ and the slip velocity $\ell_U^+$ are caused by different non-zero Reynolds shear stresses $\overline{u^{+\prime}v^{+\prime}}$ at the interface plane. In their absence, the shear in viscous units would be $\partial_{y^+}U^+=1$, and both quantities would have equal value. The comparison of figures \ref{fig:Us_vs_All6Len}$(f)$ and \ref{fig:Ls_vs_All6Len}$(f)$ suggests that the substrate depth plays a key role in this difference.

\begin{figure}
  \centerline{\includegraphics[width=1.0\linewidth]{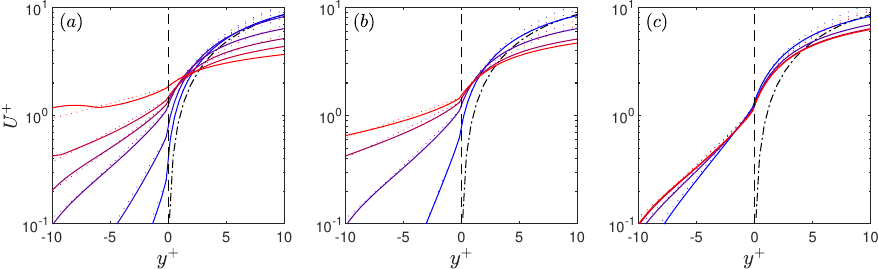}}
  \caption{Mean velocity profiles near the interface. $(a)$, $(b)$, and $(c)$ are for the cases in figures \ref{fig:prfls_L24}, \ref{fig:prfls_Gr50}, and \ref{fig:prfls_PR2450}, respectively, with line styles as in the respective figure.}
\label{fig:prfls_U_surf}
\end{figure}

\begin{figure}
  \centerline{\includegraphics[width=1.0\linewidth]{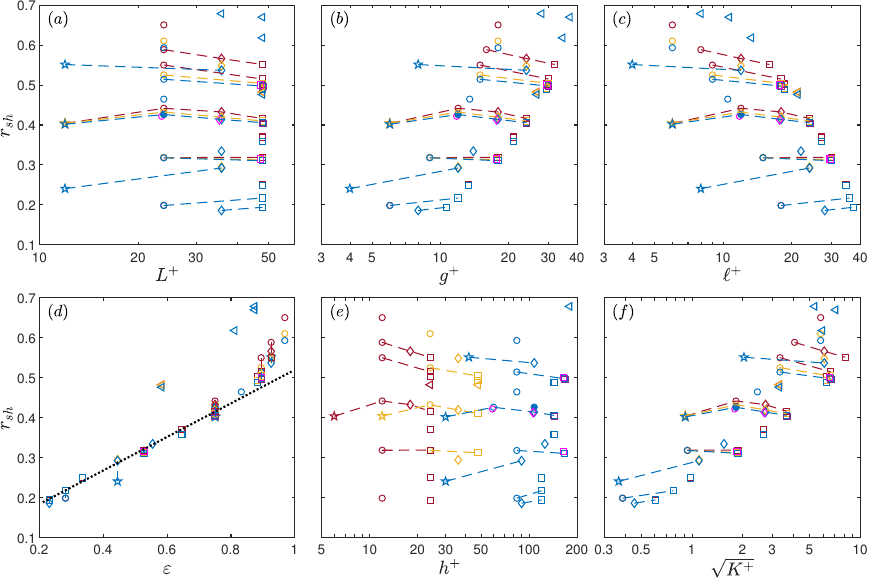}}
  \caption{Ratio of inner to outer shear across the substrate interface, $r_\textit{sh}$, versus ($a$) pitch $L^+$, ($b$) gap size $g^+$, ($c$) inclusion size \rgm{$\ell^+$}, ($d$) porosity $\varepsilon$, ($e$) depth $h^+$, and ($f$) permeability $\sqrt{K^+}$. Symbols and colours are as in figure \ref{fig:Us_vs_All6Len}. The dotted line in $(d)$ is for $r_{sh}=0.42\,\varepsilon+0.10$.}
\label{fig:Rsh0_vs_All6Len}
\end{figure}

Another feature of our porous and rough substrates in terms of tangential velocities is the discontinuity of shear $\partial_yU$ across the interface, shown in figure \ref{fig:prfls_U_surf}. \rgm{A force balance in a thin volume containing the interface
shows that the mean shear stress just above and below are different, as the shear stress above is partly balanced by the shear force between the fluid and the flat top surface of the substrate elements. This effect is in our case concentrated at the element tips, but} can be expected to be more diffuse in substrates composed of rounder grains and with less even interfaces. Figure \ref{fig:Rsh0_vs_All6Len} portrays the ratio of inner to outer shear, $r_{sh}=\partial_yU|_{0^-}\,/\,\partial_yU|_0$, where $\partial_yU|_{0^-}$ is the mean shear approaching $y=0$ from the substrate side, for all the cases. None of the length scales $L^+$, $g^+$, $L^+\!-\!g^+$, $h^+$, and $\sqrt{K^+}$ scale the ratio $r_{sh}$. Instead, $r_{sh}$ appears to correlate with the porosity $\varepsilon$, suggesting that a denser substrate with lower $\varepsilon$ tends to have a stronger jump of shear, i.e.~smaller $r_{sh}$.
This is consistent with the above observation that the discontinuity is caused by
the shear absorbed at the exposed flat faces of the grains, as their surface area is a larger fraction of the interface plane for lower $\varepsilon$.
The ratio $r_\textit{sh}$ is roughly linear with $\varepsilon$ except for highly porous cases ($\varepsilon\gtrsim0.9$), of which $r_\textit{sh}$ adjusts to approach the no-jump asymptotic limit, $\lim_{\varepsilon\to1}r_\textit{sh}=1$.

The observations in this subsection suggest that, for both porous and rough substrates, the interfacial shear jump and the
slip length are mainly influenced by the porosity and the permeability, respectively, with no significant direct influence
of the substrate microscale details or depth. The scaling of slip length with permeability applies also to slip velocities
only for small permeabilities, $\sqrt{K^+}\lesssim2$. For larger ones, the slip velocity also depends on he substrate depth, as the shear Reynolds stress at the interface becomes increasingly significant for the deeper substrates. In any event, we note that
the values of $U_s^+$ are significantly smaller than those of $\Delta U^+$, which implies that the slip plays only a small role in determining the drag.
The effect of substrate granularity and depth are further investigated in \S\ref{subSec:deepPorous} and \S\ref{Sec:unification}.

\subsection{Drag increase and near-interface flow for deep substrates}
\label{subSec:deepPorous}

\begin{figure}
  \vspace*{1.5mm}
  \centerline{\includegraphics[width=1.0\linewidth]{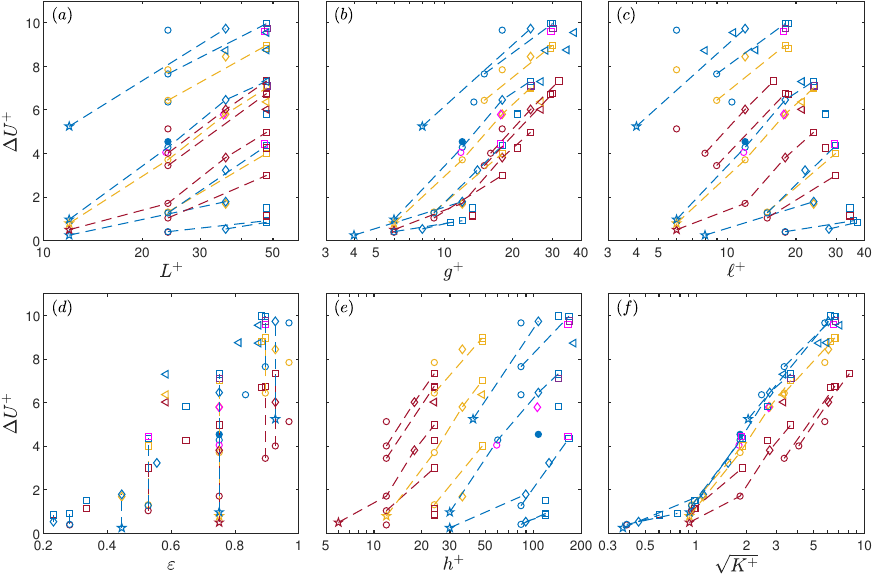}}
  \vspace*{-1.5mm}
  \caption{Velocity deficit $\Delta U^+$ versus ($a$) pitch $L^+$, ($b$) gap size $g^+$, ($c$) inclusion size \rgm{$\ell^+$}, ($d$) porosity $\varepsilon$, ($e$) depth $h^+$, and ($f$) permeability $\sqrt{K^+}$. Symbols and colours are as in figure \ref{fig:Us_vs_All6Len}.}
  \label{fig:DU_vs_All6Len}
\end{figure}

Let us now focus on the drag increase of a substrate, given by $\Delta U^+$ as defined in \S\ref{SubSec:ResultProcessing}. 
Figure \ref{fig:DU_vs_All6Len} portrays $\Delta U^+$ for all the cases simulated versus different substrate parameters. The values of $\Delta U^+$ do not correlate well with any of $L^+$, $g^+$, $L^+\!-\!g^+$, $\varepsilon$, and $h^+$, individually. However, $\Delta U^+$ for deep porous substrates shows a good correlation with $\sqrt{K^+}$, extending to shallow porous (Ps) and rough (Ro) substrates for small permeability, $\sqrt{K^+}\lesssim1$. Beyond this, $\Delta U^+$ for shallow and rough substrates is lower and exhibits more scatter than that for deep ones. In general, the discrepancies between deep and shallow substrates are considerably smaller than those between shallow and rough ones, implying an asymptotic behaviour of $\Delta U^+$ as the depth $h$ increases, similar to the observations in \S\ref{subSec:statistics3}. These results suggest that, in essence, permeability alone determines the drag for sufficiently deep porous substrates.

\begin{figure}
  \centerline{\includegraphics[width=1.0\linewidth]{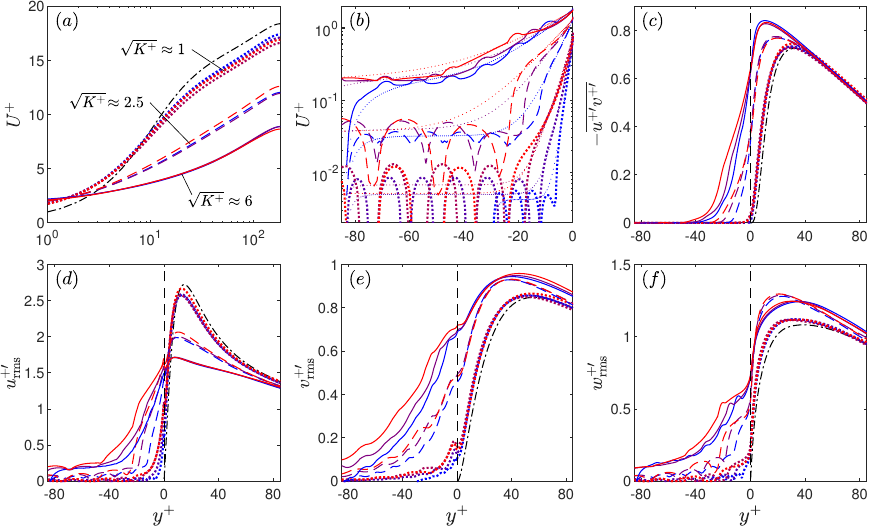}}
  \vspace*{-1mm}
  \caption{($a$-$b$) Mean velocity profile, ($c$) Reynolds shear stress, and ($d$-$f$) RMS velocity fluctuations for deep porous substrates. Blue to red dotted lines are for substrates with different $L^+$ and $\varepsilon$ but similar $\sqrt{K^+}\approx1$, cases Pd-12-50, Pd-24-38, Pd-36-33, and Pd-48-28; dashed for similar $\sqrt{K^+}\approx2.5$, cases Pd-24-56, Pd-36-50, and Pd-48-44; and solid for similar $\sqrt{K^+}\approx6$, Pd-24-75, Pd-36-67, and Pd-48-61. The dash-dotted lines are for smooth-wall data, and the vertical dashed lines mark the location of the free-flow/substrate interface.}
\label{fig:prfls_ReKTri}
\end{figure}

Focusing for now on deep substrates, we observe that the turbulent statistics in general also depend essentially on permeability alone. Figure \ref{fig:prfls_ReKTri} shows that the substrates with different $L^+$ and $\varepsilon$ but similar $\sqrt{K^+}$ have fairly similar mean velocity profiles, turbulent shear stress, and RMS fluctuations in the overlying flow. Differences in the turbulent stress and RMS occur below the interface, where cases with larger $L^+$ or lower $\varepsilon$ tend to have larger magnitudes. These differences are likely attributable to the different substrate geometries causing different dispersive or grain-coherent stresses, although this would require more in-depth analysis.

\begin{figure}
  \vspace{11pt}
  \begin{overpic}[width=1.0\textwidth]{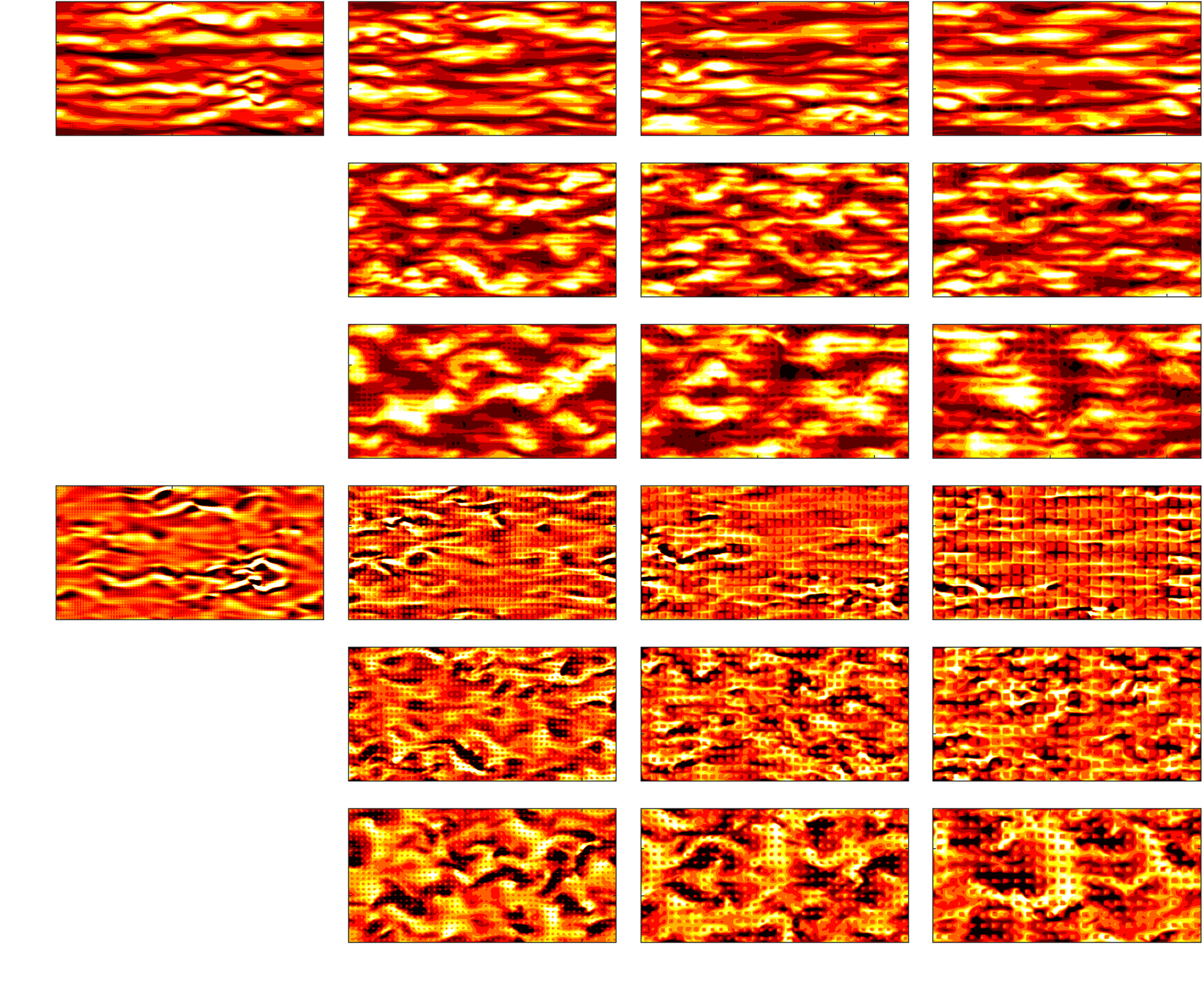}

  \put (12.4, 82.6) {\footnotesize$L^+\approx12$}
  \put (36.5, 82.6) {\footnotesize$L^+\approx24$}
  \put (60.6, 82.6) {\footnotesize$L^+\approx36$}
  \put (84.7, 82.6) {\footnotesize$L^+\approx48$}

  \put ( 2.3, 80.6) {\scriptsize$(\!a\!)$}
  \put (26.9, 80.6) {\scriptsize$(\!b\!)$}
  \put (51.2, 80.6) {\scriptsize$(\!c\!)$}
  \put (75.4, 80.6) {\scriptsize$(\!d\!)$}
  \put (26.9, 67.2) {\scriptsize$(\!e\!)$}
  \put (51.1, 67.2) {\scriptsize$(\!f\!)$}
  \put (75.4, 67.2) {\scriptsize$(\!g\!)$}
  \put (26.8, 53.8) {\scriptsize$(\!h\!)$}
  \put (51.2, 53.8) {\scriptsize$(\!i\!)$}
  \put (75.4, 53.8) {\scriptsize$(\!j\!)$}

  \put ( 2.3, 40.4) {\scriptsize$(\!k\!)$}
  \put (26.9, 40.4) {\scriptsize$(\!\ell\!)$}
  \put (51.1, 40.4) {\scriptsize$(\!\!m\!\!)$}
  \put (75.4, 40.4) {\scriptsize$(\!n\!\!)$}
  \put (26.9, 27.0) {\scriptsize$(\!o\!)$}
  \put (51.1, 27.0) {\scriptsize$(\!p\!)$}
  \put (75.4, 27.0) {\scriptsize$(\!q\!)$}
  \put (26.8, 13.6) {\scriptsize$(\!r\!)$}
  \put (51.2, 13.6) {\scriptsize$(\!s\!)$}
  \put (75.5, 13.6) {\scriptsize$(\!t\!)$}

  \put (-0.7, 75.8) {\footnotesize\rotatebox{90}{$z^+$}}
  \put ( 3.4, 70.3) {\scriptsize$0$}
  \put ( 1.3, 74.3) {\scriptsize$200$}
  \put ( 1.3, 78.2) {\scriptsize$400$}
  \put (23.6, 62.4) {\footnotesize\rotatebox{90}{$z^+$}}
  \put (27.7, 56.9) {\scriptsize$0$}
  \put (25.6, 60.9) {\scriptsize$200$}
  \put (25.6, 64.8) {\scriptsize$400$}
  \put (23.6, 49.0) {\footnotesize\rotatebox{90}{$z^+$}}
  \put (27.7, 43.5) {\scriptsize$0$}
  \put (25.6, 47.5) {\scriptsize$200$}
  \put (25.6, 51.4) {\scriptsize$400$}
  
  \put (-0.7, 35.6) {\footnotesize\rotatebox{90}{$z^+$}}
  \put ( 3.4, 30.1) {\scriptsize$0$}
  \put ( 1.3, 34.1) {\scriptsize$200$}
  \put ( 1.3, 38.0) {\scriptsize$400$}
  \put (23.6, 22.2) {\footnotesize\rotatebox{90}{$z^+$}}
  \put (27.7, 16.7) {\scriptsize$0$}
  \put (25.6, 20.7) {\scriptsize$200$}
  \put (25.6, 24.6) {\scriptsize$400$}
  \put (23.6,  8.8) {\footnotesize\rotatebox{90}{$z^+$}}
  \put (27.7,  3.3) {\scriptsize$0$}
  \put (25.6,  7.3) {\scriptsize$200$}
  \put (25.6, 11.2) {\scriptsize$400$}

  \put (15.0, 67.4) {\footnotesize{$x^+$}}
  \put (12.7, 69.4) {\scriptsize$500$}
  \put (21.9, 69.4) {\scriptsize$1000$}

  \put (15.0, 27.2) {\footnotesize{$x^+$}}
  \put (12.7, 29.2) {\scriptsize$500$}
  \put (21.9, 29.2) {\scriptsize$1000$}

  \put (39.3,  0.4) {\footnotesize{$x^+$}}
  \put (37.0,  2.4) {\scriptsize$500$}
  \put (46.3,  2.4) {\scriptsize$1000$}

  \put (63.7,  0.4) {\footnotesize{$x^+$}}
  \put (52.6,  2.4) {\scriptsize$0$}
  \put (61.4,  2.4) {\scriptsize$500$}
  \put (70.7,  2.4) {\scriptsize$1000$}

  \put (88.1,  0.4) {\footnotesize{$x^+$}}
  \put (77.0,  2.4) {\scriptsize$0$}
  \put (85.8,  2.4) {\scriptsize$500$}
  \put (95.1,  2.4) {\scriptsize$1000$}

  \end{overpic}
  \caption{Instantaneous fields of ($a$-$j$) $u^\prime$  and ($k$-$t$) $v^\prime$ at $y^+\!\approx\!3$ for the same deep porous substrates of figure \ref{fig:prfls_ReKTri}.  Columns from left to right correspond to substrates with $L^+\approx12$, $24$, $36$, and $48$, respectively.
  $(a,b,c,d)$ and $(k,l,m,n)$, substrates with $\sqrt{K^+}\!\approx\!1$; $(e,f,g)$ and $(o,p,q)$, with $\sqrt{K^+}\!\approx\!2.5$; $(h,i,j)$ and $(r,s,t)$, with $\sqrt{K^+}\!\approx\!6$. Colours from dark to clear are for the value range $[-2:2]$ relative to the RMS value of the variable at that plane.}
\label{fig:Contours_AllReKs_UV}
\end{figure}

Some more details of the structure of turbulence near the interface can be illustrated by instantaneous flow fields and energy spectral densities. The flow fields portrayed in figure \ref{fig:Contours_AllReKs_UV} exhibit a signature of the grain-coherent flow with a characteristic length scale $L^+$. Superimposed with this signature, we can observe the grain-incoherent features of the background turbulence. Just as the flow statistics in figure \ref{fig:prfls_ReKTri}, the background turbulence is visually similar for cases with different $L^+$ and $\varepsilon$ but similar $\sqrt{K^+}$, while the grain-coherent flow varies greatly with $L^+$. The four cases with $\sqrt{K^+}\approx1$ exhibit the typical features of smooth-wall turbulence in the streamwise elongated shapes in $u^\prime$, shown in figures \ref{fig:Contours_AllReKs_UV}($a$-$d$), and $v^\prime$, shown in figures \ref{fig:Contours_AllReKs_UV}($k$-$n$). The streamwise elongation of these structures is disrupted for $\sqrt{K^+}\approx2.5$, as shown in figures \ref{fig:Contours_AllReKs_UV}($e$-$g,o$-$q$), and entirely lost for $\sqrt{K^+}\approx6$, figures \ref{fig:Contours_AllReKs_UV}($h$-$j$,$r$-$t$), for which eddies have an $x$-$z$ aspect ratio closer to unity.

\begin{figure}
\vspace*{2mm}
  \centerline{\includegraphics[width=1.0\linewidth]{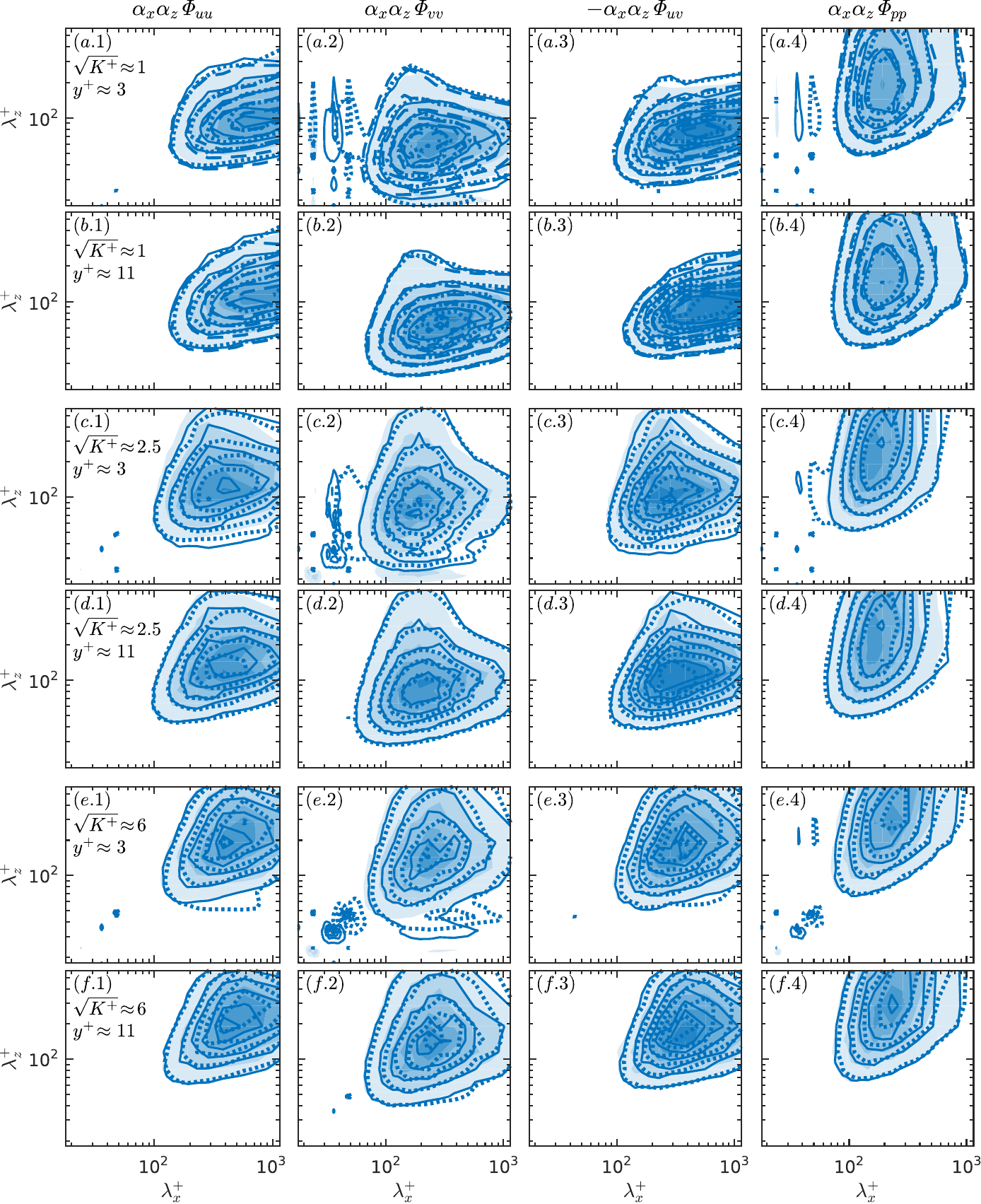}}
  \caption{Pre-multiplied spectra $\alpha_x\alpha_z\Phi_{**}$ at $(a,c,e)$ $y^+\!\approx\!3$ and $(b,d,f)$ $y^+\!\approx\!11$ for the same deep porous substrates of figure \ref{fig:prfls_ReKTri}.
  $(a,b)$, substrates with $\sqrt{K^+}\!\approx\!1$; $(c,d)$, with $\sqrt{K^+}\!\approx\!2.5$;
  $(e,f)$, with $\sqrt{K^+}\!\approx\!6$. Dashed lines are for $L^+\!\approx\!12$, shaded contours for $L^+\!\approx\!24$, solid lines for $L^+\!\approx\!36$, and dotted lines for $L^+\!\approx\!48$. The contours mark values [0.044:0.044:0.264] relative to the corresponding variance or covariance.}
\label{fig:Spectra4_AllReKs}
\end{figure}

The above discussion is also consistent with the statistical information displayed in the spectral density maps of figure \ref{fig:Spectra4_AllReKs}. At $y^+\approx3$, these maps present two distinct features: a main spectral region attributable to the background turbulence, and smaller lobes centred about the grain-spacing wavelengths, caused by the grain-coherent flow. These lobes extend beyond the mere harmonics of the texture because of the amplitude modulation of the grain-coherent flow by the background turbulence \rgm{\citep{Abderrahaman19,Khorasani2024}}. The cases with similar $\sqrt{K^+}$ show good agreement for the background turbulence, but differ in the regions produced by the grain-coherent flow due to their different $L^+$. The grain-coherent flow quickly decays away from the interface, as evidenced in the maps at $y^+\approx11$. For the background turbulence, as $\sqrt{K^+}$ increases, the spectral densities become lower in $x$-elongated wavelengths but higher in wider $z$-wavelengths.

The discussion in this subsection suggests that the effect of deep porous substrates on the overlying turbulence is essentially governed by the permeability, a characteristic not directly associated with the geometric microscale detail of individual grains in a porous medium. The grain-coherent flow near the interface, in turn, manifests the effect of the granularity, but decays quickly away from the substrate, at least for the grain \rgm{pitches} here considered, $L^+\lesssim50$ and $g^+\lesssim30$.

\section{A unified characterisation from porous to rough substrates}
\label{Sec:unification}

The mean velocity deficit $\Delta U^+$ in figure \ref{fig:DU_vs_All6Len}$(d)$ correlates well with permeability $K^+$ only for deep porous substrates, but not for rough surfaces. Unfortunately, a general approach to determine the scaling of drag for rough surfaces remains elusive \citep[see][for a review]{Chung21}. The substrates in this study have been designed to transition continuously from deep-porous to rough-but-impermeable geometry as the depth $h$ decreases, while retaining the same grain and interface topology. In this section, our aim is to identify a scaling law for $\Delta U^+$ that captures this continuous transition. With this aim, we now focus on the effect of substrate depth.

\subsection{An equivalent permeability incorporating the effect of depth}
\label{subSec:Keq}

\begin{figure}
\vspace*{2mm}
  \centerline{\includegraphics[width=1.0\linewidth]{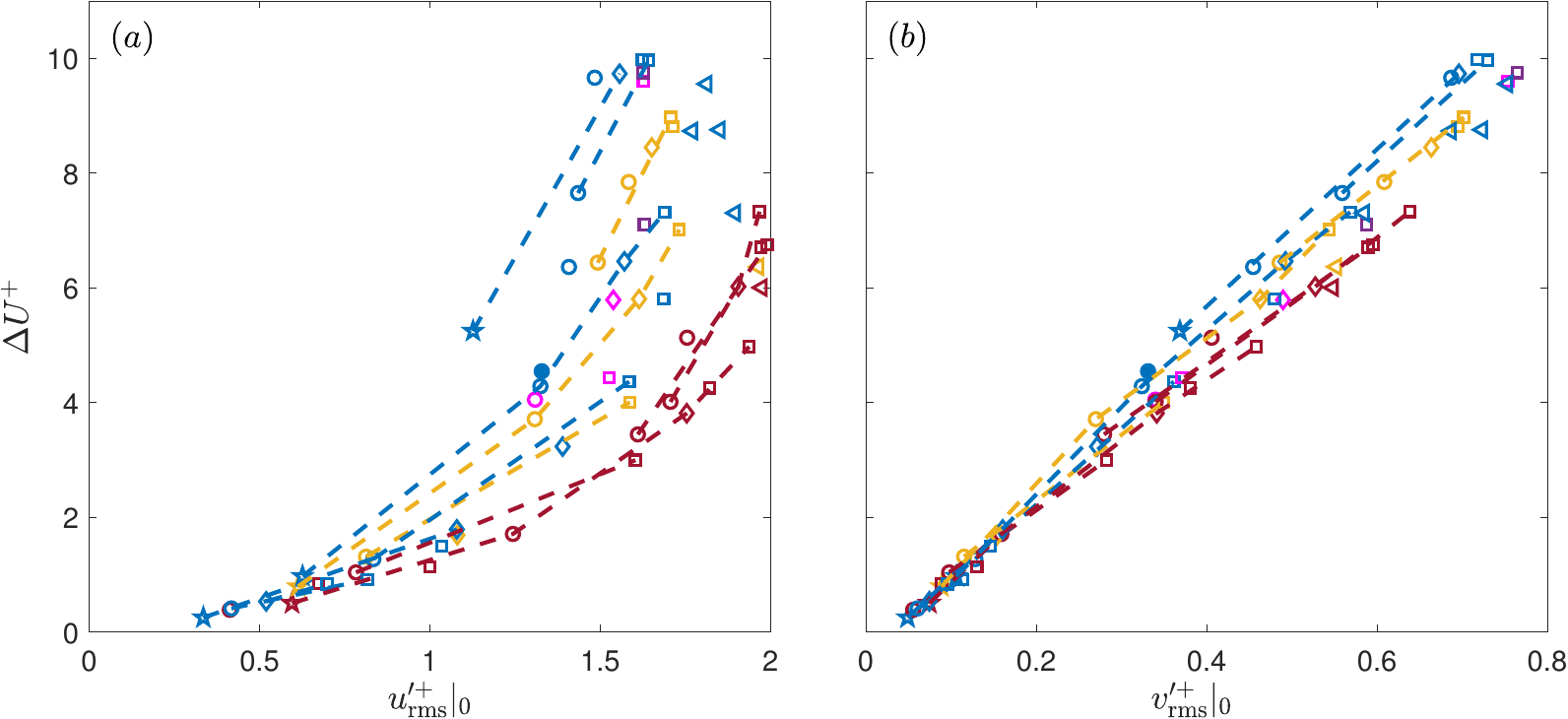}}
  \vspace*{2mm}
  \centerline{\includegraphics[width=1.0\linewidth]{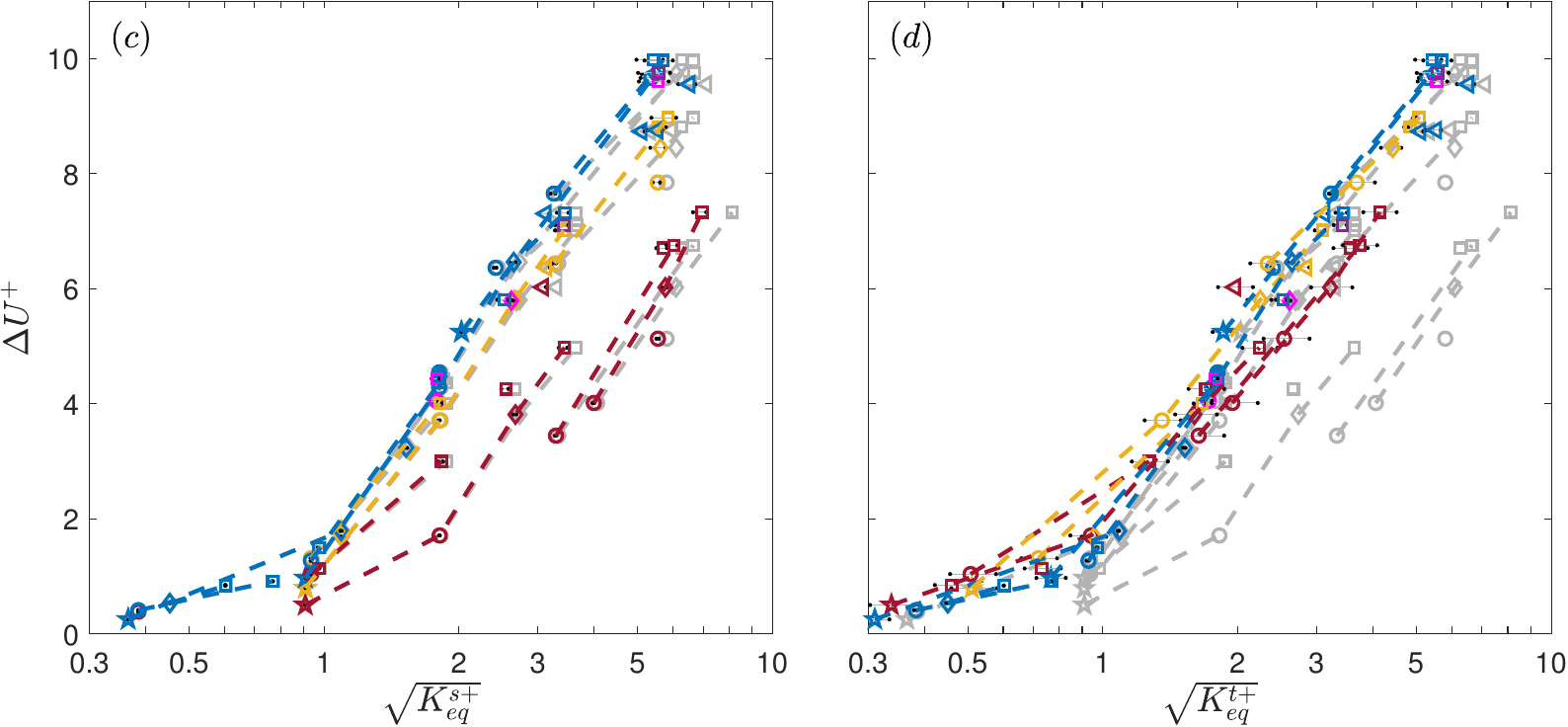}}
  \caption{Velocity deficit $\Delta U^+$ for all the cases studied as a function of ($a$) the RMS of the interfacial  $u^\prime$; ($b$) the RMS of the interfacial $v^\prime$; ($c$) the slip-based equivalent permeability $K_{eq}^{s+}$; and ($d$) the transpiration-based equivalent permeability $K_{eq}^{t+}$ (right). Lines and symbols are as in figure \ref{fig:Us_vs_All6Len}. 
  In ($d$), $K_{eq}^{t+}$ has been calculated for a characteristic near-wall pressure lengthscale $\lambda_p^+=200$, and the error bars represent the range $\lambda_p^+=150$-$250$.
  The values of $\Delta U^+$ versus $\sqrt{K^+}$ from figure \ref{fig:DU_vs_All6Len}$(f)$ are displayed in grey for comparison.}
\label{fig:DU_vs_uvKeq}
\end{figure}

In \S\ref{subSec:statistics3} and \S\ref{subSec:SurfShear}, we have discussed the relatively small influence of depth on the tangential velocity and interfacial shear and slip, i.e.~a porous substrate and a typical rough surface with identical grain geometry have fairly similar subsurface decay of the tangential mean and fluctuating velocities, as shown in figure \ref{fig:prfls_PR2450}$(b,d,f)$, and also similar interfacial slip and shear properties, as shown in figures \ref{fig:Ls_vs_All6Len}$(f)$ and \ref{fig:Rsh0_vs_All6Len}$(d)$. Therefore, we can infer that the apparent differences in drag increase between porous and rough surfaces in figure \ref{fig:DU_vs_All6Len}$(f)$ principally originate from their differences in interfacial transpiration. This is further supported by figures \ref{fig:DU_vs_uvKeq}($a$) and ($b$), which show that, for all the substrates studied, the drag increase is highly correlated with the intensity of the interfacial wall-normal velocity fluctuation, rather than with the tangential one, as is the case also for rough surfaces \rgm{\citep{Orlandi06,Orlandi06JoT,Orlandi08}}.

The effect of substrate depth on the interfacial slip and transpiration can be observed in the analytical solution of the homogenised Darcy-Brinkman model for the flow within the substrate \citep{GmezdeSegura2018,Gomez19}. Appendix \ref{Sec:Brinkman} presents this model for the case of isotropic substrates and discusses the effect of depth as deduced from its solution. The model ultimately results in \rgm{an admittance,} linear relationship, equation (\ref{eq:BrkmConstitutive}), which gives the interfacial slip and transpiration velocities in response to the interfacial shear and normal stresses. This relationship involves five coefficients,
$\mathcal{L}_{slip}$, $\mathcal{K}_{slip}$, $\mathcal{N}_{trsp}$, $\mathcal{K}_{trsp}$, and $\mathcal{L}_{slip}^{\bot}$, which characterise the effect of the substrate on the overlying free flow. Each coefficient can be written as a product of two parts: a component that is a function of permeability $K$, and a dimensionless attenuating function $f_{*}(\alpha,h)$ that depends on the wavenumber $\alpha$ of the exciting stress and the substrate depth $h$. 

The function $f_{*}(\alpha,h)$ tends to unity for deep substrates and long exciting waves, and to vanish for shallow substrates or for small waves, making the substrates effectively impermeable and smooth in the latter cases. The first component can be thus identified as the admittance coefficient relating interfacial velocity and stress for large-scale flows over sufficiently deep substrates. One exception to this, though, is the coefficient $\mathcal{K}_{trsp}=K f_{\mathcal{K}t}(\alpha,h)$ relating the transpiration velocity $v^\prime$ and the pressure fluctuation $p^\prime$, which also vanishes for very long waves, and is maximum for an intermediate wavelength comparable to the depth. The discussion in \ref{subSec:BrkmInfer} highlights that, as $h$ increases, $\mathcal{K}_{trsp}$ reaches its asymptotic value significantly more slowly than the other four coefficients for the typical wavelengths in wall turbulence. This indicates that the principal effect of a finite depth is to decrease $\mathcal{K}_{trsp}$, i.e.~to suppress the pressure-excited transpiration at the interface.
Given the strong correlation of this transpiration with $\Delta U^+$ discussed above, we would expect this suppression effect to play a leading role in determining the effect of substrate depth on drag.

The above discussion is based on Darcy-Brinkman solutions, which are after all a mere model for the homogenised subsurface flow \rgm{-- they fail to capture, for instance, the discontinuity in macroscopic velocity shear across the interface, or the direct effect of granularity, which grows with the texture size $L^+$}. We thus simply use these solutions to guide us in proposing an empirical `equivalent permeability' $K_{eq}^t$ to incorporate the effect of substrate depth on transpiration,
\begin{equation}
\label{eq:DefKeq4}
    K_{eq}^t \equiv K \, f_{\mathcal{K}t}(\tilde{\alpha},\tilde{h}),
\end{equation}
where the attenuating function $f_{\mathcal{K}t}(\tilde{\alpha},\tilde{h})$ is calculated from equation (\ref{eq:fh}$d$), and where the dimensionless wavenumber $\tilde{\alpha}$ and depth $\tilde{h}$ are defined by (\ref{eq:dimlessAlpha}) and (\ref{eq:dimlessH}). The value of $\alpha$ in (\ref{eq:DefKeq4}) should be chosen to represent the characteristic scale of the typical near-wall pressure fluctuations that excite transpiration at the interface. Informed by the spectra in figure \ref{fig:Spectra4_AllReKs}, we assume a characteristic wavelength $\lambda_p=150$ - $250\,\nu/u_\tau$ and thus $\alpha=\alpha_p=2\pi/\lambda_p$. In addition to $K_{eq}^t$ defined by (\ref{eq:DefKeq4}), for comparison we also define another `equivalent permeability' that incorporates the effect of substrate depth on slip,
\begin{equation}
\label{eq:DefKeq1}
    K_{eq}^s \equiv K \, f_{\mathcal{L}s}(\tilde{\alpha},\tilde{h}),
\end{equation}
where the attenuating function $f_{\mathcal{L}s}(\tilde{\alpha},\tilde{h})$ is calculated from equation (\ref{eq:fh}$a$). 
\rgm{This $K_{eq}^s$ gives a measure of the response of the substrate to an overlying shear, and is therefore an inherently
interfacial quantity, while $K_{eq}^t$ gives a measure of the penetration of wall-normal flow due to pressure
fluctuations, and therefore accounts for the properties of the substrate not just near the interface but deeper within.
}
\rgm{An alternative set of admittance relationships to equation (\ref{eq:BrkmConstitutive}) can be derived from homogenisation
\citep{Bottaro2020,Naqvi21}. This gives a set of upscaled coefficients that play the same role of $K_{eq}^t$, $K_{eq}^s$, and the other
coefficients derived from a Darcy-Brinkman model. The scaling
of the roughness function with those upscaled coefficients is portrayed in Appendix~\ref{Sec:homog}, and the collapse is generally not as good as the one discussed below with $\sqrt{K_{eq}^{t+}}$. \finalrev{In the authors' view,} the reason is likely the lack of separation of scales between turbulence and surface texture, which is a central assumption in homogenisation.
}

The values of $\Delta U^+$ against the newly defined $\sqrt{K_{eq}^{s+}}$ and $\sqrt{K_{eq}^{t+}}$ are portrayed in  \ref{fig:DU_vs_uvKeq} for all porous and rough cases. Figure \ref{fig:DU_vs_uvKeq}$(c)$ shows that the values of $\sqrt{K_{eq}^{s+}}$ are close to those of $\sqrt{K^+}$. This is consistent with the observations in \S\ref{subSec:SurfShear} and the analysis in \S\ref{subSec:BrkmInfer}, which suggests that
the effect of depth on interfacial slip is small. As a result,
$\sqrt{K_{eq}^{s+}}$ and $\sqrt{K^+}$ produce a similarly poor collapse for $\Delta U^+$ across all the substrates studied.
Meanwhile, the values of $\Delta U^+$ across all substrates collapse well with $\sqrt{K_{eq}^{t+}}$, as shown in figure \ref{fig:DU_vs_uvKeq}$(d)$. For rough surfaces, $\sqrt{K_{eq}^{t+}}$ is significantly smaller than the original $\sqrt{K^+}$, indicating that their small depths suppress significantly the interfacial transpiration. For deep porous substrates, such suppression tends to vanish and the differences between $\sqrt{K_{eq}^{t+}}$ and $\sqrt{K^+}$ are small.

\begin{figure}
\vspace*{0.5mm}
  \centerline{\includegraphics[width=1.0\linewidth]{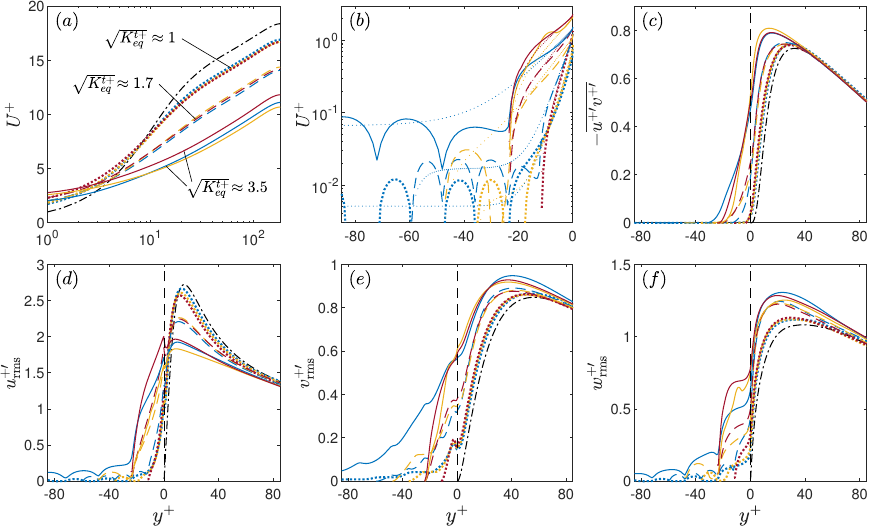}}
\vspace*{0mm}
  \caption{($a$-$b$) Mean velocity profile, ($c$) Reynolds shear stress, and ($d$-$f$) RMS velocity fluctuations for cases with similar $\sqrt{K_{eq}^{t+}}$. Blue, yellow, and red are for deep porous (Pd) substrates ($h/D\geq5$), shallow porous (Ps) substrates ($h/D=2$), and rough surfaces ($h/D=1$), respectively. Dotted lines are for the cases with different $L^+$, $\varepsilon$, and $h/D$ but similar $\sqrt{K_{eq}^{t+}}\approx1$: Pd-48-28 ($\sqrt{K^+}\!\!=\!0.98$, $\sqrt{K_{eq}^{t+}}\!=\!0.97$), Ps-36-33 ($\sqrt{K^+}\!\!=\!1.10$, $\sqrt{K_{eq}^{t+}}\!=\!0.96$), and Ro-24-50 ($\sqrt{K^+}\!\!=\!1.82$, $\sqrt{K_{eq}^{t+}}\!=\!0.94$). Dashed lines are for the cases with similar $\sqrt{K_{eq}^{t+}}\approx1.7$: Pd-24-50 ($\sqrt{K^+}\!\!=\!1.82$, $\sqrt{K_{eq}^{t+}}\!=\!1.75$), Ps-48-38 ($\sqrt{K^+}\!\!=\!1.88$, $\sqrt{K_{eq}^{t+}}\!=\!1.66$), and Ro-48-44 ($\sqrt{K^+}\!\!=\!2.66$, $\sqrt{K_{eq}^{t+}}\!=\!1.70$). Solid lines are for the cases with similar $\sqrt{K_{eq}^{t+}}\approx3.5$: Pd-48-50 ($\sqrt{K^+}\!\!=\!3.64$, $\sqrt{K_{eq}^{t+}}\!=\!3.44$), Ps-24-75 ($\sqrt{K^+}\!\!=\!5.80$, $\sqrt{K_{eq}^{t+}}\!=\!3.68$), and Ro-48-61 ($\sqrt{K^+}\!\!=\!6.28$, $\sqrt{K_{eq}^{t+}}\!=\!3.55$).
  The dash-dotted lines are for smooth-wall data, and the vertical dashed lines mark the location of the free-flow/substrate interface.}
\label{fig:prfls_ReKeqTri}
\vspace*{-2mm}
\end{figure}

\begin{figure}
\centering
\vspace*{4mm}
  \begin{overpic}[width=1.0\textwidth]{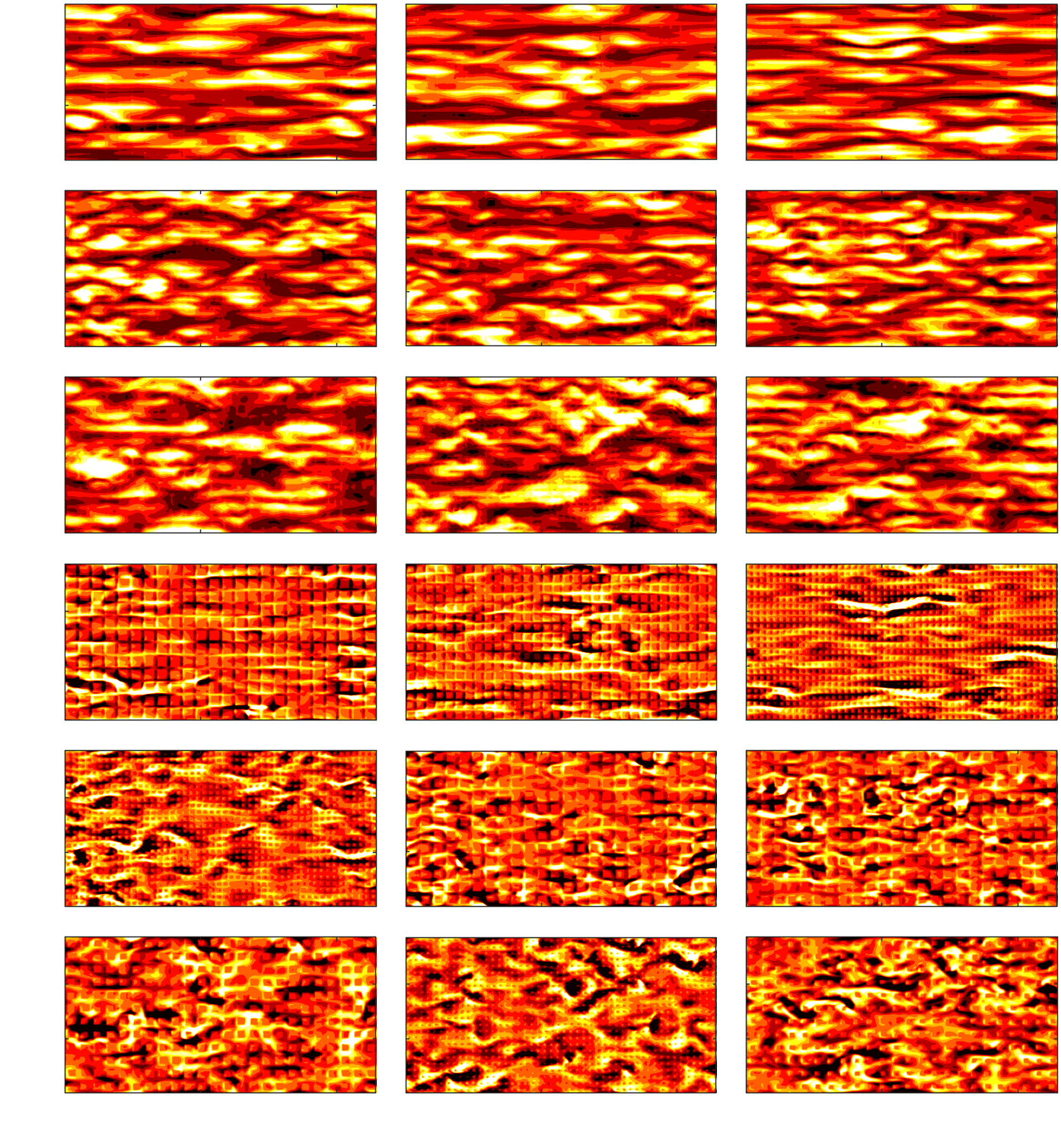}

  \put (11.1, 100.3) {\footnotesize$\text{Deep porous (Pd)}$}
  \put (39.4, 100.3) {\footnotesize$\text{Shallow porous (Ps)}$}
  \put (72.6, 100.3) {\footnotesize$\text{Rough (Ro)}$}

  \put ( 3.5, 98.3) {\scriptsize$(\!a\!)$}
  \put (33.4, 98.3) {\scriptsize$(\!b\!)$}
  \put (63.0, 98.3) {\scriptsize$(\!c\!)$}
  \put ( 3.5, 82.0) {\scriptsize$(\!d\!)$}
  \put (33.4, 82.0) {\scriptsize$(\!e\!)$}
  \put (62.8, 82.0) {\scriptsize$(\!f\!)$}
  \put ( 3.5, 65.7) {\scriptsize$(\!g\!)$}
  \put (33.2, 65.7) {\scriptsize$(\!h\!)$}
  \put (63.1, 65.7) {\scriptsize$(\!i\!)$}
  \put ( 3.6, 49.4) {\scriptsize$(\!j\!)$}
  \put (33.2, 49.4) {\scriptsize$(\!k\!)$}
  \put (63.0, 49.4) {\scriptsize$(\!\ell\!)$}
  \put ( 3.1, 33.2) {\scriptsize$(\!m\!)$}
  \put (33.2, 33.2) {\scriptsize$(\!n\!)$}
  \put (63.0, 33.2) {\scriptsize$(\!o\!)$}
  \put ( 3.5, 16.9) {\scriptsize$(\!p\!)$}
  \put (33.3, 16.9) {\scriptsize$(\!q\!)$}
  \put (63.0, 16.9) {\scriptsize$(\!r\!)$}

  \put ( 0.0, 92.2) {\footnotesize\rotatebox{90}{$z^+$}}
  \put ( 2.5, 95.2) {\scriptsize$400$}
  \put ( 2.5, 90.3) {\scriptsize$200$}
  \put ( 4.4, 85.4) {\scriptsize$0$}

  \put ( 0.0, 75.9) {\footnotesize\rotatebox{90}{$z^+$}}
  \put ( 2.5, 78.8) {\scriptsize$400$}
  \put ( 2.5, 74.0) {\scriptsize$200$}
  \put ( 4.4, 69.1) {\scriptsize$0$}

  \put ( 0.0, 59.6) {\footnotesize\rotatebox{90}{$z^+$}}
  \put ( 2.5, 62.5) {\scriptsize$400$}
  \put ( 2.5, 57.7) {\scriptsize$200$}
  \put ( 4.4, 52.9) {\scriptsize$0$}

  \put ( 0.0, 43.3) {\footnotesize\rotatebox{90}{$z^+$}}
  \put ( 2.5, 46.2) {\scriptsize$400$}
  \put ( 2.5, 41.4) {\scriptsize$200$}
  \put ( 4.4, 36.6) {\scriptsize$0$}

  \put ( 0.0, 27.0) {\footnotesize\rotatebox{90}{$z^+$}}
  \put ( 2.5, 29.9) {\scriptsize$400$}
  \put ( 2.5, 25.1) {\scriptsize$200$}
  \put ( 4.4, 20.3) {\scriptsize$0$}

  \put ( 0.0, 10.8) {\footnotesize\rotatebox{90}{$z^+$}}
  \put ( 2.5, 13.7) {\scriptsize$400$}
  \put ( 2.5,  8.9) {\scriptsize$200$}
  \put ( 4.4,  4.1) {\scriptsize$0$}

  \put (18.0, 1.0) {\footnotesize{$x^+$}}
  \put (15.9, 3.3) {\scriptsize$500$}
  \put (27.4, 3.3) {\scriptsize$1000$}

  \put (47.7, 1.0) {\footnotesize{$x^+$}}
  \put (34.8, 3.3) {\scriptsize$0$}
  \put (45.7, 3.3) {\scriptsize$500$}
  \put (57.0, 3.3) {\scriptsize$1000$}

  \put (77.3, 1.0) {\footnotesize{$x^+$}}
  \put (64.4, 3.3) {\scriptsize$0$}
  \put (75.3, 3.3) {\scriptsize$500$}
  \put (86.7, 3.3) {\scriptsize$1000$}

  \end{overpic}
\vspace*{-5mm}
  \caption{Instantaneous fields of ($a$-$i$) $u^\prime$  and ($j$-$r$) $v^\prime$ at $y^+\!\approx\!3$ for the same substrates of figure \ref{fig:prfls_ReKeqTri}.  Columns from left to right correspond to deep porous (Pd), shallow porous (Ps), and rough (Ro) substrates, respectively. $(a,b,c)$ and $(j,k,l)$, substrates with $\sqrt{K_{eq}^{t+}}\!\approx\!1$;
  $(d,e,f)$ and $(m,n,o)$, with $\sqrt{K_{eq}^{t+}}\!\approx\!1.7$; $(g,h,i)$ and $(p,q,r)$, with $\sqrt{K_{eq}^{t+}}\!\approx\!3.5$. Colours from dark to clear are for the value range $[-2:2]$ relative to the RMS value of the variable at that plane.}
\label{fig:Contours_AllReKeqs_UV-200}
\vspace*{-2mm}
\end{figure}

\begin{figure}
\vspace*{4mm}
  \centerline{\includegraphics[width=1.0\linewidth]{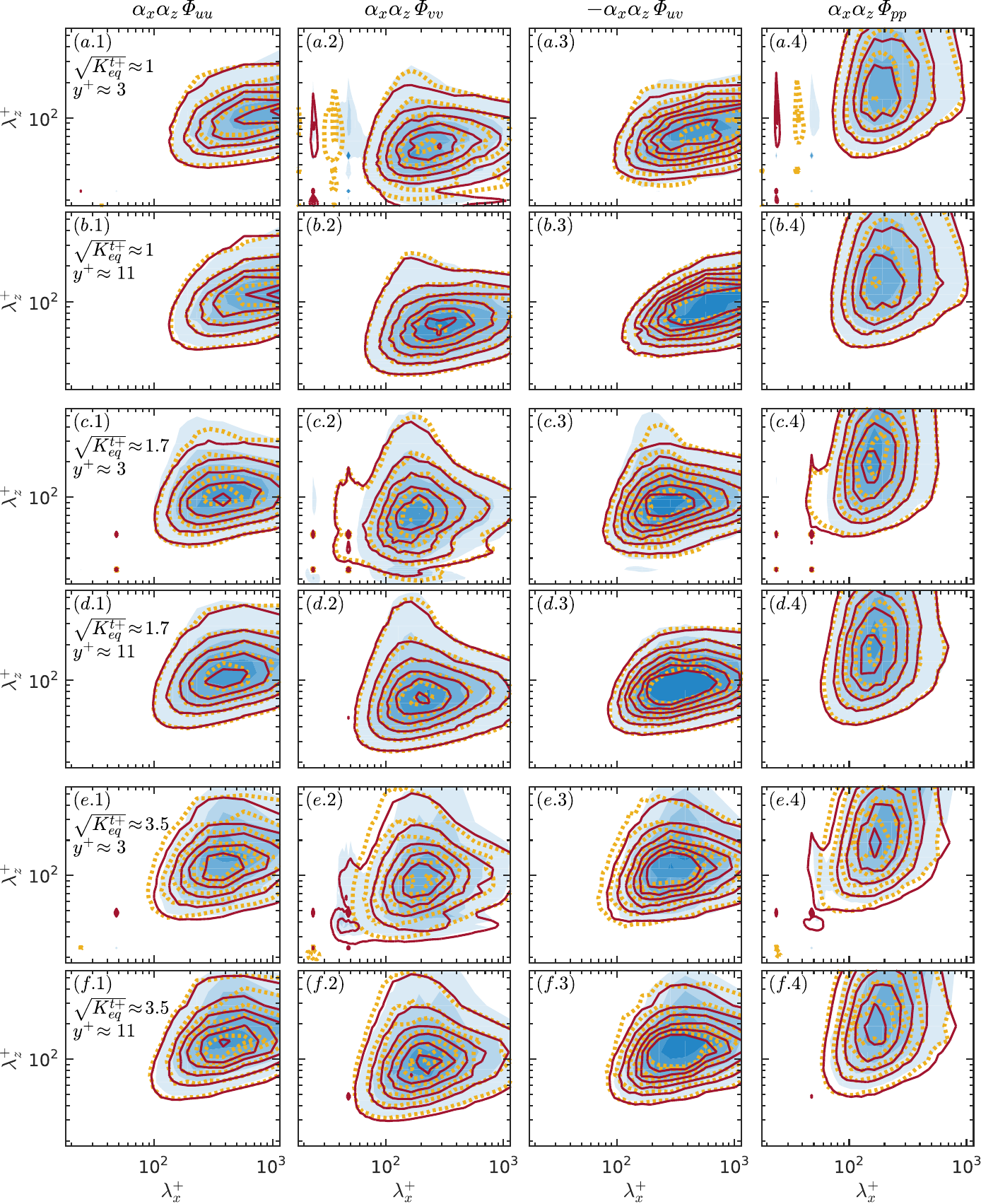}}
  \caption{Pre-multiplied spectra $\alpha_x\alpha_z\Phi_{**}$ at $(a,c,e)$ $y^+\!\approx\!3$ and $(b,d,f)$ $y^+\!\approx\!11$ for the same deep porous substrates of figure \ref{fig:prfls_ReKeqTri}.
  $(a,b)$, substrates with $\sqrt{K_{eq}^{t+}}\!\approx\!1$; $(c,d)$, with $\sqrt{K_{eq}^{t+}}\!\approx\!1.7$;
  $(e,f)$, with $\sqrt{K_{eq}^{t+}}\!\approx\!3.5$. Shaded contours are for deep porous (Pd) substrates, yellow dotted lines for shallow porous (Ps) substrates, and red solid lines for rough (Ro) surfaces. The contours mark values [0.044:0.044:0.264] relative to the corresponding variance or covariance.} 
\vspace*{3mm}
\label{fig:Spectra4_AllReKeqsDualYs}
\end{figure}

The transpiration-based equivalent permeability, $\sqrt{K_{eq}^{t+}}$, also characterises well the turbulence for substrates with different depths. In figures \ref{fig:prfls_ReKeqTri} to  \ref{fig:Spectra4_AllReKeqsDualYs}, we consider three groups of substrates, each including a deep porous (Pd), a shallow porous (Ps), and a rough (Ro) case with different depth $h/D$
 but similar $\sqrt{K_{eq}^{t+}}$.
Figure \ref{fig:prfls_ReKeqTri} shows that substrates with similar $\sqrt{K_{eq}^{t+}}$ also share similarity in their free-flow mean velocity profiles, RMS velocity fluctuations and Reynolds stress, in spite of their differences in subsurface flow, to be expected given the different depths and granularities.
This similarity can also be observed in instantaneous realisations for the flow just above the interface, as illustrated in figure \ref{fig:Contours_AllReKeqs_UV-200}. The flows with similar $\sqrt{K_{eq}^{t+}}$ exhibit similar features for the background turbulence, but differences in the grain-coherent flow, directly attributable to the different grain topologies.
The same effects can be observed in the spectral density maps of figure \ref{fig:Spectra4_AllReKeqsDualYs}, which quantify statistically the similarities in the fluctuations at different lengthscales for the background turbulence, and also display the different intensities and lengthscales of the grain-coherent flow for different topologies.

\rgm{The collapse observed for $\sqrt{K_{eq}^{t+}}$, and the comparatively lack of collapse observed for $\sqrt{K_{eq}^{s+}}$, suggest that effect of the substrate on the overlying turbulence is mainly governed by the wall-normal transpiration triggered by overlying pressure fluctuations. The effect of the interfacial slip, which is characterised by $\sqrt{K_{eq}^{s+}}$, plays in contrast a minor role.}
\rev{The good collapse with $\sqrt{K_{eq}^{t+}}$ may also suggest that the Darcy-Brinkman model captures reasonably well the fluctuating transpiration induced by 
overlying pressure waves, even if it fails to capture other features of the flow such as the tangential shear and velocity near the interface \citep{SanchezP1982}. It is somewhat surprising that this is the case even for substrates with one or two layers of grains,
for which the lack of sufficient depth would in principle preclude the type of homogenisation in $y$ implicit in Darcy-Brinkman.
An answer to this is however out of the scope of the present paper, where we limit ourselves, as stated above, to using the Darcy-Brinkman
solutions to guide us in proposing an otherwise empirical coefficient to collapse the DNS results.}
\rgm{We note that the present analysis extends to substrates that exhibit a rough interface due to their intrinsic granularity, and leaves out surfaces that are rough but would not have an obvious permeable substrate counterpart, e.g. wavy walls or smoothly curved surfaces.}
\rev{For those surfaces, we could expect the slip dynamics to play a more dominant role, as occurs also for drag-reducing textures
\citep{Ibrahim21,GG_CTR_2018}. For porous substrates like the present ones,}
\rgm{the interfacial roughness can
potentially be eliminated by making the surface artificially flat, which would partially suppress the resulting increase in drag \citep[see][or Figure 1 and 
the corresponding discussion in \citeauthor{Rosti15} \citeyear{Rosti15}]{Kim20}. It is also possible to generate an interfacial roughness completely unrelated
to the substrate topology and decouple their effects, as in \cite{Wangsawijaya23}. Both cases are beyond the scope of this study. Likewise,
we have also only considered isotropic substrates. Anisotropic substrates such as those of \cite{Khorasani2024} would be beyond the present scope. In the
latter, the permeability is tensorial, and the expression in \ref{eq:DefKeq4} for an effective $\sqrt{K_{eq}^{t+}}$ would not be valid. It is possible,
however, to find an analogous solution for the anisotropic Darcy-Brinkman equations. We have done so in the past for the case of principal directions
aligned with $x$, $y$ and $z$ \citep{GmezdeSegura2018,Gomez19,Sharma17}; the solution presented in Appendix \ref{Sec:Brinkman} is actually a
particularisation of such solutions. A simple explicit expression is not possible in that case, but from our previous work we would expect that the role
of $K$ in equation \ref{eq:DefKeq4} was played by $K_y$ for streamwise preferential substrates \citep{Gomez19}, and by
$\sqrt{K_x K_y}$ for wall normal preferential ones \citep{Sharma17}. Here, $K_x$ and $K_y$ are the streamwise and the wall-normal permeability, respectively.
}

\rgm{The reader may wonder if the present findings apply only to the regular staggered-cube and mesh topologies for which DNSs have been
conducted in this paper, with relatively small grain pitch $L^+\lesssim 50$. In Appendix \ref{Sec:OtherData} we compare our results with
data compiled for different substrate layouts and from previous studies, namely the randomly packed spheres of \citet{Zippe83} 
and \citet{Karra2023}, the staggered cubes of \citet{Kuwata16a}, the collocated spheres of \citet{Kim20}, the reticulated foams
of \citet{Esteban22}, and the isotropic mesh lattices of \citet{Khorasani2024}.
The substrates in those studies have generally larger grain and therefore larger $K^+$,
and data for the roughness function was sometimes obtained indirectly, especially for experimental
works, but the agreement is generally good. We note that some of those studies were for irregular
substrate topologies. We expect our conclusions to hold beyond the homogeneous substrates of our
DNSs, so long as the heterogeneity occurs on a scale small enough not to interact with individual turbulent
eddies. If heterogeneities were present over lengthscales of $\sim100$ viscous units or larger,
they would need to be accounted for and would probably require a unique ad hoc analysis for each instance.
}

\subsection{A conceptual regime diagram for finite-depth substrates}
\label{subSec:RegimeDiagram}

\begin{figure}
\vspace*{-.5mm}
  \centerline{\includegraphics[width=0.78\linewidth]{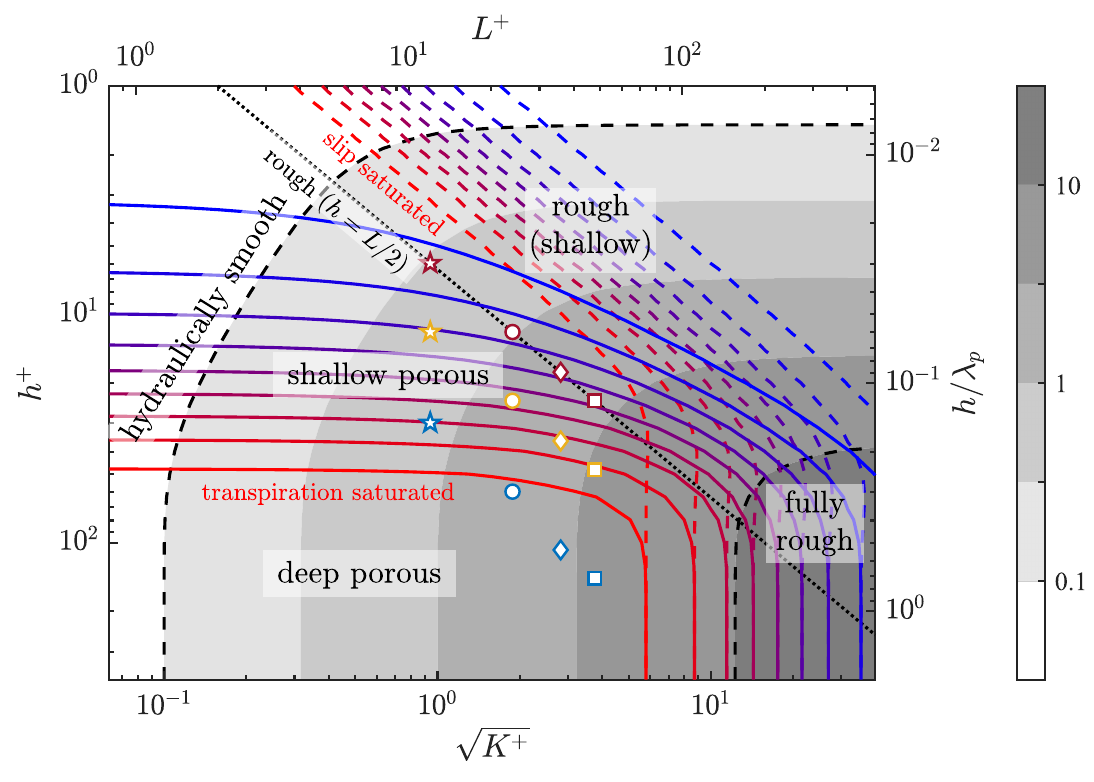}}
\vspace*{-2.5mm}
  \caption{Substrate regime diagram for a staggered-cube topology with $g/L=1/2$. The shaded contours are for the transpiration-based equivalent permeability $\sqrt{K_{eq}^{t+}}$, taken as a surrogate for the drag increase.  
  The dashed and solid contour lines are for  $f_{\mathcal{L}s}$ and $f_{\mathcal{K}t}$, respectively, for values [0.1:0.1:0.9] from red to blue. The results have been obtained for a characteristic wavelength for the overlying turbulence $\lambda_p^+=200$.
   The dotted straight line represents impermeable rough surfaces with $h=L/2$. The dashed lines represent boundaries between regimes.  The markers represent the present twelve DNSs with this topology, $g/L\!=\!1/2$.}
\label{fig:Brkm_Regimes}
\vspace*{-1mm}
\end{figure}

\rgm{The above results and discussion suggest that, for any given substrate topology, its effect on the 
overlying turbulence can be essentially characterised by $K_{eq}^{t+}$ alone. This coefficient is
proportional to the substrate permeability $K^+$, which encodes the relevant information on the grain topology,
but also depends on the substrate depth through $f_{\mathcal{K}t}$. Based on this, we propose a regime
diagram to conceptually illustrate the relationship between porous and rough surfaces for a given grain
topology. As an example, in figure \ref{fig:Brkm_Regimes} we consider our staggered-cube topology with gap-to-pitch
ratio $g/L=1/2$, corresponding to a constant $\sqrt{K}\approx0.076 L$. For other topologies, the
regime diagram would be qualitatively similar.}

\rgm{The diagram portrays, as a function of $\sqrt{K^+}$ and $h^+$, the equivalent permeability $K_{eq}^{t+}$, taken
as a surrogate for the roughness function $\Delta U^+$. To illustrate the effect of substrate depth on slip and 
transpiration, we also portray the corresponding values of the attenuation coefficients $f_{\mathcal{L}s}$ and
$f_{\mathcal{K}t}$.
To construct the diagram, these have been calculated
for a characteristic exciting wavelength $\lambda_p^+\approx200$, as discussed in \S\ref{subSec:Keq}.}
\rev{Note that this would be valid so long as the signature of the
overlying pressure fluctuations at the interface plane retained roughly its characteristic size for smooth-wall flows. This is the
case for the texture sizes studied here, $L^+\lesssim50$, as shown in figures \ref{fig:Spectra4_AllReKs} and
\ref{fig:Spectra4_AllReKeqsDualYs}, but would eventually break down for larger sizes. Therefore,}
\rgm{in the following, we \rev{only} consider substrates with 
$\sqrt{K^+}\lesssim10$ or $L^+\lesssim100$, which are the scope of the present study and for which we would expect
the diagram to be representative.
}

\rgm{In the diagram, a dotted line marks rough surfaces with cubic roughness elements of pitch
$L$ and height $h=L/2$, which includes our four rough DNSs with the topology under consideration, $g/L=1/2$.
Following this line for increasing $L^+$ is equivalent to increasing the Reynolds number for a fixed surface, thus
increasing its dimensions in viscous units, and its roughness function accordingly, as if following a Moody
chart \citep{Moody1944}. Substrates to the right of this line would also be rough, featuring roughness elements with
the same pitch and element width and gap but smaller depth, $h<L/2$. The isocontours for
$f_{\mathcal{L}s}$ indicate that, in terms of depth, the slip properties of the surfaces with $h=L/2$ would already
be maxed out, i.e. beyond the line labeled `slip saturated' -- no further increases would take place when increasing $h$ further. Meanwhile, the shallower
rough substrates with $h<L/2$ would experience comparatively reduced slip.}

\rgm{As discussed in \S\ref{subSec:Keq},
however, the effect of slip through $K_{eq}^{s+}$ is secondary, and the drag is mainly governed by
$K_{eq}^{t+}$, for which the relevant attenuating function is the transpiration one, $f_{\mathcal{K}t}$.
The latter is a continuous, smooth function
of the substrate depth, which tends to zero for vanishing depth and to unity for deep substrates. 
Its isocontours show that, while $f_{\mathcal{L}s}$ saturates at much shallower depths $h$, i.e. already for simply
rough, impermeable surfaces, $f_{\mathcal{K}t}$ saturates only for deeper substrates. Its isocontours
indicate that, for the present textures with $\sqrt{K^+}\lesssim10$, the saturation depth is essentially independent of $K^+$ or $L^+$, and is
roughly $h^+\approx 50$. This can be traced to the
\rev{fact that $h$ is scaled by $\lambda_p$ in the expression for $f_{\mathcal{K}t}$, and that
$\lambda_p^+\approx200$ independently of the
substrate, as discussed above}. Whether a substrate is shallow or sufficiently deep is governed by whether pressure fluctuations of this wavelength would penetrate deeper in the absence of an impermeable floor at a finite depth $y^+=-h^+$.
}
\rev{Perhaps counterintuitively, this depends on the depth in viscous units, rather than in permeability lengths or number of
inclusion layers.}

\rgm{From this, we can distinguish a series of regions in the diagram which are characterised by different flow
regimes. For substrates deeper than $h^+\approx 50$, the flow would perceive the substrate as deep enough to
exhibit its permeable character fully. We refer to these as `sufficiently deep porous substrates', for which any further increase in depth would not result in any significant changes
to the flow. For substrates shallower than this but deeper than the rough surfaces with cubic inclusions, $h=L/2$,
the substrate would exhibit an intermediate character, where the transpiration is partially suppressed compared to a deeper substrate with the same $K^+$. We refer to these as `shallow porous substrates'. Substrates with $h=L/2$ would be so shallow that they are conventional rough, impermeable surfaces. Substrates even shallower than $h=L/2$ would likewise be rough, but with shorter-height roughness elements, further suppression of  transpiration, and additionally suppression of slip.}

\section{Conclusions}
\label{Sec:Conclusions}

This study has focused on understanding the effect of different characteristics of a porous substrate on the properties of the overlying turbulent flow. Specifically, we have aimed to investigate the effects of three characteristics: i) the permeability of a porous medium, characterised by the bulk permeability $K$, ii) the granularity of the porous medium, whose length scale is characterised by the grain \rgm{pitch} $L$, and iii) the depth of the substrate, $h$. To this object, we have systematically explored the parameter space of $K$, $L$, and $h$ for porous substrates using DNSs. We have used staggered-cube arrays with various porosities $\varepsilon=0.23$ - $0.97$ and permeabilities $\sqrt{K}/L=0.013$ - $0.243$, and studied substrate depths ranging from 
$h=1D=L/2$ to $h\geq5D=5L/2$, corresponding to the transition from typical impermeable rough surfaces
to deep porous substrates. The \rgm{values of grain pitch} we have considered are in the range $10\lesssim L^+\lesssim50$. For this range, the overlying turbulence soon deviates from smooth-wall-like behaviour, but the grains remain small enough that the grain-coherent flow is not sufficiently intense to directly interact with the background, grain-incoherent turbulence, other than perhaps for the largest $L^+\approx50$.

Through the discussion on the scaling of flow properties with substrate parameters, two major insights have been provided on the effect of the above three.
First, the permeability of a porous medium has significantly greater relevance than its granularity to most of the main properties of the overlying turbulence. 
For all substrates with $h\geq1D$, the mean-flow slip length $\ell_U^+$  is essentially proportional to $\sqrt{K^+}$, and the inner/outer shear ratio across the interface $r_{sh}$  correlates well with $\varepsilon$. For deep substrates with $h\geq5D$, the mean velocity deficit $\Delta U^+$ and the statistics and energy spectra of the background turbulence are essentially determined by $\sqrt{K^+}$ alone.
In contrast, the effect of $L^+$ on the overlying turbulence is essentially not significant,
and can only be observed in flow realisations and  spectral density maps very near the interface. This implies that, at least for grain \rgm{pitch values} $L^+\lesssim50$, a porous substrate can be reasonably approximated as a continuum represented only by macroscale parameters like $K$, $h$, and $\varepsilon$.

Second, the principal mechanism that distinguishes porous substrates from surfaces with analogous topology that are rough, but are otherwise impermeable, is the effect of the substrate depth on the interfacial transpiration excited by the overlying pressure fluctuations. We propose an empirical `equivalent permeability', $K_{eq}^t$, with some theoretical support based on Darcy-Brinkman models, that incorporates this effect of depth on transpiration. For substrates with different depths, the values of $\sqrt{K_{eq}^{t+}}$ alone collapse $\Delta U^+$ well, and different substrates with similar $\sqrt{K_{eq}^{t+}}$ exhibit similar properties for the overlying turbulence in general. A conceptual $h^+$-$\sqrt{K^+}$ diagram illustrates the regime transition between sufficiently deep porous substrates, with $h^+\gtrsim50$, and typical impermeable rough surfaces with $h=O(L)$.

The scope of this study is substrates composed of relatively small grains, $L^+\lesssim50$. In the fully-rough regime, where $L^+\gtrsim70$, the flow coherent with individual grains could be energetic enough to generate grain-coherent eddies that interacted strongly with and modify directly the overlying turbulence. In that situation, the effects of substrate permeability and granularity may be intrinsically indistinguishable. Understanding the relationship between the substrate parameters and the overlying turbulence in the fully rough regime may thus require a different framework. The present work covers sizes just up to the onset of that regime.


\backsection[Acknowledgements]
{The authors thank Dr.~Garazi G{\'o}mez-de-Segura for her advice on flow physics and simulations.}

\backsection[Funding]{This work is supported by the UK Engineering and Physical Sciences Research
Council (EPSRC) under grant EP/S013083/1, and by the Air Force Office of Scientific Research -- European
Office of Aerospace Research and Development (AFOSR--EOARD), under grant FA8655-22-1-7062. Computational
resources were provided by the University of Cambridge Research Computing Service under EPSRC Tier-2 grant
EP/P020259/1 (projects cs066 and cs155), and by the UK 'ARCHER2' system under PRACE project pr1u1702 and
EPSRC project e776. 

For the purpose of open access, the authors have applied a Creative Commons Attribution
(CC BY) licence to any Author Accepted Manuscript version arising from this submission.
}

\backsection[Declaration of interests]{The authors report no conflict of interest.}

%
%

\backsection[Author ORCIDs]{\\
R. Garc{\'i}a-Mayoral, https://orcid.org/0000-0001-5572-2607;\\
Z. Hao, https://orcid.org/0000-0003-2199-6938.}

\appendix

\section{\rgm{Comparison of the flow in interconnected and frozen-suspension substrates}}
\label{Sec:MicroFlow}

\begin{figure}
\vspace*{2mm}
\centerline{\includegraphics[width=\linewidth]{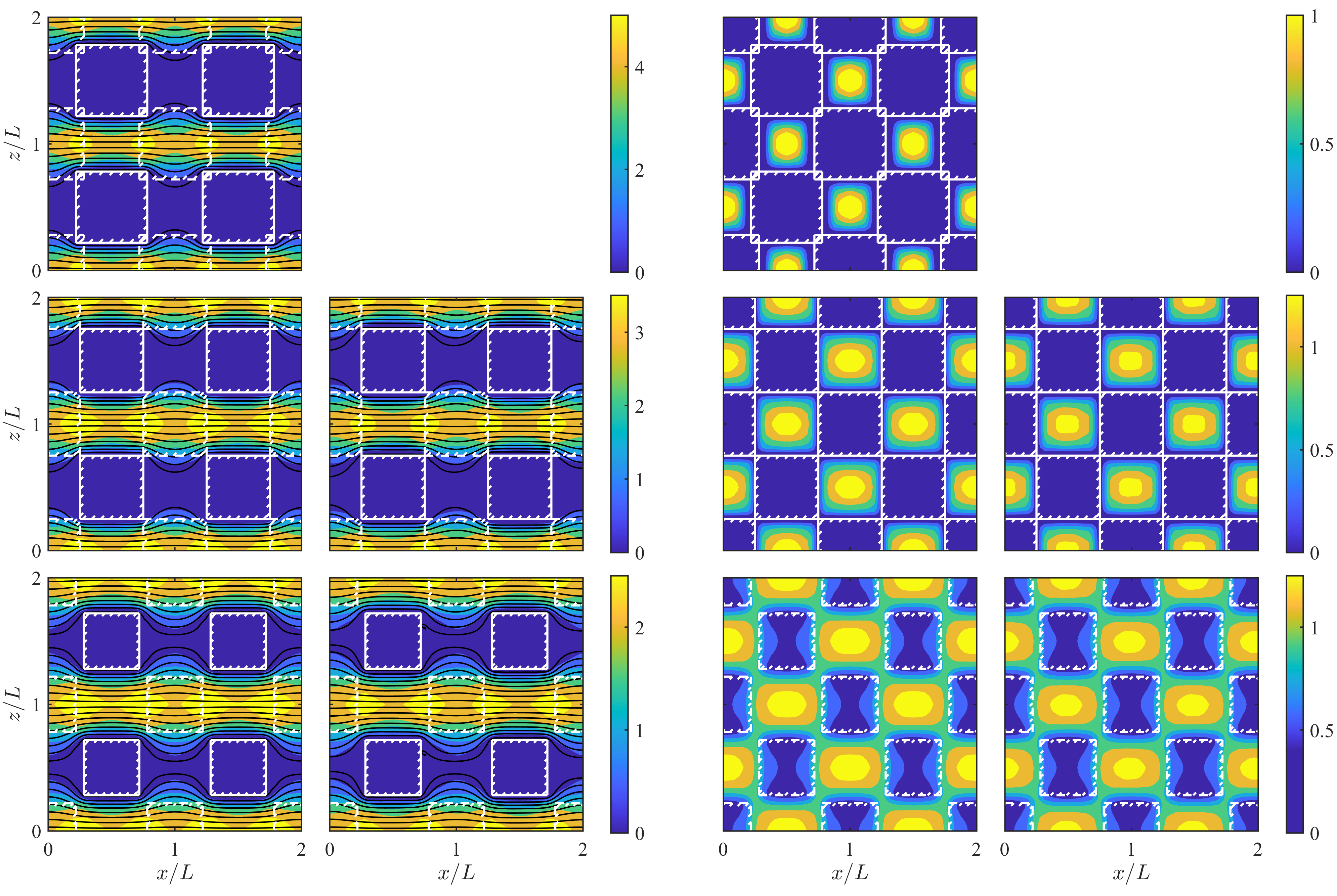}}
\mylab{-2mm}{90mm}{$(a)$}
\mylab{66mm}{90mm}{$(b)$}
\mylab{-4mm}{61mm}{$(c)$}
\mylab{64mm}{61mm}{$(d)$}
\mylab{-6mm}{32mm}{$(e)$}
\mylab{62mm}{32mm}{$(f)$}

\caption{
\rgm{Detail of the interstitial microscale flow in substrates made up of interconnected and disconnected
staggered cubes. 
Results are shown for the velocity magnitude,  normalised by the Darcy velocity, in $x$-$z$ sections; through the middle of a row of cubes in $(a,c,e)$,
and through a plane at the half-height between centres of consecutive rows in $(b,d,f)$.
The location of the cubes is marked in white, with dashed lines for cubes not intersected by the section.
In $(a,c,e)$, streamlines of the locally 2D flow are also portrayed in black.
$(a,b)$, overlapping cubes with $g/L=7/16$; $(c,d)$ just-touching cubes with $g/L=1/2$; $(e,f)$, suspended cubes with
$g/L=9/16$. For each panel pair, results on the left are for Stokes-flow simulations, and results on the right for
ensemble averages from DNSs with $L^+\approx24$, when available (cases Pd-24-50 and Pd-24-56).}
}
\label{fig:microflow}
\end{figure}

\rgm{Here we present evidence supporting the use of idealised substrates made up of `frozen-suspensions' of inclusions to study the effect of substrates with higher permeability. These
would not be realisable as solid permeable surfaces in the real world. Such is the case of our staggered-cube
arrangements with gap size $g>L/2$. The reader may wonder if this would not give rise to fundamental differences in
the interstitial flow, compared to interconnected topologies.
Figure \ref{fig:microflow} shows that this is not the case. The figure portrays the detail of the microscale flow as the gap
size changes and the topology shifts from the substrate being just interconnected to just disconnected. The figure shows that the flow evolves smoothly from one case to the other, with no intrinsic difference or abrupt changes in its structure.}

\begin{figure}
\vspace*{2mm}
\centerline{\includegraphics[width=0.48\linewidth]{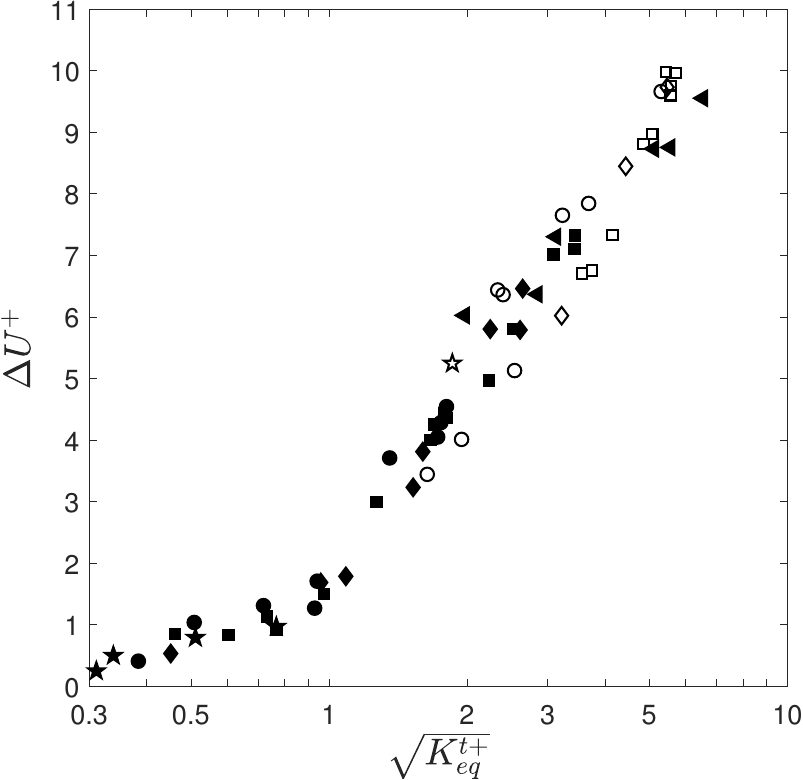}}
\vspace*{0mm}
\caption{
\rgm{Results from figure \ref{fig:DU_vs_uvKeq}($d$) for the roughness function $\Delta U^+$ as a function of the transpiration equivalent permeability $K_{eq}^{t+}$. Black symbols,
interconnected substrates of staggered cubes or mesh-like lattices. White symbols, frozen suspensions of staggered cubes.} 
}
\label{fig:DUOPT_vs_Keq4}
\vspace*{6mm}
  \centerline{\includegraphics[width=1.0\linewidth]{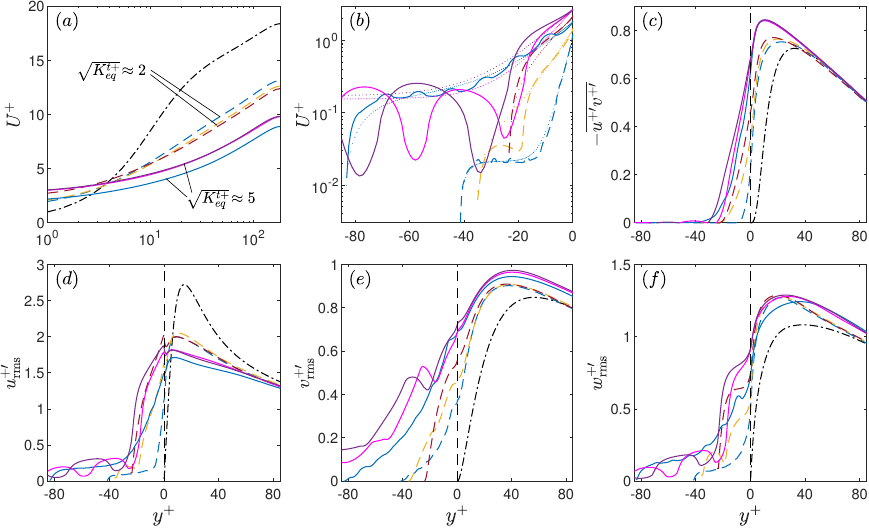}}
\vspace*{0mm}
  \caption{\rgm{Comparison of turbulent statistics for interconnected and frozen-suspension substrate topologies. ($a$-$b$) Mean velocity profile, 
  ($c$) Reynolds shear stress, and ($d$-$f$) RMS velocity fluctuations.
  Dashed lines, substrates with $\sqrt{K^{t+}_{eq}}\approx2$; blue, Pd-12-67 ($\sqrt{K^+}\!\!=\!2.03$, $\sqrt{K_{eq}^{t+}}\!=\!1.86$);
  yellow, Ps-36-50 ($\sqrt{K^+}\!\!=\!2.73$, $\sqrt{K_{eq}^{t+}}\!=\!2.25$); red, MRo-48-56 ($\sqrt{K^+}\!\!=\!3.30$,
  $\sqrt{K_{eq}^{t+}}\!=\!1.98$). Solid lines, substrates with $\sqrt{K^{t+}_{eq}}\approx5$; blue,  Pd-24-75 ($\sqrt{K^+}\!\!=\!5.80$,
  $\sqrt{K_{eq}^{t+}}\!=\!5.30$); magenta, MPd-36-78 ($\sqrt{K^+}\!\!=\!5.34$, $\sqrt{K_{eq}^{t+}}\!=\!5.11$); purple, MPd-48-72 ($\sqrt{K^+}\!\!=\!5.94$, $\sqrt{K_{eq}^{t+}}\!=\!5.54$).}} 
  \label{fig:prfls_ReKeqScafDual}
\end{figure}

\rgm{The results presented in this paper suggest that the overlying turbulence depends essentially on the macroscopic properties of substrates, and not on their microscopic detail, regardless of whether substrate inclusions are interconnected or not.
Figure \ref{fig:DUOPT_vs_Keq4} shows the same data of figure \ref{fig:DU_vs_uvKeq}($d$), but highlighting which substrates are interconnected and which are frozen suspensions.
Data from both groups spans with substantial overlap the full range of permeabilities studied, showing that there is indeed no essential difference.
}

\rgm{Integral quantities like the roughness function can conceal more subtle differences in the flow,
but the similarity across the two topology groups extends also to more detailed flow properties. Figure \ref{fig:prfls_ReKeqScafDual} compares turbulent statistics for interconnected and frozen-suspension substrates for similar values of $K_{eq}^{t+}$, and shows again that there is good agreement in the results across different substrates, regardless of their microstructure topology, for similar values of the macroscopic effective permeability.
}

\rev{
\section{Spatial resolution and grid convergence}
\label{Sec:GridDependence}

Section \ref{SubSec:Numerical} details the grid resolution used for the various DNSs conducted in this work. The grid convergence study
in \citet{Sharma20dense} concluded that, for the present DNS code and discretisation, 13-15 fluid points were sufficient to resolve the gaps in-between obstacles through which
the fluid would flow, and that when a lower resolution of 7-8 fluid points per gap was used, errors of up to $\sim20$\% were observed
in the fluctuating velocity within the substrate, mainly in the wall-normal velocity, although the error in the fluctuating flow above
was reduced to $\sim4$\%. \citet{Sharma20dense} argued that the latter marginal resolutions were acceptable for small gaps $g^+
\lesssim5$, where the flow would be predominantly viscous and instantaneously unidirectional. The present DNSs follow these 
general criteria, with 13 points or more per gap except for simulations Pd-12-33, Ro/Pd-24-25, Pd-36-22, Ro/Pd-48-22,  Pd-48-25,
and Ro/Pd-48-28, with 9-11 points per gap. Of these, the most critical case is Pd-48-25, since it has the largest gap $g^+
\approx12$ with the lowest resolution, 9 points per gap and $\Delta z^+\approx1.5$. To quantify the effect of this marginal resolution, we have run an additional 
simulation doubling the resolution in $x$ and $z$, i.e. 64 points per pitch $L$ and 17 points per gap.

\begin{figure}
\vspace*{-.5mm}
  \centering
  \includegraphics[width=1.0\linewidth]{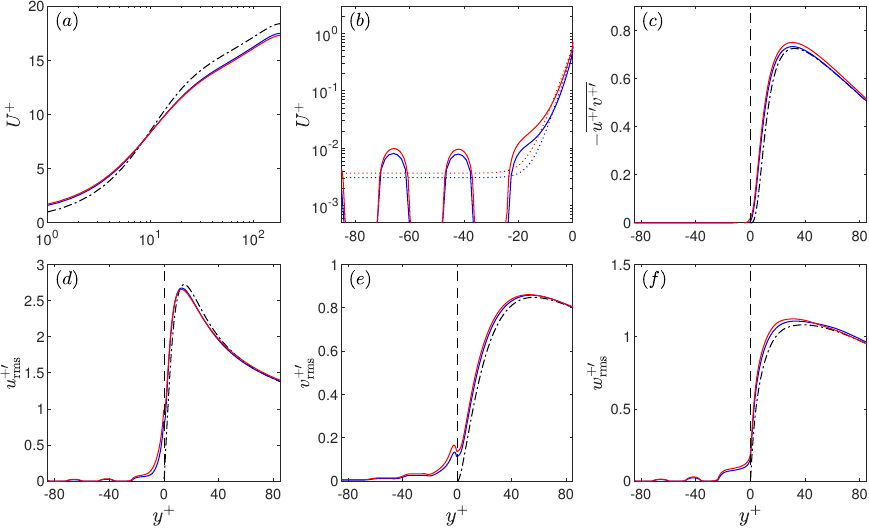}
\vspace*{-4mm}
  \caption{\rev{($a$-$b$) Mean velocity profile, ($c$) Reynolds shear stress, and ($d$-$f$) RMS velocity fluctuations for case Pd-48-25 with different resolutions. Blue and red lines are for the coarser and finer resolutions, 32 and 64 points per pitch $L$, respectively. The dashed lines mark the location of the free-flow/substrate interface.}}
  \label{fig:prfls_GridResol}
\vspace*{-1mm}
\end{figure}

The flow statistics from the DNSs with the coarser and finer  resolutions are compared in figure \ref{fig:prfls_GridResol}, and
are consistent with the resolution analysis of \citet{Sharma20dense}. The results are very close in the free flow region, with somewhat
larger deviations for the mean velocity $U^+$ and the RMS velocity fluctuations $u^{+\prime}_\mathrm{rms}$, $v^{+\prime}_\mathrm{rms}$,
and $w^{+\prime}_\mathrm{rms}$ below the interface, with the lower resolution resulting in slightly underpredicted values. Among them,
the differences for $U^+$ and $v^{+\prime}_\mathrm{rms}$ seem more apparent than the others, and can be directly
linked to the difference in \textit{a posteriori} $\sqrt{K^+}$ for the two simulations, of order $\sim8$\%.

\begin{figure}
    \centering
    \includegraphics[width=1.00\linewidth]{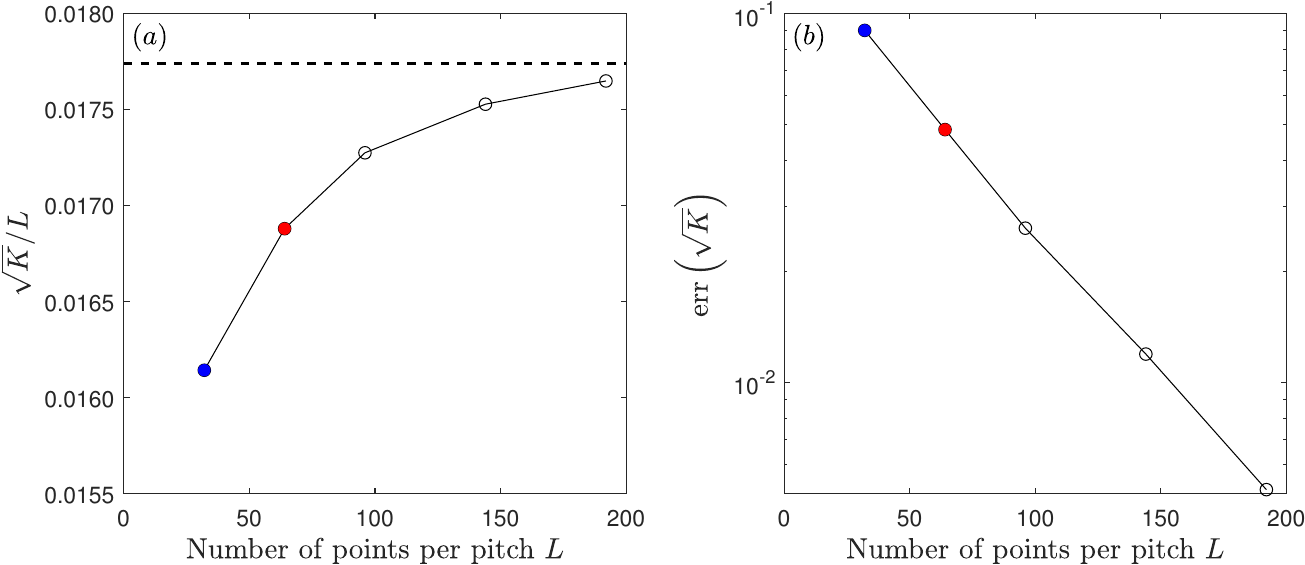}
    \caption{
    \rev{\textit{A priori} values of $\sqrt{K}$ for the staggered-cube configuration with $g/L=1/4$, obtained from Stokes-flow simulations with different grid resolutions. The blue and red makers represent the resolutions compared in DNSs for case Pd-48-25. The dashed line marks the asymtotic value $\sqrt{K^+}\approx0.0177$, used in (\textit{b}) to estimate the relative error.}}
    \label{fig:sqrtK_GridResol}
\end{figure}

In order to avoid grid-related discrepancies between the \textit{a posteriori} and \textit{a priori} values of $\sqrt{K^+}$, as mentioned in section
\ref{SubSec:Numerical} we have computed both with the same resolution. This is the same strategy we followed in \citet{Garcia11},
\citet{Fairhall2018} and \citet{Sharma20dense} to compute macroscopic surface parameters such as protrusion heights, slip lengths and
permeabilities consistent with the DNSs. The deviation from the grid-converged values of $\sqrt{K^+}$ is in any event small, with 
underpredictions $\lesssim10$\% for the topologies with narrowest gaps, $g \leq L/4$, and $\lesssim5$\% for the topologies with
$g \geq L/3$. These deviations are estimated from the values kindly provided by a reviewer and the grid-convergence results portrayed in
figure \ref{fig:sqrtK_epsilon}. For the grid-convergence study, we have computed the \textit{a priori} $\sqrt{K^+}$ for the staggered-cube
configuration with $g=L/4$, the topology of Pd-48-25, for our baseline resolution of 36 points per $L$ plus for 64, 96, 144 and 192
points per $L$. The results are shown in figure~\ref{fig:sqrtK_GridResol}, and are consistent with a ground-truth
value $\sqrt{K^+}\approx0.0177$, to which they tend asymptotically -- we note that this entails an overprediction of $\sim2$\% in the 
reviewer's value of 0.0181. Relative to the asymptotic value, the values of $\sqrt{K}$ corresponding to 32 and 64 points per pitch have errors of $\sim9$\% and $\sim5$\%, respectively.
}

\section{Effect of the Reynolds number}
\label{Sec:ResHR}

\begin{figure}
  \centering
\vspace{14pt}
\begin{overpic}[width=1.0\textwidth]{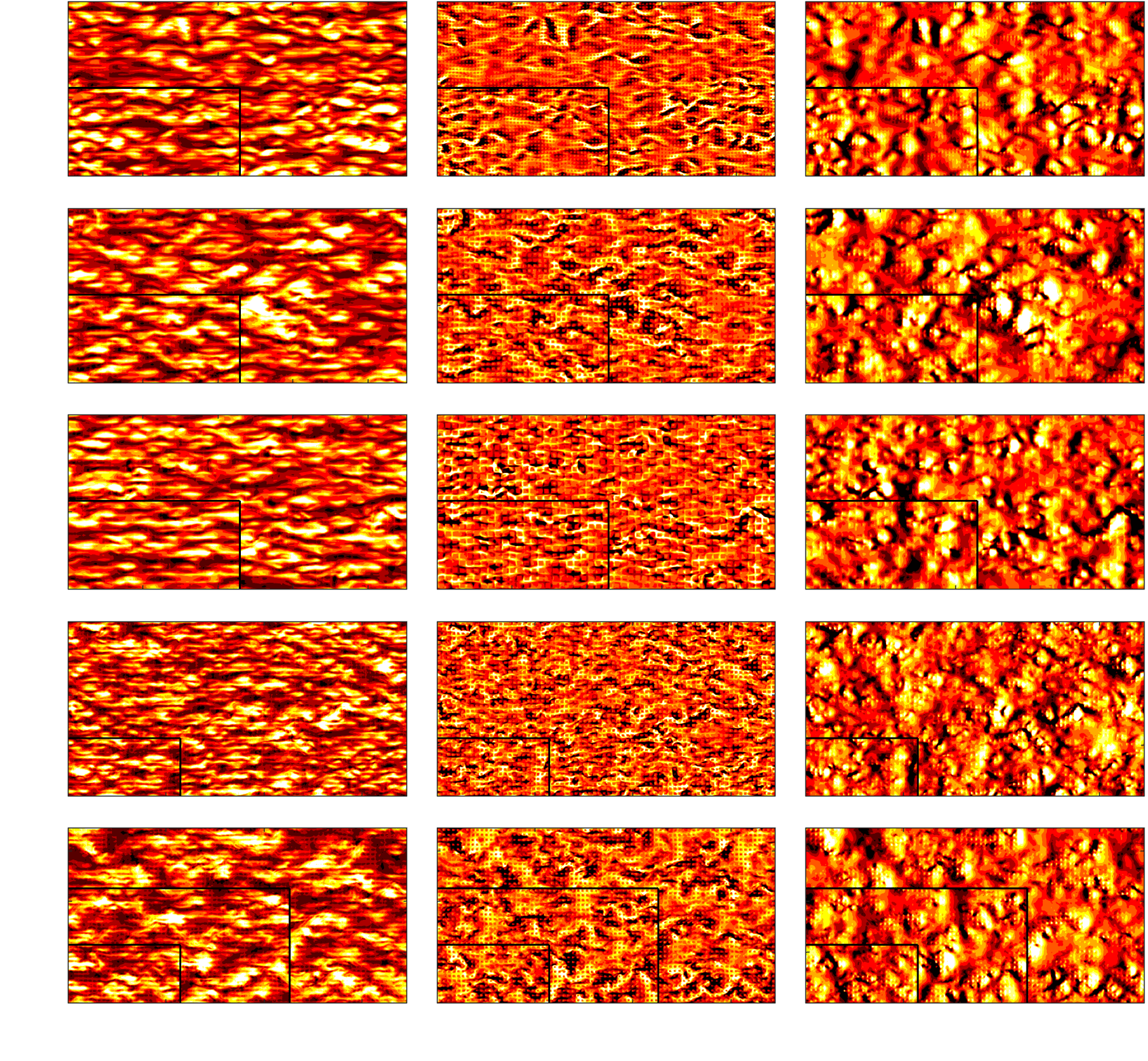}

 \put (19.4, 93.2) {\footnotesize$u^\prime$}
 \put (51.5, 93.2) {\footnotesize$v^\prime$}
 \put (83.6, 93.2) {\footnotesize$p^\prime$}

 \put ( 3.4, 90.8) {\scriptsize$(\!a\!)$}
 \put (35.8, 90.8) {\scriptsize$(\!b\!)$}
 \put (67.9, 90.8) {\scriptsize$(\!c\!)$}
 \put ( 3.4, 72.7) {\scriptsize$(\!d\!)$}
 \put (35.8, 72.7) {\scriptsize$(\!e\!)$}
 \put (67.8, 72.7) {\scriptsize$(\!f\!)$}
 \put ( 3.4, 54.8) {\scriptsize$(\!g\!)$}
 \put (35.6, 54.8) {\scriptsize$(\!h\!)$}
 \put (68.0, 54.8) {\scriptsize$(\!i\!)$}
 \put ( 3.4, 36.6) {\scriptsize$(\!j\!)$}
 \put (35.7, 36.6) {\scriptsize$(\!k\!)$}
 \put (68.0, 36.6) {\scriptsize$(\!\ell\!)$}
 \put ( 2.9, 18.6) {\scriptsize$(\!m\!)$}
 \put (35.6, 18.6) {\scriptsize$(\!n\!)$}
 \put (67.9, 18.6) {\scriptsize$(\!o\!)$}

 \put (-0.4, 84.4) {\footnotesize\rotatebox{90}{$z^+$}}
 \put ( 4.4, 76.6) {\scriptsize$0$}
 \put ( 2.3, 83.3) {\scriptsize$500$}
 \put (10.7, 75.6) {\scriptsize$500$}
 \put (16.8, 75.6) {\scriptsize$1000$}
 \put (23.4, 75.6) {\scriptsize$1500$}
 \put (30.0, 75.6) {\scriptsize$2000$}
 \put (37.3, 75.6) {\scriptsize$0$}
 \put (42.8, 75.6) {\scriptsize$500$}
 \put (48.9, 75.6) {\scriptsize$1000$}
 \put (55.5, 75.6) {\scriptsize$1500$}
 \put (62.1, 75.6) {\scriptsize$2000$}
 \put (69.4, 75.6) {\scriptsize$0$}
 \put (74.9, 75.6) {\scriptsize$500$}
 \put (81.0, 75.6) {\scriptsize$1000$}
 \put (87.6, 75.6) {\scriptsize$1500$}
 \put (94.2, 75.6) {\scriptsize$2000$}

 \put (-0.4, 66.3) {\footnotesize\rotatebox{90}{$z^+$}}
 \put ( 4.4, 58.5) {\scriptsize$0$}
 \put ( 2.3, 65.2) {\scriptsize$500$}
 \put (10.7, 57.5) {\scriptsize$500$}
 \put (16.8, 57.5) {\scriptsize$1000$}
 \put (23.4, 57.5) {\scriptsize$1500$}
 \put (30.0, 57.5) {\scriptsize$2000$}
 \put (37.3, 57.5) {\scriptsize$0$}
 \put (42.8, 57.5) {\scriptsize$500$}
 \put (48.9, 57.5) {\scriptsize$1000$}
 \put (55.5, 57.5) {\scriptsize$1500$}
 \put (62.1, 57.5) {\scriptsize$2000$}
 \put (69.4, 57.5) {\scriptsize$0$}
 \put (74.9, 57.5) {\scriptsize$500$}
 \put (81.0, 57.5) {\scriptsize$1000$}
 \put (87.6, 57.5) {\scriptsize$1500$}
 \put (94.2, 57.5) {\scriptsize$2000$}

 \put (-0.4, 48.4) {\footnotesize\rotatebox{90}{$z^+$}}
 \put ( 4.4, 40.6) {\scriptsize$0$}
 \put ( 2.3, 47.3) {\scriptsize$500$}
 \put (10.7, 39.6) {\scriptsize$500$}
 \put (16.8, 39.6) {\scriptsize$1000$}
 \put (23.4, 39.6) {\scriptsize$1500$}
 \put (30.0, 39.6) {\scriptsize$2000$}
 \put (37.3, 39.6) {\scriptsize$0$}
 \put (42.8, 39.6) {\scriptsize$500$}
 \put (48.9, 39.6) {\scriptsize$1000$}
 \put (55.5, 39.6) {\scriptsize$1500$}
 \put (62.1, 39.6) {\scriptsize$2000$}
 \put (69.4, 39.6) {\scriptsize$0$}
 \put (74.9, 39.6) {\scriptsize$500$}
 \put (81.0, 39.6) {\scriptsize$1000$}
 \put (87.6, 39.6) {\scriptsize$1500$}
 \put (94.2, 39.6) {\scriptsize$2000$}

 \put (-0.4, 30.2) {\footnotesize\rotatebox{90}{$z^+$}}
 \put ( 4.4, 22.6) {\scriptsize$0$}
 \put ( 2.3, 27.0) {\scriptsize$500$}
 \put ( 1.4, 31.4) {\scriptsize$1000$}
 \put (12.2, 21.6) {\scriptsize$1000$}
 \put (20.9, 21.6) {\scriptsize$2000$}
 \put (29.6, 21.6) {\scriptsize$3000$}
 \put (37.3, 21.6) {\scriptsize$0$}
 \put (44.3, 21.6) {\scriptsize$1000$}
 \put (53.0, 21.6) {\scriptsize$2000$}
 \put (61.7, 21.6) {\scriptsize$3000$}
 \put (69.4, 21.6) {\scriptsize$0$}
 \put (76.4, 21.6) {\scriptsize$1000$}
 \put (85.1, 21.6) {\scriptsize$2000$}
 \put (93.8, 21.6) {\scriptsize$3000$}

 \put (-0.4, 12.0) {\footnotesize\rotatebox{90}{$z^+$}}
 \put ( 4.4,  4.4) {\scriptsize$0$}
 \put ( 2.3,  8.8) {\scriptsize$500$}
 \put ( 1.4, 13.2) {\scriptsize$1000$}
 \put (12.2,  3.4) {\scriptsize$1000$}
 \put (20.9,  3.4) {\scriptsize$2000$}
 \put (29.6,  3.4) {\scriptsize$3000$}
 \put (37.3,  3.4) {\scriptsize$0$}
 \put (44.3,  3.4) {\scriptsize$1000$}
 \put (53.0,  3.4) {\scriptsize$2000$}
 \put (61.7,  3.4) {\scriptsize$3000$}
 \put (69.4,  3.4) {\scriptsize$0$}
 \put (76.4,  3.4) {\scriptsize$1000$}
 \put (85.1,  3.4) {\scriptsize$2000$}
 \put (93.8,  3.4) {\scriptsize$3000$}
 
 \put (19.0,  0.9) {\footnotesize$x^+$}
 \put (51.1,  0.9) {\footnotesize$x^+$}
 \put (83.2,  0.9) {\footnotesize$x^+$}
 
  \end{overpic}
  \vspace*{-5mm}
  \caption{Instantaneous fields $u^\prime$ (left), $v^\prime$ (middle), and $p^\prime$ (right) at $y^+\!\approx\!3$ for cases with different $Re_\tau$. From top to bottom: Pd-24-50(-HR), Pd-36-50(-HR), Pd-48-38(-HR), Pd-48-50(-HHR), and Pd-48-62(-HR,-HHR). The subplot at the bottom-left corner of each panel is for the case with lower $Re_\tau$ than the corresponding case with higher $Re_\tau$ in the remainder of the panel. Colour range black-red-yellow-white corresponds to $[-2:2]$ times the RMS of the variable on the plane.}
  \vspace*{-1mm}
\label{fig:Contours_AllRetauComps}
\end{figure}
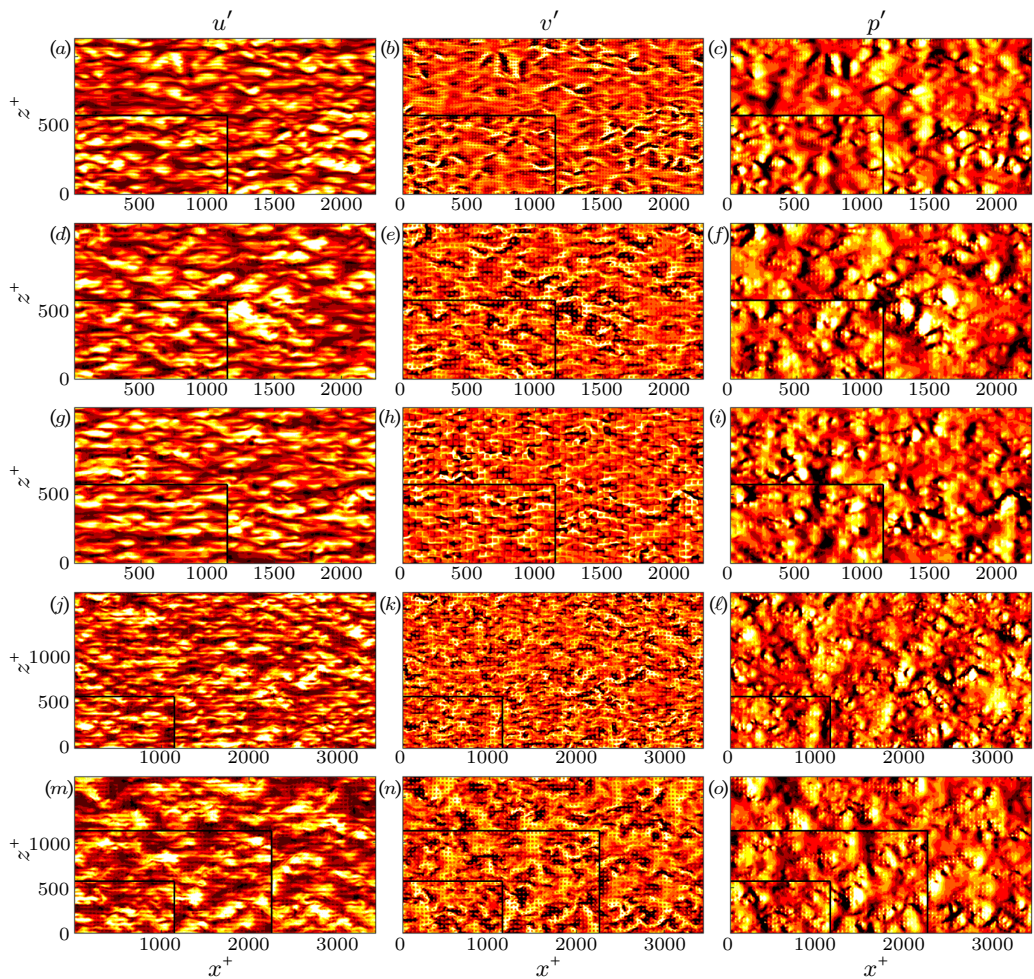

Most of the analysis on the scaling of flow properties in this paper is for the data non-dimensionalised in wall units. This is based on the assumption that these flow properties are dominated by the dynamics of near-wall turbulence in the buffer layer \citep{Garcia12}. In this section, to check the dependence of the analysis on Reynolds number, we compare some results of simulations run at different $Re_\tau\approx180$, $360$, and $550$, while matching the substrate dimensions in wall units. Five sets of cases are considered, in each set the cases sharing the same $L^+$, $g/L$, and $h/D$: Pd-24-50(-HR), Pd-36-50(-HR), Pd-48-38(-HR), Pd-48-50(-HHR), and Pd-48-62(-HR,-HHR).
Near the interface, the visual similarity between the cases in a set can be observed in the instantaneous flow fields in figure \ref{fig:Contours_AllRetauComps}.

\begin{figure}
  \centerline{\includegraphics[width=1.0\linewidth]{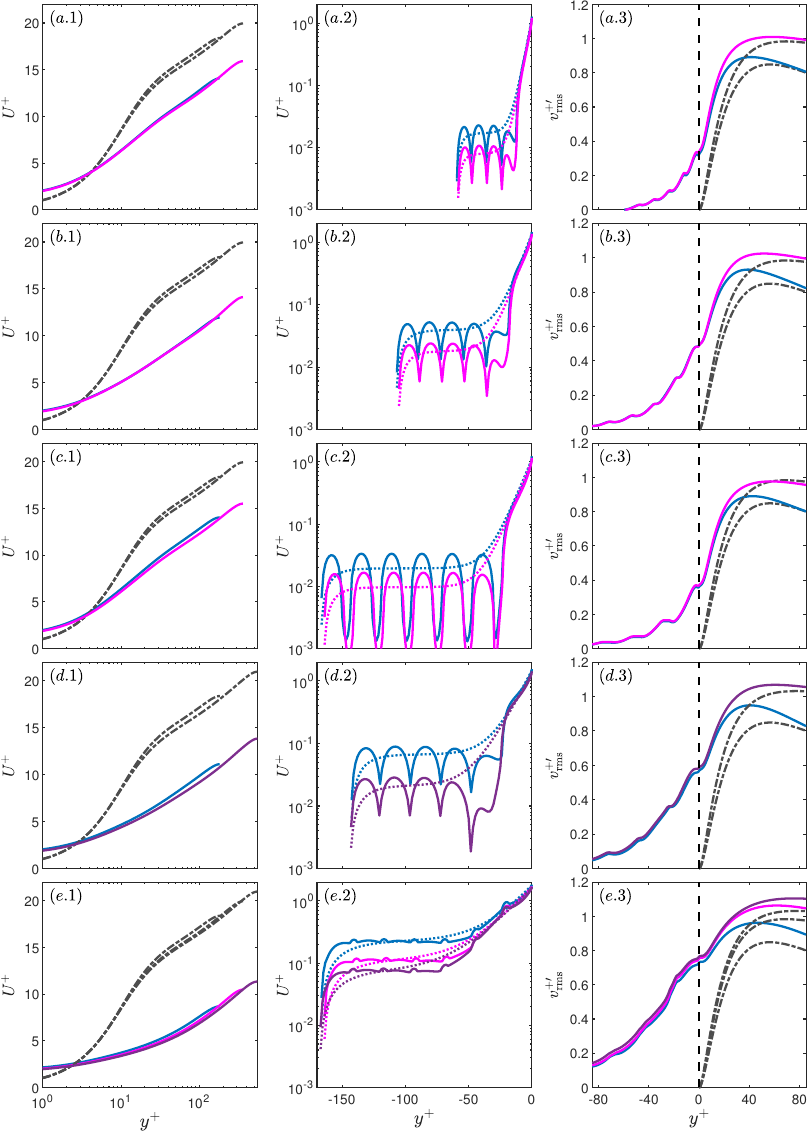}}
  \caption{Mean velocity profile and wall-normal velocity RMS fluctuation at different $Re_\tau$. From top to bottom, cases Pd-24-50(-HR), Pd-36-50(-HR), Pd-48-38(-HR), Pd-48-50(-HHR), and Pd-48-62(-HR,-HHR). Blue lines are for $Re_\tau\!\approx\!180$, magenta lines for $Re_\tau\!\approx\!360$, and purple lines for $Re_\tau\!\approx\!550$.}
\label{fig:prfls_AllRetauComps}
\end{figure}

\begin{figure}
  \centerline{\includegraphics[width=1.0\linewidth]{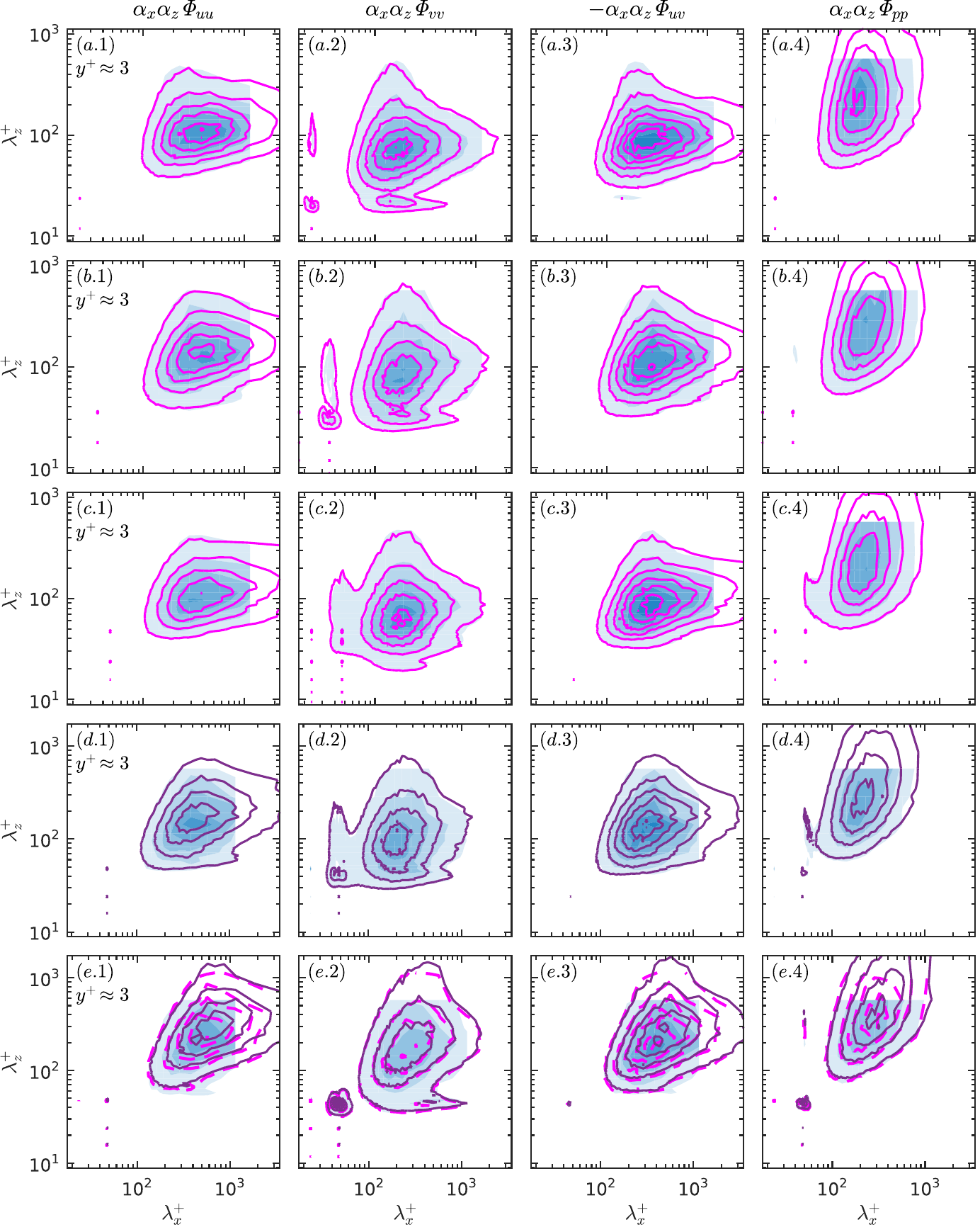}}
  \caption{Pre-multiplied spectra $\alpha_x\alpha_z\Phi_{**}$ at $y^+\approx3$ for the cases with different $Re_\tau$. Blue shades are for $Re_\tau\approx180$, magenta lines for $Re_\tau\approx360$, and purple lines for $Re_\tau\approx550$. From top to bottom: Pd-24-50(-HR), Pd-36-50(-HR), Pd-48-38(-HR), Pd-48-50(-HHR), and Pd-48-62(-HR,-HHR). Contours represent six equidistant levels (0.044, 0.088, 0.132, 0.176, 0.220, and 0.264) relative to the corresponding variance or covariance.}
\label{fig:Spectra4_AllRetauComps}
\end{figure}

Figure \ref{fig:prfls_AllRetauComps} shows the profiles of the mean velocity $U^+$ and the RMS fluctuations of the wall-normal velocity $v_{\rms}^{+\prime}$ for the five sets of cases, together with the corresponding smooth-wall results for reference, with figure \ref{fig:Spectra4_AllRetauComps} showing the corresponding spectral energy densities.
For each set, above the interface, the profiles of $U^+$ for substrates show good agreement, with small discrepancies in the wake region similar to those observed for smooth walls. Such an agreement extends below the interface until the Darcy regions are reached, below where the $U^+$ ratio is proportional to $1/Re_\tau$. This is a result of the mean pressure gradient being different in wall units. The cases in a set have similar profiles of $v_{\rms}^{+\prime}$ for $y^+\lesssim15$, above which the values gradually approach the corresponding smooth-wall results at each given $Re_\tau$. Although not shown, the profiles of the RMS fluctuations $u_{\rms}^{+\prime}$ and $w_{\rms}^{+\prime}$ and the Reynolds shear stress $\overline{u^{+\prime}v^{+\prime}}$ exhibit similar trends to the ones of $v_{\rms}^{+\prime}$, as in \citet{Sharma20sparse,Sharma20dense}. The spectral energy densities of the velocity fluctuations and the Reynolds shear stress show the same agreement in figure \ref{fig:Spectra4_AllRetauComps}, with further information at larger wavelengths available for the higher-$Re_\tau$ cases due to their larger domain in viscous units. In summary, the comparison in this section indicates that the scaling of flow properties presented in this study is essentially Reynolds number-independent.


\section{Analytical solutions of the Darcy-Brinkman equation for isotropic permeable substrates}
\label{Sec:Brinkman}

\rev{As in \citet{Taylor71}, we interpret the Darcy-Brinkman equation as a constitutive model for the volume-averaged (i.e.~homogenised) flow in permeable media, rather than an empirical remedy to the Darcy equation near interfaces to satisfy boundary conditions.} Compared to the classic Darcy model \citep{Darcy56}, which accounts for the effect of pressure gradients, the Darcy-Brinkman model \rev{\citep{Brinkman1949}} also incorporates the stresses that are induced by gradients of the volume-averaged velocity over length scales larger than the pore size.
\rev{A formal derivation of the Darcy-Brinkman equation for the case of isotropic highly porous media can be found in \citet{SanchezP1982} and \citet{Levy1983}.}
\citet{GmezdeSegura2018} and \citet{Gomez19} derived analytical solutions to the Darcy-Brinkman equation for permeable substrates with anisotropic permeability tensors whose principal orientations coincide with $x$, $y$, and $z$. For isotropic substrates, these solutions are considerably simplified and provide direct insight into how the substrate characteristics can affect the overlying flow. This section presents these analytical solutions with specific emphasis on their particularisation at the interface, which then serves as a boundary condition for the overlying free flow. This is complemented by a brief discussion on the role of substrate depth in these solutions.

The Darcy-Brinkman equation in an isotropic permeable substrate with permeability $K$ and depth $h$ reads
\begin{equation}
\label{eq:BrkmPhys}
    \nabla p = - \nu K^\text{-1} \boldsymbol{u} + \nu_\textit{eff} \nabla^2 \boldsymbol{u}, \qquad y\in[-h,0],
\end{equation}
where $p$ is the fluid-volume-average (`intrinsic average') pressure and $\boldsymbol{u}$ is the total-volume-average (`superficial average') velocity. $\nu_\textit{eff}=\nu/r_\nu$ is an effective viscosity perceived by the macroscale shear of the substrate flow, where the viscosity ratio $r_\nu$ is usually regarded as a constant.


\subsection{Solution for the mean flow}
\label{subSec:BrkmMean}

Let us first consider the mean flow field $\Phi(y)$ averaged in $x$ and $z$. Assume a constant pressure gradient $\partial_x P$, a mean velocity shear at the interface $\partial_y U|_0$ (approaching from below), and a no-slip condition on the floor $U|_{\text{-}h}=0$. Then the mean-flow solution to equation~(\ref{eq:BrkmPhys}) can be directly obtained as the sum of a pressure-driven and a shear-driven part,
\begin{equation*}
    U(\tilde{y}) = U_\mathrm{Pr}(\tilde{y}) + U_\mathrm{Sh}(\tilde{y}),
\end{equation*}
\begin{subeqnarray}
\label{eq:BrkmMeanU}
   & & U_\mathrm{Pr}(\tilde{y}) =  (-\partial_xP)\; \nu^\text{-1}\!K\left(1-\frac{\cosh(\tilde{y})}{\cosh(\tilde{h})}\right),  \\ 
   & & U_\mathrm{Sh}(\tilde{y}) =  (\partial_y U|_0)\;\sqrt{K/r_\nu}\left(\frac{\sinh(\tilde{y}+\tilde{h})}{\cosh(\tilde{h})}\right),
\end{subeqnarray}
where the dimensionless coordinate $\tilde{y}$ and depth $\tilde{h}$ are
\begin{equation}
\label{eq:dimlessH}
    \tilde{y}\equiv y/\sqrt{K/r_\nu}, \quad \tilde{h}\equiv h/\sqrt{K/r_\nu}\;.
\end{equation}

For the DNS cases in this study, $U_\mathrm{Pr}$ is significantly smaller than $U_\mathrm{Sh}$, so $U\approx U_\mathrm{Sh}$. Also, if the depth $h$ is large enough, $\tilde{h}\gg1$ , we can approximate (\ref{eq:BrkmMeanU}$b$) by an exponential decay
\begin{equation}
\label{eq:BrkmMeanUApprox}
    U_\mathrm{Sh}(\tilde{y}) \propto e^{\tilde{y}} = e^{y/\sqrt{K/r_\nu}}
\end{equation}
in the region not too close to the floor. The decaying exponent in equation (\ref{eq:BrkmMeanUApprox}) can thus be used to estimate the viscosity ratio $r_\nu$. In this paper, $r_\nu$ is estimated by fitting the mean velocity profile from DNSs between $y=0$ and the location where $U\approx0.1 U|_0$, while ensuring (\ref{eq:BrkmMeanU}) roughly holds. The resulting values of $r_\nu$ in this study are in the range $0.05$ - $0.98$, and the corresponding Darcy-Brinkman solutions are shown for comparison with the DNS mean velocity profiles in figures
\ref{fig:prfls_L24}, \ref{fig:prfls_Gr50}, \ref{fig:prfls_Gr50_Similar}, \ref{fig:prfls_PR2450}, \ref{fig:prfls_ReKTri} and \ref{fig:prfls_ReKeqTri}.

\subsection{Solutions for the fluctuations at the interface}
\label{subSec:BrkmFluc}

With the mean-flow part subtracted, a zero-mean field $\phi(x,z,y)$ can be represented by a Fourier series in $x$ and $z$ such that $\phi(x,z,y)=\sum_{\boldsymbol{\alpha}\neq\boldsymbol{0}}\hat{\phi}(\alpha_x,\alpha_z,y)e^{\mathrm{i}(\alpha_x x + \alpha_z z)}$ where $\alpha_x$ and $\alpha_z$ are the $x$- and $z$- components of the planar wavevector $\boldsymbol{\alpha}$. Each Fourier mode $\hat{\phi}(\alpha_x,\alpha_z,y)$ follows the transform of Eq.~(\ref{eq:BrkmPhys}),
\begin{subeqnarray}
\label{eq:BrkmFour}
    \mathrm{i}\alpha_x\hat{p} & = &  r_\nu^\text{-1}\nu\left(\partial^2_{yy}-\alpha^2-r_\nu K^{\text{-1}}\right)\!\hat{u}\,, \\
    \partial_y\hat{p} & = &  r_\nu^\text{-1}\nu\left(\partial^2_{yy}-\alpha^2-r_\nu K^{\text{-1}}\right)\!\hat{v}\,, \\
    \mathrm{i}\alpha_z\hat{p} & = &  r_\nu^\text{-1}\nu\left(\partial^2_{yy}-\alpha^2-r_\nu K^{\text{-1}}\right)\!\hat{w},
\end{subeqnarray}
where $\alpha\equiv\sqrt{\alpha_x^2+\alpha_z^2}$.

We define a wavevector-aligned coordinate system, of which the column basis vectors $\boldsymbol{e}_{/\!\!/}$, $\boldsymbol{e}_y$, and $\boldsymbol{e}_{\bot}$ are obtained by the rotation transform
\begin{equation}
    \left[                  \;
    \boldsymbol{e}_{/\!\!/} \
    \boldsymbol{e}_y        \
    \boldsymbol{e}_{\bot}   \;
    \right] = 
    \mathsfbi{Q}\cdot
    \left[                  \;
    \boldsymbol{e}_x        \
    \boldsymbol{e}_y        \
    \boldsymbol{e}_z        \;
    \right], \qquad 
    \mathsfbi{Q}\equiv
    \left[\begin{array}{ccc}
        \alpha_x/\alpha\ & 0 &  -\alpha_z/\alpha \\
               0       \ & 1 & \        0        \\
        \alpha_z/\alpha\ & 0 & \ \alpha_x/\alpha
    \end{array}\right],
\end{equation}
where $\boldsymbol{e}_x$, $\boldsymbol{e}_y$ and $\boldsymbol{e}_z$ are the unit vectors in $x$, $y$ and $z$. Eq.~(\ref{eq:BrkmFour}) can then be written as
\begin{subeqnarray}
\label{eq:BrkmWVsys}
    \mathrm{i}\alpha\:\hat{p} & = &  r_\nu^\text{-1}\nu\left(\partial^2_{yy}-\alpha^2-r_\nu K^{\text{-1}}\right)\!\hat{u}_{/\!\!/}, \\
    \partial_y\hat{p} & = &  r_\nu^\text{-1}\nu\left(\partial^2_{yy}-\alpha^2-r_\nu K^{\text{-1}}\right)\!\hat{v}\;\;, \\
    0 & = &  r_\nu^\text{-1}\nu\left(\partial^2_{yy}-\alpha^2-r_\nu K^{\text{-1}}\right)\!\hat{u}_\bot,
\end{subeqnarray}
where $[\ \hat{u}_{/\!\!/}\ \ \hat{v}\quad \hat{u}_\bot]^\mathrm{T}=\mathsfbi{Q}^\mathrm{T}\cdot[\ \hat{u}\quad \hat{v}\quad \hat{w} \ ]^\mathrm{T}$. The equation for $\hat{u}_\bot$, (\ref{eq:BrkmWVsys}c), decouples from the rest of the problem and can be solved by itself under appropriate boundary conditions. Separately, equations (\ref{eq:BrkmWVsys}a,b) combined with the continuity equation,
\begin{equation}
\label{eq:ContinuityWVsys}
    \mathrm{i}\alpha\:\hat{u}_{/\!\!/} + \partial_y\hat{v} = 0,
\end{equation}
form the equation system for $\hat{u}_{/\!\!/}$, $\hat{v}$, and $\hat{p}$. 

At the free-flow interface, $y=0$, let us assume that the values of velocity shear rates $\partial_y\hat{u}_{/\!\!/}|_0$ and  $\partial_y\hat{u}_\bot|_0$ and pressure $\hat{p}|_0$ are given. At the floor boundary, $y=-h$, no-slip, no-penetration conditions, $\hat{u}_{/\!\!/}|_{\text{-}h}=\hat{v}|_{\text{-}h}=\hat{u}_\bot|_{\text{-}h}=0$, are imposed. The solutions to equations (\ref{eq:BrkmWVsys})-(\ref{eq:ContinuityWVsys}) yield then a constitutive relationship between velocity and stress at the interface:
\begin{equation}
\label{eq:BrkmConstitutive}
    \left[\begin{array}{r}
             \hat{u}_{/\!\!/}|_0 \\
        \mathrm{i}\:\hat{v}\:|_0 \\
                 \hat{u}_\bot|_0
    \end{array}\right] = 
    \mathsfbi{H}_{Br}\cdot
    \left[\begin{array}{r}
                 \partial_y\hat{u}_{/\!\!/}|_0 \\
        -\mathrm{i}\:\nu^\text{-1}\hat{p}\:|_0 \\
                     \partial_y\hat{u}_\bot|_0
    \end{array}\right], \quad
    \mathsfbi{H}_{Br} \equiv
    \left[\begin{array}{ccc}
        \ \    \mathcal{L}_{slip} & \alpha \mathcal{K}_{slip} & \:  0    \\
        \alpha \mathcal{N}_{trsp} & \alpha \mathcal{K}_{trsp} & \:  0    \\
        \          0    & \         0     & \mathcal{L}_{slip}^{\bot}
    \end{array}\right] \: ,
\end{equation}
in which $\mathsfbi{H}_{Br}$ is a real-valued matrix with dimension of length, and is an inherent characteristic of a substrate. It contains five independent coefficients, which are functions of $\alpha$ and $h$,
\begin{subeqnarray}
\label{eq:HMat}
    \mathcal{L}_{slip} & = & \sqrt{(K/r_\nu)\: f_{\mathcal{L}s}(\tilde{\alpha},\tilde{h})}, \\
    \mathcal{K}_{slip} & = & \qquad K \quad \; \, f_{\mathcal{K}s}(\tilde{\alpha},\tilde{h}), \\
    \mathcal{N}_{trsp} & = & \quad (K/r_\nu) \: f_{\mathcal{N}t}(\tilde{\alpha},\tilde{h}), \\
    \mathcal{K}_{trsp} & = & \qquad K \quad \; \, f_{\mathcal{K}t}(\tilde{\alpha},\tilde{h}), \\
    \mathcal{L}_{slip}^{\bot} & = & \sqrt{(K/r_\nu)\: f_{\mathcal{L}s}^{\bot}(\tilde{\alpha},\tilde{h})},
\end{subeqnarray}
where
\begin{subeqnarray}
\label{eq:fh}
    f_{\mathcal{L}s} & = & \frac{\left(\tanh(\tilde{\alpha}_K\tilde{h})-(\tilde{\alpha}/\tilde{\alpha}_K)\tanh(\tilde{\alpha}\tilde{h})\right)^2}{\tilde{\alpha}_K^2\:\xi^2}, \qquad \\
    f_{\mathcal{K}s} & = & \frac{ \left(1\!+\!(\tilde{\alpha}/\tilde{\alpha}_K\!)^2\right)\!\left(1\!-\!\sech(\tilde{\alpha}\tilde{h})\sech(\tilde{\alpha}_K\tilde{h})\right)
    - 2(\tilde{\alpha}/\tilde{\alpha}_K\!)\tanh(\tilde{\alpha}\tilde{h})\tanh(\tilde{\alpha}_K\tilde{h})}{\xi}, \qquad \\
    f_{\mathcal{N}t} & = & \frac{1-\sech(\tilde{\alpha}\tilde{h})\sech(\tilde{\alpha}_K\tilde{h})-(\tilde{\alpha}/\tilde{\alpha}_K)\tanh(\tilde{\alpha}\tilde{h})\tanh(\tilde{\alpha}_K\tilde{h})}{\tilde{\alpha}_K^2\:\xi}, \\
    f_{\mathcal{K}t} & = & \frac{\tanh(\tilde{\alpha}\tilde{h})-(\tilde{\alpha}/\tilde{\alpha}_K)\tanh(\tilde{\alpha}_K\tilde{h})}{\tilde{\alpha}_K^2\:\xi}, \qquad \\
    f_{\mathcal{L}s}^{\bot} & = & \frac{\tanh^2(\tilde{\alpha}_K\tilde{h})}{\tilde{\alpha}_K^2}, \qquad 
\end{subeqnarray}
and where $\tilde{\alpha}$ is the dimensionless wavenumber defined by
\begin{equation}
\label{eq:dimlessAlpha}
    \tilde{\alpha}\equiv\alpha\sqrt{K/r_\nu},
\end{equation}
\begin{equation}
\tilde{\alpha}_K = \sqrt{1+\tilde{\alpha}^2}, \nonumber
\end{equation}
\begin{equation}
\xi = 1-(\tilde{\alpha}/\tilde{\alpha}_K)\tanh(\tilde{\alpha}\tilde{h})\tanh(\tilde{\alpha}_K\tilde{h})-(\tilde{\alpha}/\tilde{\alpha}_K)^2\sech(\tilde{\alpha}\tilde{h})\sech(\tilde{\alpha}_K\tilde{h}). \nonumber
\end{equation}\\
The five attenuating functions $f_{*}(\tilde{\alpha},\tilde{h})$ in (\ref{eq:fh}) have values in the range of $0$ - $1$ and account for the effect of substrate depth in (\ref{eq:HMat}).

\subsection{On the effect of depth}
\label{subSec:BrkmInfer}

\begin{figure}
  \centerline{\includegraphics[width=1.0\linewidth]{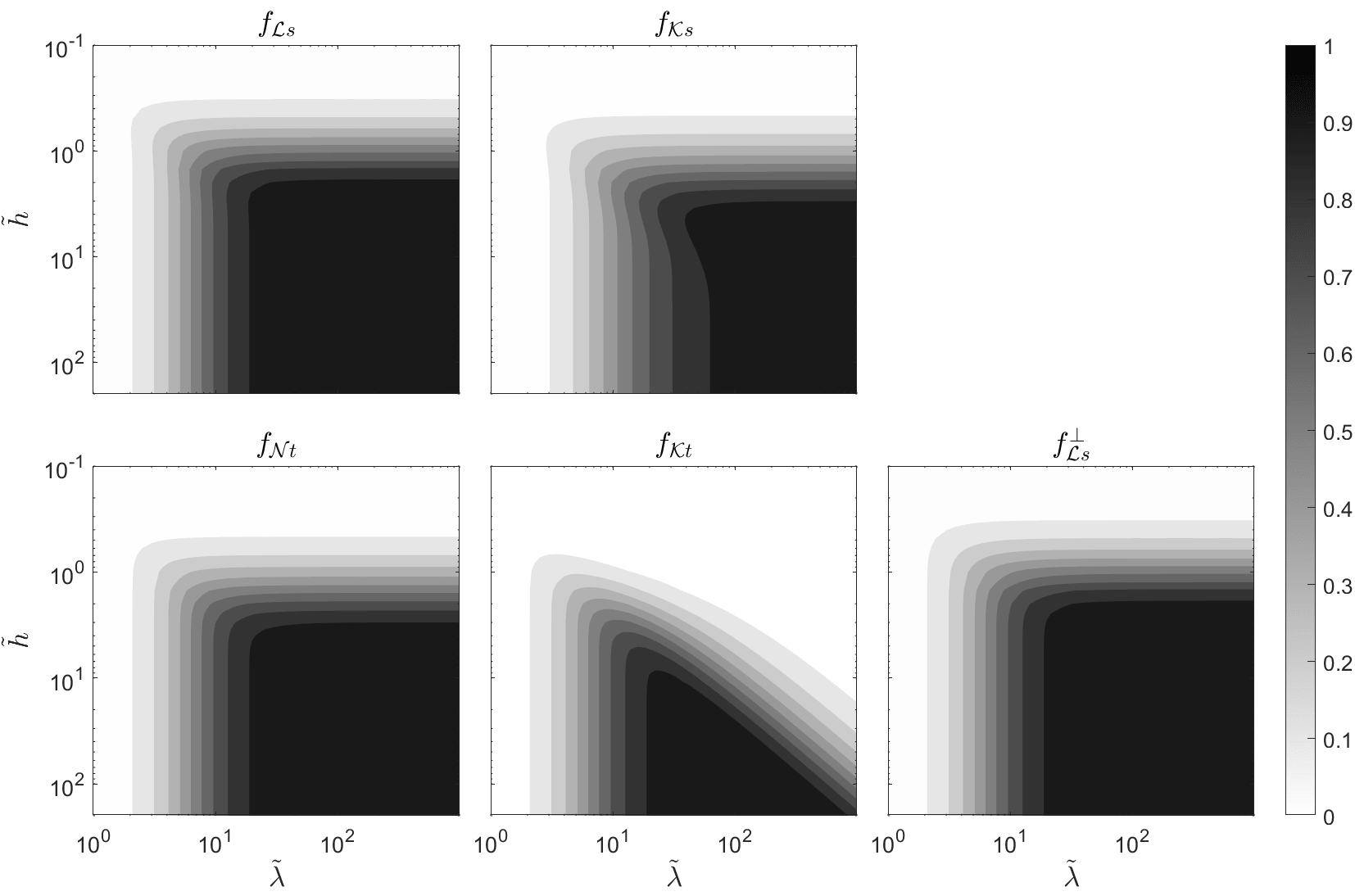}}
  \caption{Darcy-Brinkman-based attenuating coefficients $f_{*}$ as functions of dimensionless wavelength $\tilde{\lambda}\equiv2\pi/\tilde{\alpha}$ and depth $\tilde{h}$.}
\label{fig:fh_brkm}
\end{figure}

Figure \ref{fig:fh_brkm} shows the values of the five attenuating functions $f_{*}$ against $\tilde{h}$ and the dimensionless wavelength $\tilde{\lambda}=2\pi/\tilde{\alpha}$. For $f_{\mathcal{L}s}$, $f_{\mathcal{K}s}$, $f_{\mathcal{N}t}$ and $f_{\mathcal{L}s}^{\bot}$, a threshold $\tilde{h}\approx2$ - $3$  can be established, beyond which $f_{*}$ becomes essentially depth-independent, and thus the substrate can be seen as `sufficiently deep' with regards to $\mathcal{L}_{slip}$, $\mathcal{K}_{slip}$, $\mathcal{N}_{trsp}$ and $\mathcal{L}_{slip}^{\bot}$. Such a threshold, however, cannot be established for $f_{\mathcal{K}t}$. The contour lines of $f_{\mathcal{K}t}$ for large waves follow $\tilde{\lambda}\sim\tilde{h}$, with $\tilde{\lambda}\approx5\tilde{h}$ for $f_{\mathcal{K}t}=0.9$. This scaling behaviour indicates that pressure fluctuations with wavelength $\tilde{\lambda}\gtrsim5\tilde{h}$ will always perceive the presence of the floor impeding them from inducing any significant transpiration across the interface, regardless of how deep the substrate may be relative to $\sqrt{K}$. Therefore, `sufficiently deep' for $\mathcal{K}_{trsp}$ is not a property exclusively inherent to the substrate, but instead depends on the typical length scales of the overlying pressure fluctuations, $\lambda_{p}$.

For near-wall turbulence, the scale $\lambda_{p}$ in wall units is typically $\lambda_{p}^+\approx200$ for active eddies (see the figures of spectra in \S\ref{subSec:deepPorous} and \S\ref{subSec:Keq}), so for a permeable substrate with $\sqrt{K^+}\lesssim10$, we have $\tilde{\lambda}_{p}=\lambda_{p}/\sqrt{K/r_\nu}\gtrsim10$. In such a scenario, figure \ref{fig:fh_brkm} shows that, for $f_{\mathcal{K}t}$, a sufficiently deep substrate must have at least $\tilde{h}\gtrsim5$. Consequently, as an initially deep substrate becomes gradually shallower, the overlying turbulence will first perceive the suppression of pressure-induced transpiration, i.e.~$\mathcal{K}_{trsp}$ decreasing at large scales, whereas the effect of the substrate depth on the other coefficients will only be perceived later, when $\tilde{h}\approx2$ - $3$.

\section{\rgm{Roughness function versus upscaled coefficients from homogenisation}}
\label{Sec:homog}

\begin{figure}
\vspace*{4mm}
  \centerline{\includegraphics[width=1.0\linewidth]{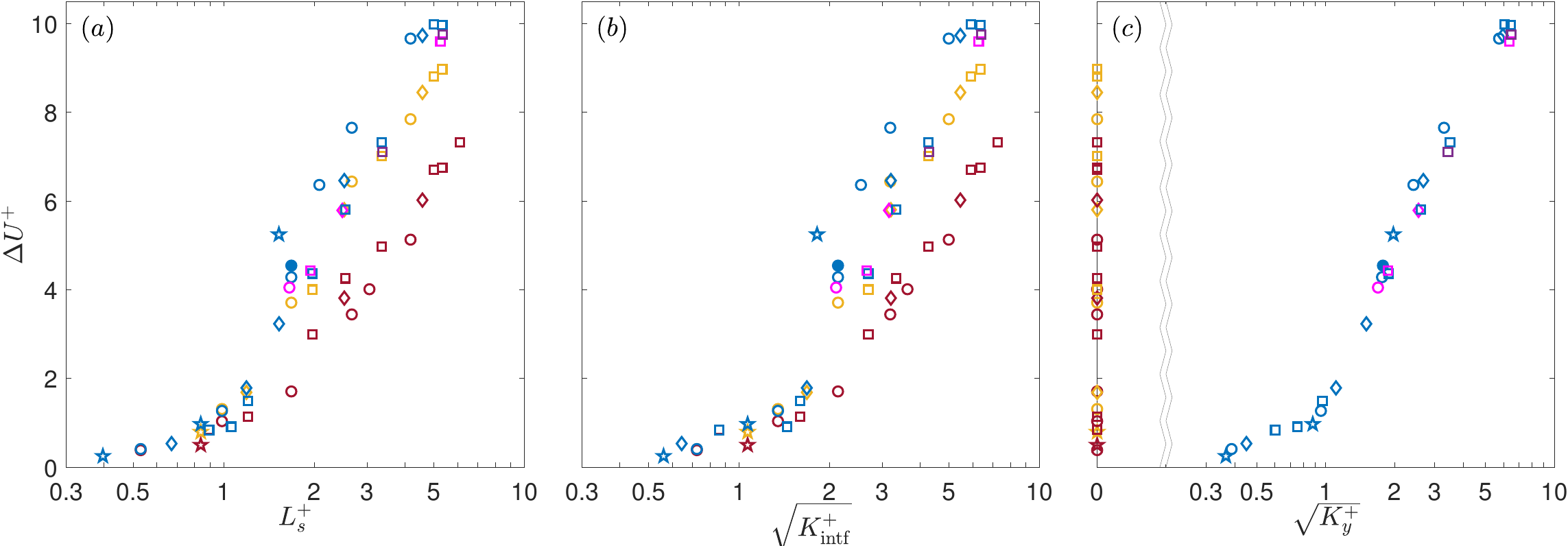}}
\vspace*{2mm}
  \caption{
  \rgm{Velocity deficit $\Delta U^+$ for the present DNS substrates as a function of their upscaled coefficients derived from homogenisation
  \citep{Bottaro2020,Naqvi21}. Symbols are as in figure~\ref{fig:Us_vs_All6Len}. $(a)$~ slip length,~$L_s$; $(b)$ interface permeability, $K_\mathrm{intf}$; $(c)$ intrinsic permeability, $K_y$.
  }
  }
  \label{fig:DUOPT_vs_Battaros}
\end{figure}

\rgm{
Here we report the scaling of the roughness function obtained in our DNSs with the admittance coefficients
for our substrate geometries obtained following the second-order homogenisation model of \citet{Bottaro2020} and \citet{Naqvi21}.
The coefficients are obtained \textit{a priori} from a series of simple auxiliary problems on the substrate geometry
based on a Stokes-flow assumption. The first coefficient is the conventional slip length, $L_s$, which
essentially gives the $u$-$\partial u/\partial y$ admittance. The second is the `interfacial permeability',
$K_\mathrm{intf}$, which gives the $u$-$p$ admittance, and from the problem symmetries also the
$v$-$\partial u/\partial y$ admittance. The final coefficient is the intrinsic permeability, $K_y$, which
gives the $v$-$p$ admittance, and is thus analogue to the $K_{eq}^t$ of equation \ref{eq:DefKeq4}. Results for the DNS-measured roughness function $\Delta U^+$ as a function
of the pre-computed $L_s$, $K_\mathrm{intf}$ and $K_y$ are shown in figure \ref{fig:DUOPT_vs_Battaros}.
Overall, the collapse is generally better with $K_{eq}^t$, as shown in figure \ref{fig:DU_vs_uvKeq}($d$). We note
that the methods used to obtain the coefficients in \citet{Bottaro2020} and \citet{Naqvi21} differ for
rough and porous substrates. For rough substrates, it is assumed that there is a bottom impermeable wall blocking
the flow, and this results in $K_y=0$ \citep{Bottaro2020}. For permeable substrates, the auxiliary problems are solved in a mesoscale domain that penetrates into the substrate a few grains, typically of order $\sim 5L$. A Darcy-like transpiration boundary condition is set at the bottom of the mesoscale domain. As a result, this Darcy transpiration is transferred to the solution, yielding $K_y=K$. For our present substrates, we would assume that at least for substrate depths $h=1D$-$2D$, if not for $h=5D$-$7D$, the rough-wall framework is more appropriate, and thus $K_y=0$, as portrayed in \ref{fig:DUOPT_vs_Battaros}($c$). If we overlooked the presence of a bottom wall in our substrates and assumed $K_y=K$ for all of them, the result would be as shown in figure \ref{fig:DU_vs_All6Len}($f$).
}

\section{\rgm{Roughness function for other substrate topologies}}
\label{Sec:OtherData}

\begin{figure}
\vspace*{1mm}
  \centerline{\includegraphics[width=1.0\linewidth]{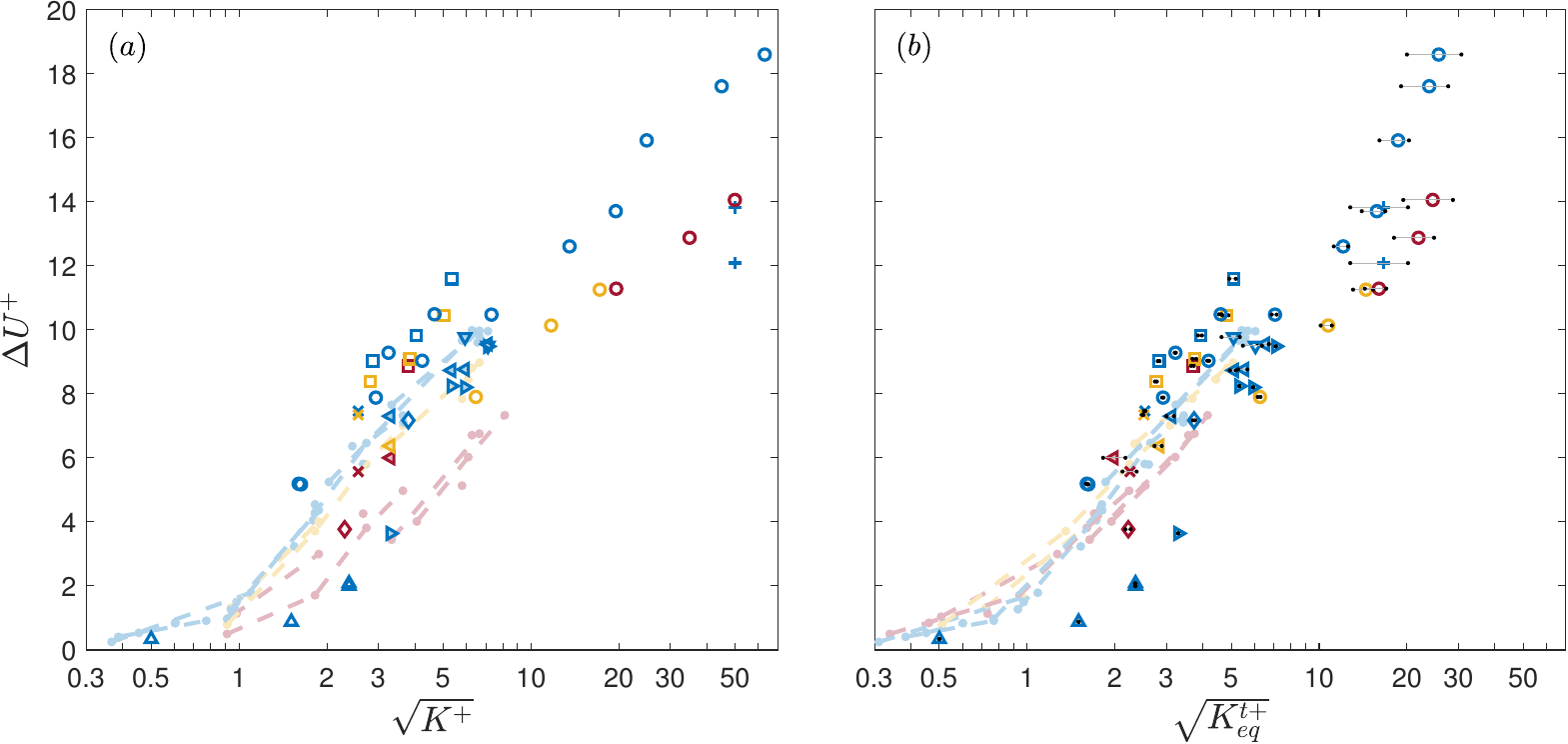}}
  \caption{
  \rgm{Velocity deficit $\Delta U^+$ as a function of $(a)$ the permeability $K^+$ and $(b)$ the equivalent
  permeability $K_{eq}^{t+}$ for the present DNSs and other substrates in the literature. $\bullet$ and lines faded in the background, present
  staggered cubes; $\medtriangleleft$ meshes of cases MPd; $\medtriangleright$, meshes of cases MFPd;
  $\medtriangledown$, meshes of cases MTPd; $\square$, randomly packed spheres from \citet{Zippe83};
  $\meddiamond$, staggered cubes from \citet{Kuwata16a}; $+$, collocated spheres from \citet{Kim20};
  $\medcircle$, reticulated foams from \citet{Esteban22}; $\times$, randomly packed spheres
  from \citet{Karra2023}; $\medtriangleup$, mesh lattices from \citet{Khorasani2024}.}
  }
  \label{fig:DUext_vs_Keq}
\end{figure}

\rgm{
In this final appendix we compile roughness function results for other substrate topologies from both
experimental and numerical literature. Most of these have larger permeabilities $K^+$ than the scope of the
present work. As a result, we would expect some of our underlying assumptions to fail, such as assuming Stokes
flow for the microscale flow when deriving expression \ref{eq:DefKeq4} for $K_{eq}^t$, or that the direct effect of texture granularity on the overlying turbulence is small. The collapse of roughness
function data with $K_{eq}^{t+}$, and with $K^+$ for reference, is portrayed in \ref{fig:DUext_vs_Keq}, showing
reasonably good agreement even at these larger $K^+$ values. We note that the substrates of 
\citet{Khorasani2024} appear to follow a lower $\Delta U^+$ trend, especially for small $K_{eq}^{t+}$. These mesh
substrates are different from the rest in that their exposed interface is perfectly flat. This implies an
absence of interfacial roughness, probably suppressing partially the increase in drag compared to
substrates with roughness. Following this, we have conducted an additional set of simulations for
our mesh topologies, where we made the interface flat as in \citet{Khorasani2024} or, conversely, extended the protruding ligaments from their original height $\ell/2$ to
$\ell$. The parameters of these simulations are summarised in table~\ref{tab:cases_mesh}. The flat-interface cases
follow the same trend of the substrates of \citet{Khorasani2024}, while the taller-protrusion cases
show a trend similar to all the other substrates with rough interfaces.
}
\rev{We note that this figure includes results for various irregular grain topologies
\citep{Zippe83,Esteban22,Karra2023}, and they follow the same trend of regular ones.}

\begin{table}
  \begin{center}
\def~{\hphantom{0}}
\fontsize{8}{9}\selectfont
\renewcommand{\arraystretch}{1.08}
\setlength{\tabcolsep}{5.3pt}
\rgm{
  \begin{tabular}{lcccccccc}
    $\qquad$Case & $\ Re_\tau\ $ & $L^+$ & $g/L$ & $h/L$ & $\quad\varepsilon\quad$ & $\sqrt{K^+}\ $ & $\sqrt{K_{eq}^{t+}}$ & $\Delta U^+$ \\ [3pt]
    MFPd-36-78    & 182.7 & 35.9 &  7/9  & 5 & 0.87 & 5.34 & 5.33 & 8.24 \\
    MFPd-48-56    & 182.7 & 47.8 &  5/9  & 5 & 0.58 & 3.30 & 3.29 & 3.64 \\
    MFPd-48-72    & 182.7 & 47.8 & 13/18 & 5 & 0.81 & 5.94 & 5.94 & 8.20 \\
    MFPd-48-78    & 182.7 & 47.8 &  7/9  & 5 & 0.87 & 7.12 & 7.11 & 9.48 \\
    MTPd-48-72    & 182.7 & 47.8 & 13/18 & 5 & 0.81 & 5.94 & 5.09 & 9.77 \\
    MTPd-48-78    & 182.7 & 47.8 &  7/9  & 5 & 0.87 & 7.12 & 6.05 & 9.50 \\
  \end{tabular}
}
  \caption{
  \rgm{Parameters for mesh substrates with flat interfaces (MFPd) and tall interfacial protrusions (MTPd). $L$ is the
  grain pitch, $g$ the gap size, $h$ the substrate depth, $\varepsilon$ the porosity, $K$ the permeability, $K_{eq}^{t}$ the transpiration equivalent permeability, and $\Delta U^+$ the resulting roughness function. The `$+$' superscripts indicate viscous scaling.}
  }
  \label{tab:cases_mesh}
  \end{center}
\end{table}

\bibliographystyle{jfm}
\bibliography{porous}

\begin{thebibliography}{72}
\expandafter\ifx\csname natexlab\endcsname\relax\def\natexlab#1{#1}\fi
\def\au#1{#1} \def\ed#1{#1} \def\yr#1{#1}\def\at#1{#1}\def\jt#1{\textit{#1}}
  \def\bt#1{#1}\def\bvol#1{\textbf{#1}} \def\vol#1{#1} \def\pg#1{#1}
  \def\publ#1{#1}\def\arxiv#1{#1}\def\org#1{#1}\def\st#1{\textit{#1}}

\bibitem[Abderrahaman-Elena {\em et~al.\/}(2019)Abderrahaman-Elena, Fairhall \&
  Garc{\'i}a-Mayoral]{Abderrahaman19}
{\sc \au{Abderrahaman-Elena, Nabil}, \au{Fairhall, Chris~T} \&
  \au{Garc{\'i}a-Mayoral, Ricardo}} \yr{2019}  \at{Modulation of near-wall
  turbulence in the transitionally rough regime}.  \jt{J. Fluid Mech.}
  \bvol{865},  \pg{1042--1071}.

\bibitem[Abderrahaman-Elena \& Garc{\'i}a-Mayoral(2017)]{Abderrahaman17}
{\sc \au{Abderrahaman-Elena, Nabil} \& \au{Garc{\'i}a-Mayoral, Ricardo}}
  \yr{2017}  \at{Analysis of anisotropically permeable surfaces for turbulent
  drag reduction}.  \jt{Phys. Rev. Fluids}  \bvol{2}~(11),  \pg{114609}.

\bibitem[Aghaei-Jouybari {\em et~al.\/}(2024)Aghaei-Jouybari, Seo, Pinto,
  Cattafesta, Meneveau \& Mittal]{AghaeiJouybari2024}
{\sc \au{Aghaei-Jouybari, Mostafa}, \au{Seo, Jung-Hee}, \au{Pinto, Sasindu},
  \au{Cattafesta, Louis}, \au{Meneveau, Charles} \& \au{Mittal, Rajat}}
  \yr{2024}  \at{Extended darcy–forchheimer law including inertial flow
  deflection effects}.  \jt{J. Fluid Mech.}  \bvol{980},  \pg{A13}.

\bibitem[Bottaro \& Naqvi(2020)]{Bottaro2020}
{\sc \au{Bottaro, Alessandro} \& \au{Naqvi, Sahrish~B.}} \yr{2020}
  \at{Effective boundary conditions at a rough wall: a high-order
  homogenization approach}.  \jt{Meccanica}  \bvol{55}~(9),  \pg{1781--1800}.

\bibitem[Bou-Zeid {\em et~al.\/}(2020)Bou-Zeid, Anderson, Katul \&
  Mahrt]{BouZeid2020}
{\sc \au{Bou-Zeid, Elie}, \au{Anderson, William}, \au{Katul, Gabriel~G.} \&
  \au{Mahrt, Larry}} \yr{2020}  \at{The persistent challenge of surface
  heterogeneity in boundary-layer meteorology: A review}.  \jt{Bound.-Layer
  Meteorol.}  \bvol{177}~(2-3),  \pg{227--245}.

\bibitem[Breugem {\em et~al.\/}(2006)Breugem, Boersma \&
  Uittenbogaard]{Breugem06}
{\sc \au{Breugem, WP}, \au{Boersma, BJ} \& \au{Uittenbogaard, RE}} \yr{2006}
  \at{The influence of wall permeability on turbulent channel flow}.  \jt{J.
  Fluid Mech.}  \bvol{562},  \pg{35--72}.

\bibitem[Breugem \& Boersma(2005)]{Breugem05}
{\sc \au{Breugem, Wim-Paul} \& \au{Boersma, Bendiks-Jan}} \yr{2005}  \at{Direct
  numerical simulations of turbulent flow over a permeable wall using a direct
  and a continuum approach}.  \jt{Phys. Fluids}  \bvol{17}~(2),  \pg{025103}.

\bibitem[Brinkman(1949)]{Brinkman1949}
{\sc \au{Brinkman, H.~C.}} \yr{1949}  \at{A calculation of the viscous force
  exerted by a flowing fluid on a dense swarm of particles}.  \jt{Flow Turbul.
  Combust.}  \bvol{1}~(1).

\bibitem[Chandesris \& Jamet(2009)]{Chandesris2009}
{\sc \au{Chandesris, M.} \& \au{Jamet, D.}} \yr{2009}  \at{Derivation of jump
  conditions for the turbulence $k$--$\epsilon$ model at a fluid/porous
  interface}.  \jt{Int. J. Heat Fluid Flow}  \bvol{30}~(2),  \pg{306--318}.

\bibitem[Chen \& Garc{\'i}a-Mayoral(2023)]{Chen23}
{\sc \au{Chen, Zishen} \& \au{Garc{\'i}a-Mayoral, Ricardo}} \yr{2023}
  \at{Examination of outer-layer similarity in wall turbulence over obstructing
  surfaces}.  \jt{J.~Fluid Mech.}  \bvol{973},  \pg{A31}.

\bibitem[Chung {\em et~al.\/}(2021)Chung, Hutchins, Schultz \& Flack]{Chung21}
{\sc \au{Chung, Daniel}, \au{Hutchins, Nicholas}, \au{Schultz, Michael~P} \&
  \au{Flack, Karen~A}} \yr{2021}  \at{Predicting the drag of rough surfaces}.
  \jt{Annu. Rev. Fluid Mech.}  \bvol{53},  \pg{439--471}.

\bibitem[Cooper {\em et~al.\/}(2017)Cooper, Ockleford, Rice \&
  Powell]{Cooper2017}
{\sc \au{Cooper, James~R.}, \au{Ockleford, Annie}, \au{Rice, Stephen~P.} \&
  \au{Powell, D.~Mark}} \yr{2017}  \at{Does the permeability of gravel river
  beds affect near‐bed hydrodynamics?}  \jt{Earth Surf. Process. Landf.}
  \bvol{43}~(5),  \pg{943--955}.

\bibitem[Darcy(1856)]{Darcy56}
{\sc \au{Darcy, Henry}} \yr{1856} {\em Les fontaines publiques de la ville de
  Dijon: exposition et application des principes {\`a} suivre et des formules
  {\`a} employer dans les questions de distribution d'eau\/}, ,  \vol{vol.~1}.
  \publ{Victor dalmont}.

\bibitem[Efstathiou \& Luhar(2018)]{Efstathiou18}
{\sc \au{Efstathiou, Christoph} \& \au{Luhar, Mitul}} \yr{2018}  \at{Mean
  turbulence statistics in boundary layers over high-porosity foams}.  \jt{J.
  Fluid Mech.}  \bvol{841},  \pg{351--379}.

\bibitem[Esteban {\em et~al.\/}(2022)Esteban, Rodr{\'\i}guez-L{\'o}pez,
  Ferreira \& Ganapathisubramani]{Esteban22}
{\sc \au{Esteban, LB}, \au{Rodr{\'\i}guez-L{\'o}pez, E}, \au{Ferreira, MA} \&
  \au{Ganapathisubramani, B}} \yr{2022}  \at{Mean flow of turbulent boundary
  layers over porous substrates}.  \jt{Phys. Rev. Fluids}  \bvol{7}~(9),
  \pg{094603}.

\bibitem[Fairhall {\em et~al.\/}(2019)Fairhall, Abderrahaman-Elena \&
  Garc\'{i}a-Mayoral]{Fairhall2019}
{\sc \au{Fairhall, C.~T.}, \au{Abderrahaman-Elena, N.} \&
  \au{Garc\'{i}a-Mayoral, R.}} \yr{2019}  \at{{The effect of slip and surface
  texture on turbulence over superhydrophobic surfaces.}}  \jt{J. Fluid Mech.}
  \bvol{861},  \pg{88--118}.

\bibitem[Fairhall \& García-Mayoral(2018)]{Fairhall2018}
{\sc \au{Fairhall, C.~T.} \& \au{García-Mayoral, R.}} \yr{2018}  \at{Spectral
  analysis of the slip-length model for turbulence over textured
  superhydrophobic surfaces}.  \jt{Flow Turbul. Combust.}  \bvol{100}~(4),
  \pg{961--978}.

\bibitem[Fang {\em et~al.\/}(2018)Fang, Han, He \& Dey]{Fang18}
{\sc \au{Fang, Hongwei}, \au{Han, Xu}, \au{He, Guojian} \& \au{Dey, Subhasish}}
  \yr{2018}  \at{Influence of permeable beds on hydraulically macro-rough
  flow}.  \jt{J. Fluid Mech.}  \bvol{847},  \pg{552--590}.

\bibitem[Flores \& Jimenez(2006)]{Flores06}
{\sc \au{Flores, Oscar} \& \au{Jimenez, Javier}} \yr{2006}  \at{Effect of
  wall-boundary disturbances on turbulent channel flows}.  \jt{J. Fluid Mech.}
  \bvol{566},  \pg{357--376}.

\bibitem[Garc{\'i}a-Mayoral {\em et~al.\/}(2024)Garc{\'i}a-Mayoral, Chung,
  Durbin, Hutchins, Knopp, McKeon, Piomelli \& Sandberg]{GarciaMayoral2024}
{\sc \au{Garc{\'i}a-Mayoral, Ricardo}, \au{Chung, Daniel}, \au{Durbin, Paul},
  \au{Hutchins, Nicholas}, \au{Knopp, Tobias}, \au{McKeon, Beverley~J.},
  \au{Piomelli, Ugo} \& \au{Sandberg, Richard~D.}} \yr{2024}  \at{Challenges
  and perspective on the modelling of high-re, incompressible, non-equilibrium,
  rough-wall boundary layers}.  \jt{J. Turbul.} .

\bibitem[Garc{\'i}a-Mayoral \& Jim{\'e}nez(2011)]{Garcia11}
{\sc \au{Garc{\'i}a-Mayoral, Ricardo} \& \au{Jim{\'e}nez, Javier}} \yr{2011}
  \at{Hydrodynamic stability and breakdown of the viscous regime over riblets}.
   \jt{J. Fluid Mech.}  \bvol{678},  \pg{317--347}.

\bibitem[Garc{\'i}a-Mayoral \& Jim{\'e}nez(2012)]{Garcia12}
{\sc \au{Garc{\'i}a-Mayoral, Ricardo} \& \au{Jim{\'e}nez, Javier}} \yr{2012}
  \at{Scaling of turbulent structures in riblet channels up to
  re$\tau\approx$550}.  \jt{Phys. Fluids}  \bvol{24}~(10).

\bibitem[{G{\'o}mez-de-Segura} \& Garc{\'i}a-Mayoral(2019)]{Gomez19}
{\sc \au{{G{\'o}mez-de-Segura}, Garazi} \& \au{Garc{\'i}a-Mayoral, Ricardo}}
  \yr{2019}  \at{Turbulent drag reduction by anisotropic permeable
  substrates--analysis and direct numerical simulations}.  \jt{J. Fluid Mech.}
  \bvol{875},  \pg{124--172}.

\bibitem[{G{\'o}mez-de-Segura} {\em et~al.\/}(2018{\natexlab{{\em
  a\/}}}){G{\'o}mez-de-Segura}, Sharma \& Garc{\'i}a-Mayoral]{GmezdeSegura2018}
{\sc \au{{G{\'o}mez-de-Segura}, G.}, \au{Sharma, A.} \& \au{Garc{\'i}a-Mayoral,
  R.}} \yr{2018{\natexlab{{\em a\/}}}}  \at{Turbulent drag reduction using
  anisotropic permeable substrates}.  \jt{Flow Turbul. Combust.}
  \bvol{100}~(4),  \pg{995--1014}.

\bibitem[{G{\'o}mez-de-Segura} {\em et~al.\/}(2018{\natexlab{{\em
  b\/}}}){G{\'o}mez-de-Segura}, Sharma \& Garc{\'i}a-Mayoral]{GG_CTR_2018}
{\sc \au{{G{\'o}mez-de-Segura}, G.}, \au{Sharma, A.} \& \au{Garc{\'i}a-Mayoral,
  R.}} \yr{2018{\natexlab{{\em b\/}}}} Virtual origins in turbulent flows over
  complex surfaces.  \bt{In {\em Proc 2018 CTR Summer Program\/}},  \pg{pp.
  277--286}.

\bibitem[Habibi~Khorasani {\em et~al.\/}(2024)Habibi~Khorasani, Luhar \&
  Bagheri]{Khorasani2024}
{\sc \au{Habibi~Khorasani, Seyed~Morteza}, \au{Luhar, Mitul} \& \au{Bagheri,
  Shervin}} \yr{2024}  \at{Turbulent flows over porous lattices: alteration of
  near-wall turbulence and pore-flow amplitude modulation}.  \jt{J. Fluid
  Mech.}  \bvol{984},  \pg{A63}.

\bibitem[Ibrahim {\em et~al.\/}(2021)Ibrahim, G{\'o}mez-de Segura, Chung \&
  Garc{\'i}a-Mayoral]{Ibrahim21}
{\sc \au{Ibrahim, Joseph~I}, \au{G{\'o}mez-de Segura, Garazi}, \au{Chung,
  Daniel} \& \au{Garc{\'i}a-Mayoral, Ricardo}} \yr{2021}  \at{The
  smooth-wall-like behaviour of turbulence over drag-altering surfaces: a
  unifying virtual-origin framework}.  \jt{J. Fluid Mech.}  \bvol{915},
  \pg{A56}.

\bibitem[Jim{\'e}nez(2004)]{Jimenez04}
{\sc \au{Jim{\'e}nez, Javier}} \yr{2004}  \at{Turbulent flows over rough
  walls}.  \jt{Annu. Rev. Fluid Mech.}  \bvol{36},  \pg{173--196}.

\bibitem[Jimenez {\em et~al.\/}(2001)Jimenez, Uhlmann, Pinelli \&
  Kawahara]{Jimenez01}
{\sc \au{Jimenez, Javier}, \au{Uhlmann, Markus}, \au{Pinelli, Alfredo} \&
  \au{Kawahara, Genta}} \yr{2001}  \at{Turbulent shear flow over active and
  passive porous surfaces}.  \jt{J. Fluid Mech.}  \bvol{442},  \pg{89--117}.

\bibitem[Jin {\em et~al.\/}(2015)Jin, Uth, Kuznetsov \& Herwig]{Jin2015}
{\sc \au{Jin, Y.}, \au{Uth, M.-F.}, \au{Kuznetsov, A.~V.} \& \au{Herwig, H.}}
  \yr{2015}  \at{Numerical investigation of the possibility of macroscopic
  turbulence in porous media: a direct numerical simulation study}.  \jt{J.
  Fluid Mech.}  \bvol{766},  \pg{76--–103}.

\bibitem[Karra {\em et~al.\/}(2023)Karra, Apte, He \& Scheibe]{Karra2023}
{\sc \au{Karra, Shashank~K.}, \au{Apte, Sourabh~V.}, \au{He, Xiaoliang} \&
  \au{Scheibe, Timothy~D.}} \yr{2023}  \at{Pore-resolved investigation of
  turbulent open channel flow over a randomly packed permeable sediment bed}.
  \jt{J. Fluid Mech.}  \bvol{971}.

\bibitem[Kawano {\em et~al.\/}(2021)Kawano, Motoki, Shimizu \&
  Kawahara]{Kawano2021}
{\sc \au{Kawano, Koki}, \au{Motoki, Shingo}, \au{Shimizu, Masaki} \&
  \au{Kawahara, Genta}} \yr{2021}  \at{Ultimate heat transfer in
  ‘wall-bounded’convective turbulence}.  \jt{J. Fluid Mech.}  \bvol{914},
  \pg{A13}.

\bibitem[Kim {\em et~al.\/}(2020)Kim, Blois, Best \& Christensen]{Kim20}
{\sc \au{Kim, Taehoon}, \au{Blois, Gianluca}, \au{Best, James~L} \&
  \au{Christensen, Kenneth~T}} \yr{2020}  \at{Experimental evidence of
  amplitude modulation in permeable-wall turbulence}.  \jt{J. Fluid Mech.}
  \bvol{887}.

\bibitem[Kong \& Schetz(1982)]{Kong82}
{\sc \au{Kong, F} \& \au{Schetz, J}} \yr{1982} Turbulent boundary layer over
  porous surfaces with different surfacegeometries.  \bt{In {\em 20th Aerospace
  Sciences Meeting\/}},  \pg{p.~30}.

\bibitem[Kuwata \& Suga(2016{\natexlab{{\em a\/}}})]{Kuwata16a}
{\sc \au{Kuwata, Y} \& \au{Suga, K}} \yr{2016{\natexlab{{\em a\/}}}}
  \at{Lattice boltzmann direct numerical simulation of interface turbulence
  over porous and rough walls}.  \jt{Int. J. Heat Fluid Flow}  \bvol{61},
  \pg{145--157}.

\bibitem[Kuwata \& Suga(2016{\natexlab{{\em b\/}}})]{Kuwata16b}
{\sc \au{Kuwata, Yusuke} \& \au{Suga, Kazuhiko}} \yr{2016{\natexlab{{\em
  b\/}}}}  \at{Transport mechanism of interface turbulence over porous and
  rough walls}.  \jt{Flow Turbul. Combust.}  \bvol{97}~(4),  \pg{1071--1093}.

\bibitem[L{\=a}cis {\em et~al.\/}(2020)L{\=a}cis, Sudhakar, Pasche \&
  Bagheri]{Lacis20}
{\sc \au{L{\=a}cis, Uǧis}, \au{Sudhakar, Y}, \au{Pasche, Simon} \&
  \au{Bagheri, Shervin}} \yr{2020}  \at{Transfer of mass and momentum at rough
  and porous surfaces}.  \jt{J. Fluid Mech.}  \bvol{884},  \pg{A21}.

\bibitem[Li {\em et~al.\/}(2021)Li, de~Silva, Chung, Pullin, Marusic \&
  Hutchins]{Li2021}
{\sc \au{Li, Mogeng}, \au{de~Silva, Charitha~M}, \au{Chung, Daniel},
  \au{Pullin, Dale~I}, \au{Marusic, Ivan} \& \au{Hutchins, Nicholas}} \yr{2021}
   \at{Experimental study of a turbulent boundary layer with a rough-to-smooth
  change in surface conditions at high reynolds numbers}.  \jt{J. Fluid Mech.}
  \bvol{923},  \pg{A18}.

\bibitem[Ligrani \& Moffat(1986)]{Ligrani86}
{\sc \au{Ligrani, Phillip~M} \& \au{Moffat, Robert~J}} \yr{1986}  \at{Structure
  of transitionally rough and fully rough turbulent boundary layers}.  \jt{J.
  Fluid Mech.}  \bvol{162},  \pg{69--98}.

\bibitem[Liu \& Prosperetti(2011)]{Liu2011}
{\sc \au{Liu, Qianlong} \& \au{Prosperetti, Andrea}} \yr{2011}
  \at{Pressure-driven flow in a channel with porous walls}.  \jt{J. Fluid
  Mech.}  \bvol{679},  \pg{77--100}.

\bibitem[Luchini(1996)]{Luchini96}
{\sc \au{Luchini, P}} \yr{1996} Reducing the turbulent skin friction.  \bt{In
  {\em Computational methods in applied sciences' 96 (Paris, 9-13 September
  1996)\/}},  \pg{pp. 465--470}.

\bibitem[Lévy(1983)]{Levy1983}
{\sc \au{Lévy, Thérèse}} \yr{1983}  \at{Fluid flow through an array of fixed
  particles}.  \jt{Int. J. Eng. Sci.}  \bvol{21}~(1),  \pg{11–23}.

\bibitem[Manes {\em et~al.\/}(2011)Manes, Poggi \& Ridolfi]{Manes11}
{\sc \au{Manes, Costantino}, \au{Poggi, Davide} \& \au{Ridolfi, Luca}}
  \yr{2011}  \at{Turbulent boundary layers over permeable walls: scaling and
  near-wall structure}.  \jt{J. Fluid Mech.}  \bvol{687},  \pg{141--170}.

\bibitem[Manes {\em et~al.\/}(2009)Manes, Pokrajac, McEwan \& Nikora]{Manes09}
{\sc \au{Manes, Costantino}, \au{Pokrajac, Dubravka}, \au{McEwan, Ian} \&
  \au{Nikora, Vladimir}} \yr{2009}  \at{Turbulence structure of open channel
  flows over permeable and impermeable beds: A comparative study}.  \jt{Phys.
  Fluids}  \bvol{21}~(12),  \pg{125109}.

\bibitem[Moody(1944)]{Moody1944}
{\sc \au{Moody, Lewis~F.}} \yr{1944}  \at{Friction factors for pipe flow}.
  \jt{J. Fluids Eng.}  \bvol{66}~(8),  \pg{671--678}.

\bibitem[Naqvi \& Bottaro(2021)]{Naqvi21}
{\sc \au{Naqvi, Sahrish~B} \& \au{Bottaro, Alessandro}} \yr{2021}
  \at{Interfacial conditions between a free-fluid region and a porous medium}.
  \jt{Int. J. Multiph. Flow}  \bvol{141},  \pg{103585}.

\bibitem[Orlandi \& Leonardi(2006)]{Orlandi06JoT}
{\sc \au{Orlandi, P.} \& \au{Leonardi, S.}} \yr{2006}  \at{{DNS} of turbulent
  channel flows with two- and three-dimensional roughness}.  \jt{J. Turbul.}
  \bvol{7},  \pg{N73}.

\bibitem[Orlandi \& Leonardi(2008)]{Orlandi08}
{\sc \au{Orlandi, P.} \& \au{Leonardi, S.}} \yr{2008}  \at{Direct numerical
  simulation of three-dimensional turbulent rough channels: parameterization
  and flow physics}.  \jt{J. Fluid Mech.}  \bvol{606},  \pg{399--415}.

\bibitem[Orlandi {\em et~al.\/}(2006)Orlandi, Leonardi \& Antonia]{Orlandi06}
{\sc \au{Orlandi, Paolo}, \au{Leonardi, S} \& \au{Antonia, RA}} \yr{2006}
  \at{Turbulent channel flow with either transverse or longitudinal roughness
  elements on one wall}.  \jt{J. Fluid Mech.}  \bvol{561},  \pg{279--305}.

\bibitem[Prinos {\em et~al.\/}(2003)Prinos, Sofialidis \&
  Keramaris]{Prinos2003}
{\sc \au{Prinos, Panayotis}, \au{Sofialidis, Dimitrios} \& \au{Keramaris,
  Evangelos}} \yr{2003}  \at{Turbulent flow over and within a porous bed}.
  \jt{J. Hydraul. Eng.}  \bvol{129}~(9),  \pg{720--733}.

\bibitem[Rao \& Jin(2022)]{Rao2022}
{\sc \au{Rao, Feixiong} \& \au{Jin, Yan}} \yr{2022}  \at{Possibility for
  survival of macroscopic turbulence in porous media with high porosity}.
  \jt{J. Fluid Mech.}  \bvol{937},  \pg{A17}.

\bibitem[Rosti {\em et~al.\/}(2015)Rosti, Cortelezzi \& Quadrio]{Rosti15}
{\sc \au{Rosti, Marco~E}, \au{Cortelezzi, Luca} \& \au{Quadrio, Maurizio}}
  \yr{2015}  \at{Direct numerical simulation of turbulent channel flow over
  porous walls}.  \jt{J. Fluid Mech.}  \bvol{784},  \pg{396--442}.

\bibitem[Rousseau \& Ancey(2020)]{Rousseau2020}
{\sc \au{Rousseau, Gauthier} \& \au{Ancey, Christophe}} \yr{2020}  \at{Scanning
  piv of turbulent flows over and through rough porous beds using refractive
  index matching}.  \jt{Exp. Fluids}  \bvol{61}~(8),  \pg{172}.

\bibitem[Sanchez-Palencia(1982)]{SanchezP1982}
{\sc \au{Sanchez-Palencia, E.}} \yr{1982}  \at{On the asymptotics of the fluid
  flow past an array of fixed obstacles}.  \jt{Int. J. Eng. Sci.}
  \bvol{20}~(12),  \pg{1291--1301}.

\bibitem[Sharma(2020)]{Sharma20thesis}
{\sc \au{Sharma, Akshath}} \yr{2020}  \at{Turbulent flows over canopies}. PhD
  thesis, University of Cambridge.

\bibitem[Sharma \& Garc{\'i}a-Mayoral(2020{\natexlab{{\em
  a\/}}})]{Sharma20sparse}
{\sc \au{Sharma, Akshath} \& \au{Garc{\'i}a-Mayoral, Ricardo}}
  \yr{2020{\natexlab{{\em a\/}}}}  \at{Scaling and dynamics of turbulence over
  sparse canopies}.  \jt{J. Fluid Mech.}  \bvol{888},  \pg{A1}.

\bibitem[Sharma \& Garc{\'i}a-Mayoral(2020{\natexlab{{\em
  b\/}}})]{Sharma20dense}
{\sc \au{Sharma, Akshath} \& \au{Garc{\'i}a-Mayoral, Ricardo}}
  \yr{2020{\natexlab{{\em b\/}}}}  \at{Turbulent flows over dense filament
  canopies}.  \jt{J. Fluid Mech.}  \bvol{888},  \pg{A2}.

\bibitem[Sharma {\em et~al.\/}(2017)Sharma, G{\'o}mez-de Segura \&
  Garc{\'i}a-Mayoral]{Sharma17}
{\sc \au{Sharma, Akshath}, \au{G{\'o}mez-de Segura, Garazi} \&
  \au{Garc{\'i}a-Mayoral, Ricardo}} \yr{2017} Linear stability analysis of
  turbulent flows over dense filament canopies.  \bt{In {\em Tenth
  International Symposium on Turbulence and Shear Flow Phenomena\/}}. Begel
  House Inc.

\bibitem[Shen {\em et~al.\/}(2020)Shen, Yuan \& Phanikumar]{Shen20}
{\sc \au{Shen, Guangchen}, \au{Yuan, Junlin} \& \au{Phanikumar, Mantha~S}}
  \yr{2020}  \at{Direct numerical simulations of turbulence and hyporheic
  mixing near sediment--water interfaces}.  \jt{J. Fluid Mech.}  \bvol{892},
  \pg{A20}.

\bibitem[Stroh {\em et~al.\/}(2020)Stroh, Sch{\"a}fer, Frohnapfel \&
  Forooghi]{Stroh2020}
{\sc \au{Stroh, A}, \au{Sch{\"a}fer, K}, \au{Frohnapfel, B} \& \au{Forooghi,
  P}} \yr{2020}  \at{Rearrangement of secondary flow over spanwise
  heterogeneous roughness}.  \jt{J. Fluid Mech.}  \bvol{885},  \pg{R5}.

\bibitem[Sudhakar {\em et~al.\/}(2021)Sudhakar, Lācis, Pasche \&
  Bagheri]{Sudhakar2021}
{\sc \au{Sudhakar, Y.}, \au{Lācis, Ugis}, \au{Pasche, Simon} \& \au{Bagheri,
  Shervin}} \yr{2021}  \at{Higher-order homogenized boundary conditions for
  flows over rough and porous surfaces}.  \jt{Transp. Porous Media}
  \bvol{136}~(1),  \pg{1--42}.

\bibitem[Suga {\em et~al.\/}(2010)Suga, Matsumura, Ashitaka, Tominaga \&
  Kaneda]{Suga10}
{\sc \au{Suga, K}, \au{Matsumura, Y}, \au{Ashitaka, Y}, \au{Tominaga, S} \&
  \au{Kaneda, M}} \yr{2010}  \at{Effects of wall permeability on turbulence}.
  \jt{Int. J. Heat Fluid Flow}  \bvol{31}~(6),  \pg{974--984}.

\bibitem[Suga {\em et~al.\/}(2011)Suga, Mori \& Kaneda]{Suga11}
{\sc \au{Suga, K}, \au{Mori, M} \& \au{Kaneda, M}} \yr{2011}  \at{Vortex
  structure of turbulence over permeable walls}.  \jt{Int. J. Heat Fluid Flow}
  \bvol{32}~(3),  \pg{586--595}.

\bibitem[Taylor(1971)]{Taylor71}
{\sc \au{Taylor, GI~t}} \yr{1971}  \at{A model for the boundary condition of a
  porous material. part 1}.  \jt{J. Fluid Mech.}  \bvol{49}~(2),
  \pg{319--326}.

\bibitem[Voermans {\em et~al.\/}(2017)Voermans, Ghisalberti \&
  Ivey]{Voermans17}
{\sc \au{Voermans, JJ}, \au{Ghisalberti, M} \& \au{Ivey, GN}} \yr{2017}
  \at{The variation of flow and turbulence across the sediment--water
  interface}.  \jt{J. Fluid Mech.}  \bvol{824},  \pg{413--437}.

\bibitem[Wang {\em et~al.\/}(2021)Wang, Chu, Lozano-Dur{\'a}n, Helmig \&
  Weigand]{Wang21}
{\sc \au{Wang, Wenkang}, \au{Chu, Xu}, \au{Lozano-Dur{\'a}n, Adri{\'a}n},
  \au{Helmig, Rainer} \& \au{Weigand, Bernhard}} \yr{2021}  \at{Information
  transfer between turbulent boundary layers and porous media}.  \jt{J. Fluid
  Mech.}  \bvol{920},  \pg{A21}.

\bibitem[Wangsawijaya {\em et~al.\/}(2023)Wangsawijaya, Jaiswal \&
  Ganapathisubramani]{Wangsawijaya23}
{\sc \au{Wangsawijaya, D.D.}, \au{Jaiswal, P.} \& \au{Ganapathisubramani, B.}}
  \yr{2023}  \at{Towards decoupling the effects of permeability and roughness
  on turbulent boundary layers}.  \jt{J. Fluid Mech.}  \bvol{967},  \pg{R2}.

\bibitem[Wangsawijaya {\em et~al.\/}(2020)Wangsawijaya, Baidya, Chung, Marusic
  \& Hutchins]{Wangsawijaya2020}
{\sc \au{Wangsawijaya, D.~D.}, \au{Baidya, R.}, \au{Chung, D.}, \au{Marusic,
  I.} \& \au{Hutchins, N.}} \yr{2020}  \at{The effect of spanwise wavelength of
  surface heterogeneity on turbulent secondary flows}.  \jt{J. Fluid Mech.}
  \bvol{894},  \pg{A7}.

\bibitem[Xie {\em et~al.\/}(2024)Xie, Fairhall \& Garc\'{i}a-Mayoral]{Xie2024}
{\sc \au{Xie, W.}, \au{Fairhall, C.~T.} \& \au{Garc\'{i}a-Mayoral, R.}}
  \yr{2024}  \at{Resolving turbulence and drag over textured surfaces using
  texture-less simulations: the case of slip/no-slip textures}.  \jt{J. Fluid
  Mech.}  \bvol{1000},  \pg{A36}.

\bibitem[Zagni \& Smith(1976)]{Zagni76}
{\sc \au{Zagni, Anthony~FE} \& \au{Smith, Kenneth~VH}} \yr{1976}  \at{Channel
  flow over permeable beds of graded spheres}.  \jt{J. Hydraul. Div.}
  \bvol{102}~(2),  \pg{207--222}.

\bibitem[Zhang \& Prosperetti(2009)]{Zhang2009}
{\sc \au{Zhang, Quan} \& \au{Prosperetti, Andrea}} \yr{2009}
  \at{Pressure-driven flow in a two-dimensional channel with porous walls}.
  \jt{J. Fluid Mech.}  \bvol{631},  \pg{1--21}.

\bibitem[Zippe \& Graf(1983)]{Zippe83}
{\sc \au{Zippe, Hans~J} \& \au{Graf, Walter~H}} \yr{1983}  \at{Turbulent
  boundary-layer flow over permeable and non-permeable rough surfaces}.  \jt{J.
  Hydraul. Res.}  \bvol{21}~(1),  \pg{51--65}.

\end{thebibliography}

\end{document}